\documentclass[11pt, a4paper]{article}
\usepackage{jstyle, enumitem}
\usepackage{hyperref}
\usepackage{slashed,framed}
\usepackage{empheq}
\usetikzlibrary{calc,decorations.markings}
\usepackage[normalem]{ulem}

\usepackage{amsfonts, amssymb, amsmath, amsthm, stmaryrd, amsbsy,
  mathrsfs, genyoungtabtikz, mathdots, dsfont, relsize}
\usepackage{color, fancybox, empheq, graphicx}
\usepackage{mdframed}
\usepackage[strict]{changepage}
\newcommand{\Pd}[1]{\mathcal{P}_{#1}}
\newcommand*\Bell{\ensuremath{\boldsymbol\ell}}
\newcommand{\ez}{\epsilon_0}

\newcommand{\half}{\mathbf{\tfrac 12}}

\newcommand{\C}{\mathbb{C}}
\newcommand{\Ca}{\mathcal{C}_2}

\renewcommand{\D}{\mathcal{D}}

\newcommand{\W}{\mathcal{W}}
\newcommand{\R}{\mathbb{R}}

\newcommand{\N}{\mathbb{N}}
\newcommand{\Z}{\mathbb{Z}}

\newcommand{\V}{\mathcal{V}}

\renewcommand{\U}{\mathcal{U}}
\newcommand{\Sn}{\mathcal{S}}
\newcommand{\uv}{\mathsf{e}}
\newcommand{\id}{\mathlarger{\mathds{1}}}

\newcommand{\rac}{\mathrm{Rac}}
\newcommand{\arac}{\overline{\mathrm{Rac}}}
\newcommand{\di}{\mathrm{Di}}
\newcommand{\adi}{\overline{\mathrm{Di}}}

\newcommand{\1}{\mathbf{1}}
 \newtheorem*{theorem*}{Theorem}
 \newtheorem*{lemma*}{Lemma}
\theoremstyle{definition}

\newtheorem*{definition*}{Definition}
\definecolor{rougef}{rgb}{0.56,0,0}
\definecolor{vertf}{rgb}{0,0.5,0}
\definecolor{bleuf}{rgb}{0,0,0.8}

\newcommand{\ie}{{\it i.e.\ }}
\definecolor{shadowcolor}{RGB}{0, 0, 102}
\newcommand*\eqbox[1]{%
  \setlength\shadowsize{2pt}%
  \shadowbox*{#1}% 
}
\definecolor{myblue}{rgb}{.8, .8, 1}
\Yboxdim{8pt} 
\Ylinethick{0.9pt}
\usepackage{tikz, tikz-cd, animate}
\usetikzlibrary{arrows, matrix}
\author[a]{Thomas BASILE}
\author[b]{\qquad Xavier BEKAERT}
\author[a]{\qquad Euihun JOUNG}

\affiliation[a]{Department of Physics and Research Institute of Basic
  Science, \\
	Kyung Hee University,\\ Seoul 02447, Korea}
\affiliation[b]{Institut Denis Poisson,\\
Universit\'e de Tours, Universit\'e d'Orl\'eans, CNRS,\\
  Parc de Grandmont, 37200 Tours, France}

\emailAdd{thomas.basile@khu.ac.kr}
\emailAdd{xavier.bekaert@lmpt.univ-tours.fr}
\emailAdd{euihun.joung@khu.ac.kr}

\begin{document}

\title{\centering 
Twisted Flato-Fronsdal Theorem \\ for Higher-Spin Algebras}

\abstract{We explore the relation between the singleton and adjoint
  modules of higher- spin algebras via $so(2,d)$ characters. In order
  to relate the tensor product of the singleton and its dual to the
  adjoint module, we consider a heuristic formula involving
  symmetrization over the variables of the character. We show that our
  formula reproduces correctly the adjoint-module character for type-A
  (and its high-order extensions) and type-B higher-spin gravity
  theories in any dimension.  Implications and subtleties of this
  symmetrization prescription in other models are discussed.
}

\maketitle

%%%%%%%%%%%%%%%%%%%%%%
\section{Introduction}
\label{sec:intro}
%%%%%%%%%%%%%%%%%%%%%%

In 1963, Dirac discovered two ``remarkable representations''
\cite{Dirac:1963ta} of the isometry group of the four-dimensional anti
de Sitter spacetime $AdS_4$, which are the ultrashort modules of
$SO(2,3)$ nowadays known as the (Dirac) ``singletons''.  Fifteen years
later, Flato and Fronsdal showed that ``one massless particle equals
two Dirac singletons'' \cite{Flato:1978qz}, \ie the tensor product of
two singletons gives an infinite (direct) sum of massless particles of
all integer spins. This result, often referred to as Flato-Fronsdal
theorem, together with its generalizations (to supersymmetric and some
higher-dimensional cases
\cite{Gunaydin:1984fk,Gunaydin:1984vz,Gunaydin:1984wc,Gunaydin:1985tc},
to arbitrary dimensions \cite{Angelopoulos:1999bz, Vasiliev:2004cm},
to multilinetons \cite{Bekaert:2013zya, Basile:2014wua} as well as to
arbitrary spin singletons \cite{Dolan:2005wy}) has provided an
important guiding principle for higher-spin gravity as it dictates
consistent field contents of the theory prior to the actual
construction of its dynamical equations.
 
Another cornerstone for higher-spin gravity theories is the
higher-spin algebra playing the role of the global symmetry algebra
associated to the gauge symmetry underlying the theory. Fradkin and
Vasiliev first constructed a consistent higher-spin algebra
\cite{Fradkin:1986ka}, upon which the latter author obtained a set of
nonlinear field equations describing interacting massless higher-spin
fields propagating around $AdS_4$ \cite{Vasiliev:1990en} (see
e.g. \cite{Bekaert:2005vh, Didenko:2014dwa} for reviews of these
equations and their higher-dimensional generalizations). Later on, it
was realized that the higher-spin algebra can be viewed as the algebra
of symmetries, namely the endomorphisms, of singletons
\cite{Eastwood:2002su}. This point of view allowed a wide range of
generalizations, notably to dynamical equations for completely
symmetric tensor gauge fields in higher dimensions
\cite{Vasiliev:2003ev}.

To recapitulate, there are three key modules of the higher-spin
algebra: the singleton (which plays a role analogous to the
fundamental representation), the Hilbert space of the theory (the
so-called ``twisted-adjoint module'') and finally the vector space of
the higher-spin algebra itself (the ``adjoint module''). The second
module is the tensor product of the first one with itself, whereas the
last module corresponds to the endomorphisms of the first one. In this
sense, the adjoint module was identified in \cite{Iazeolla:2008ix}
with the tensor product of the singleton (denoted by $\rm Sng$) with
its dual (denoted by $\overline{\rm Sng}$). In \cite{Iazeolla:2008ix},
this naive relation between the tensor product module ${\rm
  Sng}\otimes \overline{\rm Sng}$ and the infinite sum of
finite-dimensional modules spanning the adjoint module was referred to
as the ``twisted Flato-Fronsdal theorem''.  We will use here the same
terminology for the refined relation that we will propose below.
 
A very convenient tool to handle various modules and their operations
is the Lie algebra character. This mathematical object is closely
related to the (one loop) partition function in physics (see
e.g. \cite{Gupta:2012he, Lal:2012ax, Giombi:2014yra, Giombi:2014iua,
  Campoleoni:2015qrh, Bae:2016hfy, Bae:2016rgm, Gunaydin:2016amv,
  Bae:2017fcs} in the context of higher-spin holography) in the
presence of chemical potentials for angular momenta. As partition
functions contain most of physical information about the system under
consideration, one can expect to be able to use characters in many
mathematical analyses about the system. Indeed, the Flato-Fronsdal
theorem was derived originally in a handy way using the $so(2,3)$
character of Dirac singletons and those of massless spin-$s$
representations.
 
In this note, we reconsider the twisted Flato-Fronsdal theorem, that
is, the relation between the adjoint module and the tensor product
module ${\rm Sng}\otimes \overline{\rm Sng}$.  Since ${\rm Sng}$ and
$\overline{\rm Sng}$ are respectively a lowest and a highest weight
module of infinite dimension, the decomposition of their tensor
product is subtle and requires a more careful treatment. In order to
study this issue more concretely, we use the characters of the
relevant modules and work with a prescription in which the characters
can be manipulated in the usual manners.  In this way, we find that
the character of the tensor product module ${\rm Sng}\otimes
\overline{\rm Sng}$ does not coincide with the character of the
adjoint module.  Analyzing in detail the four-dimensional massless
higher-spin algebra, we find that the adjoint module differs from
${\rm Sng}\otimes \overline{\rm Sng}$ and obtain the precise relation
between them by observing that the adjoint module character coincides
in fact with the symmetrization of the ${\rm Sng}\otimes \overline{\rm
  Sng}$ character over the variables of the character.  We examine
this heuristic formula over other higher-spin algebras confirming its
validity in the type-A$_\ell$ and type-B theories in any dimensions,
but mismatches by a few finite-dimensional modules in the
type-B$_{\ell\geqslant 2}$ and type-J cases.  This confirms that the
adjoint module is not given by the simple tensor product ${\rm
  Sng}\otimes \overline{\rm Sng}$, but requires proper amendments,
about which we shall discuss throughout the paper.

The organization of the paper is as follows. In the next section
(Section \ref{sec: 4d}), we sketch the issue with the classical
four-dimensional example. Then, we move to general dimensions in
Section \ref{sec: gen d}.  The cases of lower dimensions (two and
three dimensions) are presented as useful toy models in Section
\ref{low-dim}. We extend our consideration to higher-order and
higher-spin singletons and find some exceptions to our conjecture in
Section \ref{sec: exex}.  The last section contains a brief summary
and discussion of our results. In Appendix \ref{app: detail}, we
collect some technical details on (generalized) Verma modules, while
their Lie algebra characters (see e.g. \cite{Beccaria:2014jxa,
  Basile:2016aen, Bourget:2017kik} for the character formulae of all
irreps of the conformal algebra) are discussed in Appendix
\ref{app:Weyl}.

%%%%%%%%%%%%%%%%%%%%%%%%%%%%%%%%%%%%%%%%%%
\section{Four Dimensions}
\label{sec: 4d}
%%%%%%%%%%%%%%%%%%%%%%%%%%%%%%%%%%%%%%%%%%

Let us first consider four-dimensional higher-spin gravity whose
equations were initially constructed by Vasiliev in
\cite{Vasiliev:1990en}, and whose spectrum is concerned by Flato and Fronsdal's
original result \cite{Flato:1978qz}. The four-dimensional Vasiliev
equations contain an ``interaction ambiguity'' (first exhibited in
  \cite{Vasiliev:1992av} and further studied in
  e.g. \cite{Vasiliev:1999ba, Vasiliev:2001ur, Engquist:2002vr,
    Sezgin:2011hq}), given by a series of parameters.  If the theory
  is required to have a definite parity, there remain only two allowed
  values for those parameters.  These two choices are referred to as
  type A and type B, respectively. By now, it is understood that the
  type-A theory with the Neumann boundary condition\footnote{In the Poincar\'e patch of
    AdS$_4$, the bulk scalar with Dirichlet condition approaches the
    boundary as $\phi(z, \vec x) \underset{z \rightarrow 0}{\sim} z \,
    \varphi(\vec x)\,$ whereas the one with Neumann condition does as
    $\phi(z, \vec x) \underset{z \rightarrow 0}{\sim} z^2 \,
    \varphi(\vec x)\,$.} for the bulk
  scalar corresponds to the free scalar $CFT_3$ \cite{Klebanov:2002ja}, whereas the type-B theory
  with the Dirichlet condition  corresponds to the free spinor $CFT_3$ \cite{Sezgin:2002rt}.  

In the following, we shall review some details of the three modules of
the higher-spin algebra --- singleton, twisted-adjoint and adjoint
modules. As they are also modules of the isometry subalgebra
$so(2,3)$, it will be useful to treat them using $so(2,3)$ irreducible
representations (irreps). For that, the basic object to consider is
the lowest-weight module $\cV(\Delta,s)$\,, whose character is given
by
\begin{equation}
  \chi^{so(2,3)}_{\cV(\Delta,s)}(q,x) =
  \tr_{\cV(\Delta,s)}\left(q^{E}\,x^{J_3}\right) =
  \frac{q^{\Delta}}{(1-q)(1-q\,x)(1-q\,x^{-1})}\, \chi^{so(3)}_s(x)\,,
\end{equation}
where the spin-$s$ character of $so(3)$ is
\begin{equation}
  \chi^{so(3)}_s(x) := \frac{x^{s+\frac12} -
    x^{-s-\frac12}}{x^{\frac12} - x^{-\frac12}}\, .
\end{equation}
Here $E$ and $J_3$ are the Cartan generators of
$so(2,3)$\, (see Appendix \ref{app: detail} where our conventions are
summarized).  For the purpose of the current section, it is sufficient
to take the above formula for granted. Details about the derivation in any dimension
will be provided in Section \ref{sec: gen d}. In terms of the
variables $q=e^{-\b}$ and $x=e^{i\,\alpha}$, this $so(2,3)$ character
reads
\begin{equation}
  \chi^{so(2,3)}_{\cV(\Delta,s)}(\b,\a) =
  \frac{e^{-(\Delta-\frac32)\,\b}}{4\, \sinh\frac\b2\,
    (\cosh\b-\cos\a)}\,\chi^{so(3)}_s(\a)\,, \quad {\rm with} \quad
  \chi^{so(3)}_s(\a)=\frac{\sin(s+\frac12)\a}{\sin\frac\a2}\,,
  \label{LWM}
\end{equation}
%that is,
{and}, the spin-$s$ $so(3)$ character coincides with the
  Dirichlet kernel.
        
%****************************%
\subsection*{Singleton Module}
%****************************%
The free massless scalar and spinor fields in three dimensions are
nothing but the singleton representations that Dirac had found in
\cite{Dirac:1963ta}. Flato and Fronsdal named the latter and former as
``Di'' and ``Rac'', respectively. In terms of the lowest-weight module
$\cV(\Delta,s)$, the singletons Rac and Di correspond to the
quotients,
\begin{equation}
  {\rm Rac}:=\cD(\tfrac12,0)=\cV(\tfrac12,0)/\cV(\tfrac52,0)\,, \qquad
  {\rm Di}:= \cD(1,\tfrac12)=\cV(1,\tfrac12)/\cV(2,\tfrac12)\,.
\end{equation}
These representations are ``ultrashort'', even ``minimal'' in the
sense (which can be made mathematically precise \cite{Joseph1976,
  Fernando:2015tiu, Joung:2014qya}) that they can be described as
three-dimensional on-shell fields. Using the character \eqref{LWM} of
$\cV(\Delta,s)$, it is simple to derive the characters of the
singletons. They are
\begin{equation}
  \chi^{so(2,3)}_{\rm
    Rac}(\b,\a)=\frac{\cosh\frac\b2}{\cosh\b-\cos\a}\,, \qquad
  \chi^{so(2,3)}_{\rm
    Di}(\b,\a)=\frac{\cos\frac\a2}{\cosh\b-\cos\a}\,.
  \label{DiRac char}
\end{equation}
It is also instructive to study the oscillator realization of the
singletons.  Using two sets of oscillators $(a, a^\dagger)$ and
$(b,b^\dagger)$ with canonical commutation relations $[a,
  a^\dagger]=1=[b,b^\dagger]$\,, the generators of $so(2,3)$ can be
realized as \cite{Gunaydin:1983yj,Gunaydin:1983cc,Konstein1989}
\begin{eqnarray}
  & E=\frac12\left(a^\dagger\,a+b^\dagger\,b+1\right),\quad
  J_3=\frac12\left(a^\dagger\,a-b^\dagger\,b\right),\nn &
  J_+=a^\dagger\,b\,,\quad L^{-}_1=\frac12\left(a^2+b^2\right)\,,\quad
  L^{-}_2=-\frac{i}2\left(a^2-b^2\right)\,,\quad L^-_3=a\,b\,,
\end{eqnarray}
with $so(3) = {\rm span}\{J_+, J_-, J_3\}$ and where $L^-_a$
  ($a=1,2,3$) are the lowering operators with respect to $E$. The remaining
generators are the Hermitian conjugates of the above (see Appendix
\ref{app: detail} for conventions). The Fock states,
\begin{equation}
  |m,n\ra=\frac{(a^\dagger)^m\,(b^\dagger)^n}{\sqrt{m!\,n!}}\,|0,0\ra\,,
  \label{Fock}
\end{equation}
are eigenvectors of the Cartan subalgebra generators $E$ and $J_3$:
\begin{equation}\label{eigenvectso(3,2)}
  E\ket{m,n} = \tfrac12 (m+n+1) \ket{m,n} \, , \qquad
  J_3\ket{m,n} = \tfrac12 (m-n) \ket{m,n}\,.
\end{equation}
The vacuum state $|0,0\ra$ is the the lowest-energy state of the
$\rac$ module, whereas the $\di$ module has lowest-energy module spanned by the doublet
$\{|1,0\ra,|0,1\ra\}$.  Indeed, we have $\di = \D(1,\tfrac12)$,
i.e. the vacuum carries a spin-$\tfrac12$ representation of $so(3)$
and its energy is one.  The lowest-energy states of $\rac$ and $\di$
are annihilated by the lowering operators $L^-_a\,$. The full $\rac$
and $\di$ modules are then freely generated by applying the raising
operators $L^+_a$.  As a consequence $\rac$ and $\di$ are spanned by
states $|m,n\ra$ with even and odd $m+n$, respectively. Using these
results, we can calculate the characters of the singletons as
\begin{eqnarray}
  \chi^{so(2,3)}_{\rac/\di}(\beta,\alpha) =
  \tr_{\rac/\di}\left(e^{-\beta\, E + i\,\alpha\, J_3}\right) =
  \sum_{{\rm even/odd}\ m+n} e^{-\beta\,\frac{m+n+1}2 +
    i\,\alpha\,\frac{m-n}2}\,.
\end{eqnarray}
To perform the sum, we can make the change of variables,
\begin{equation}
  m+n=2s\,,\qquad m-n=2(s-k)\,,
\end{equation}
where $k=0,1,\dots,2s$ and $s\in \N$ for $\rac$ and $s\in \N+\frac12$
for $\di$. Then, we get
\begin{equation}\label{infsum}
  \chi^{so(2,3)}_{\rac/\di}(\beta,\alpha) = \sum_{s \in \N+0/\frac12}
  e^{-\beta\,(s+\frac12)} \sum_{k=0}^{2\,s} e^{i\,\alpha\, (s-k)} =
  \sum_{s \in \N+0/\frac12} e^{-\beta\,(s+\frac12)}\,
  \chi^{so(3)}_s(\alpha)\, .
\end{equation}
The infinite sum in \eqref{infsum} leads to geometric series and one
finally recovers the characters \eqref{DiRac char}.

%**********************************%
\subsection*{Twisted-Adjoint Module}
%**********************************%
All the other irreps $\cD(\Delta,s)$ of $so(2,3)$ in the unitary
region $\Delta\geqslant s+1$ are much ``longer'' and they can be
viewed as the Hilbert space of a four-dimensional on-shell
field.  In particular, the representations describing massless
spin-$s$ particles on $AdS_4$ lie at the unitary bound, and
correspond to the quotients,
\begin{equation}
  \cD(s+1,s)=\cV(s+1,s)/\cV(s+2,s-1)\,,
  \label{massless}
\end{equation}
with the characters,
\begin{equation}
  \chi^{so(2,3)}_{\cD(s+1,s)}(\b,\a)= \frac{e^{-(s-\frac12)\,\b}\,
    \sin(s+\frac12)\a - e^{-(s+\frac12)\,\b}\, \sin(s-\frac12)\a}{4\,
    \sinh\frac\b2\, \sin\frac\a2\, (\cosh\b-\cos\a)}\,.
\end{equation}
Flato and Fronsdal have shown in \cite{Flato:1978qz} the following
rule for the decomposition in irreducible $so(2,3)$-modules of the
tensor product of two Rac or Di:
\begin{equation}
  {\rm Rac}\otimes {\rm Rac} = \bigoplus_{s=0}^\infty\,\cD(s+1,s)\,,
  \qquad {\rm Di}\otimes {\rm Di} = \cD(2,0)\oplus
  \bigoplus_{s=1}^\infty\,\cD(s+1,s)\,.
  \label{FF}
\end{equation}
The right-hand-side of the above equations is nothing but the field
content --- namely, the twisted-adjoint module --- of the type-A and
type-B higher-spin gravity theories, respectively.  This suggests
that the $CFT_3$ operators bilinear in the free massless scalar (Rac)
or spinor (Di) fields --- hence fall in the tensor product of two singletons representations
--- corresponds to the $AdS_4$ massless gauge fields of higher-spin
gravity together with one bulk scalar field (with ``Neumann'' or ``Dirichlet'' boundary conditions, respectively)
\cite{Klebanov:2002ja, Sezgin:2002rt, Leigh:2003gk,
  Sezgin:2003pt}. This tensor product decomposition has been proven
with the help of the $so(2,3)$ characters by checking the following
algebraic identities,
\begin{equation}
  \left(\chi^{so(2,3)}_{\rm Rac}\right)^2=\sum_{s=0}^\infty
  \chi^{so(2,3)}_{\cD(s+1,s)}\,, \qquad \left(\chi^{so(2,3)}_{\rm
    Di}\right)^2=\chi_{\cD(2,0)}+\sum_{s=1}^\infty
  \chi^{so(2,3)}_{\cD(s+1,s)}\,.
\end{equation}

In terms of oscillators, the tensor product of two singletons is realized by doubling the oscillators: $(a_i,a^\dagger_i)$ and $(b_i, b_i^\dagger)$
with $i=1,2$\,. Hence, the twisted-adjoint module is spanned by the
states of the type,
\begin{equation}
  |m, n; p, q \ra = \frac{(a^\dagger_1)^m\, (b^\dagger_1)^n\,
    (a^\dagger_2)^p\, (b^\dagger_2)^q}{\sqrt{m!\, n!\, p!\, q!}}\,
  |0,0;0,0\ra\,.
\end{equation}
 Defining the action of an $so(2,3)$ element $X$ on the singleton Fock
 state $|m,n\ra$ as
\begin{equation}
  X\,|m,n\ra=\sum_{p,q}\,R^{m,n}{}_{p,q}(X)\,|p,q\ra\,,
\end{equation}
 the action of  $X$ on $|m, n; p, q\ra$ gives
\begin{equation}
  X\,|m,n;p,q\ra =\sum_{s,t}\,\big(\,R^{m,n}{}_{s,t}(X)\,|s,t;p,q\ra+
  R^{p,q}{}_{s,t}(X)\,|m,n;s,t\ra\,\big).
\end{equation}
For the decomposition of the twisted-adjoint module into
$so(2,3)$-irreducible ones, one can examine the lowest-weight states
--- that are annihilated by $L_1^-+i\,L^-_2$ and $J_+$ (then,
consequently all $L_a^-$ with $a=1,2,3$ annihilate the state) --- in
this doubled singleton Fock space (aka ``doubleton''),
\begin{equation}
  (a_1^2+a_2^2)\, |\Psi\ra = 0 = (a_1^\dagger\,b_1+a_2^\dagger\,b_2)\,
  |\Psi\ra\,, \qquad |\Psi\ra = \sum_{m,n,p,q}\, c_{m,n,p,q}\,
  |m,n;p,q\ra\,.
\end{equation}
It is simple (see e.g. \cite{Konstein:1989ij, Sezgin:2001zs,
  Bae:2016rgm}) to show that any such a state $|\Psi\ra$ is a linear
combination of the lowest-weight states of $\cD(s+1,s)$ (and
$\cD(2,0)$ for the case of Di) hence confirming the rule \eqref{FF}.

%**************************%
\subsection*{Adjoint Module}
%**************************%
The adjoint module, namely the higher-spin algebra, is spanned by the
higher-spin Killing tensors.  For a given spin $s$, the Killing tensor
is a finite-dimensional module of $so(2,3)$\,. In terms of Young
diagram, it corresponds to the rectangle made of two rows of length $s-1$,
\begin{equation}
  {\footnotesize \gyoung(_6{s-1},_6{s-1})}\,,
  \label{KT}
\end{equation}
whereas in terms of the lowest-weight module it corresponds to the
non-unitary module $\cD(1-s,s-1)$ defined by the following sequence of
quotients,
\begin{eqnarray}
  \cD(1-s,s-1)\eq \cV(1-s,s-1)/\cD(2-s,s)\, , \nn
  \cD(2-s,s)\eq\cV(2-s,s)/\cD(s+1,s)\,,
\end{eqnarray}
where $\cD(s+1,s)$ is defined in \eqref{massless}.  Here we used the
Bernstein-Gel'fand-Gel'fand resolution detailed in
\cite{Shaynkman:2004vu}. Another point of view on this module makes
use of the fact that it is finite-dimensional. The two real Lie
algebras $so(2,3)$ and $so(5)$ are two distinct real forms of the same
complex Lie algebra $so_{\mathbb C}(5)$. The character of the
finite-dimensional $so(5)$-module labeled by the dominant integral
weight $(s-1,s-1)$ reads
\begin{equation}
  \chi^{so(5)}_{(s-1,s-1)}(\a_1,\a_2) = \frac{\sin[(s-\frac12)\a_1]\,
    \sin[(s+\frac12)\a_2]-\sin[(s-\frac12)\a_2]\,
    \sin[(s+\frac12)\a_1]} {2\, \sin\frac{\a_1}2\, \sin\frac{\a_2} 2
    \left(\cos\a_1-\cos\a_2\right)}\,.
  \label{(r,r)}
\end{equation}
Using the above information, we can obtain the corresponding $so(2,3)$
character, which is in fact simply related to the $so(5)$ character
\eqref{(r,r)} as
\begin{equation}
  \chi^{so(2,3)}_{\cD(1-s,s-1)}(\b,\a)=\chi^{so(5)}_{(s-1,s-1)}(i\,\b,\a)\,.
\end{equation}
Collecting all these results, we can calculate the $so(2,3)$ character
of the adjoint module of the higher-spin algebra as
\begin{equation}
  \chi^{so(2,3)}_{\rm Adj}(\b,\a) = \sum_{s=1}^\infty
  \chi^{so(2,3)}_{\cD(1-s,s-1)}(\b,\a)\,.
\end{equation}
This infinite sum of characters involves a trigonometric series which
is not convergent in the classical sense, but which is convergent in
the sense of distribution theory.\footnote{See e.g. the section 6.13
  of the book \cite{Kanwal}, devoted to the summability of Fourier
  series of periodic distributions.} Accordingly, it can be evaluated
using resummation techniques,\footnote{ More precisely, this
  trigonometric series is Cesaro (thus Abel) resummable. For a proof
  of \eqref{trigo}, see e.g. \cite{Knopp} (Chap. XIII, Sect. 60,
  Ex. 5).}
\begin{equation}\label{trigo}
  \sum_{n=1}^\infty\sin(n\,x)=\frac12\,\cot(x/2)\,.
\end{equation} 
Using this formula, we obtain the character of the adjoint module as
\begin{equation}
  \chi^{so(2,3)}_{\rm Adj}(\b,\a) = \frac{\cosh^2\frac\b2 +
    \cos^2\frac\a2}{(\cosh\b - \cos\a)^2}\,.
  \label{Adj char}
\end{equation}
Now the question is whether we can obtain the above character from the
characters of the singletons \eqref{DiRac char}.  If this was
possible in general, for an unknown higher-spin theory dual to a
certain CFT with given spectrum, then we would be able to
systematically identify the corresponding higher-spin algebra.

One of the simplest descriptions of the higher-spin algebra is viewing
it as the algebra of endomorphisms of the singleton module,
\begin{equation} 
  {\rm Adj}={\rm End}({\rm Sng})\,,
  \label{endo}
\end{equation}
where ``Sng'' stands for either the Di or Rac module.  We already know
that the higher-spin algebra is identical both in type-A and type-B
theories.  Let us explore this point in the oscillator realization.
Since the singleton module is the Fock space spanned by $|m,n\ra$
\eqref{Fock}, its endomorphism algebra can be generated by the
operators,
\begin{equation}
  X^{m,n;p,q} = \frac{(a^\dagger)^m\, (b^\dagger)^n\, a^p\, b^q}{
    \sqrt{m!\, n!\, p!\, q!}}\,,
  \label{basisHS}
\end{equation}
with even $m+n+p+q$\,. The above presentation of the higher-spin
algebra is simply related to the more typical realization in terms of
the oscillators $y_\a$ and $\bar y_{\dot \a}$\footnote{The higher-spin
  algebra is the algebra of \it even \rm functions of
  $y_\a$ and $\bar y_{\dot \a}$ endowed with the Moyal star product.}
by
\begin{equation}
  y_1=a+b^\dagger\,, \quad y_2 = i\,(a^\dagger-b)\,, \quad \bar
  y_{\dot \a} = (y_\a)^\dagger\,.
\end{equation}
 The action of an $so(2,3)$ element $X$
on this state is, by definition of the adjoint representation,
\begin{equation}
  X\rhd X^{m,n;p,q} = [X\,, X^{m,n;p,q}]\,.
\end{equation}
However, this cannot be written easily in terms of the singleton
representation $R^{p,q}_{m,n}(X)$\,.  What is more naturally connected
to the singleton representation is the basis,
\begin{equation}
  T^{m,n;p,q} = |m,n\ra\la p,q| = \frac{(a^\dagger)^m\,
    (b^\dagger)^n}{\sqrt{m!\, n!}}\, |0,0\ra\la0,0|\, \frac{a^p\,
    b^q}{\sqrt{p!\, q!}}\,,
\end{equation}
on which an $so(2,3)$ element $X$ acts as
\begin{equation}
  X\rhd T^{m,n;p,q} = \sum_{s,t}\,R^{m,n}{}_{s,t}(X)\,T^{s,t;p,q}
  -\overline{R^{p,q}{}_{s,t}(X^\dagger)}\,T^{m,n;s,t}\,.
\end{equation}
Hence, in this basis, it becomes clear that the adjoint module is the
tensor product of the singleton module --- represented by
$R_{s,t}^{m,n}(X)$ --- and its dual module --- represented
by $-\overline{R_{s,t}^{p,q}(X^\dagger)}$.  In order to relate
$T^{m,n;p,q}$ to the more standard basis $X^{m,n;p,q}$, we need to
realize the vacuum projector $|0,0\ra\la0,0|$ as a function of
oscillators,
\begin{equation}
  |0,0\ra\la0,0| = \Pi_{\rm vac}(a, a^\dagger)\, \Pi_{\rm
    vac}(b,b^\dagger)\,.
\end{equation}
By imposing the conditions,
\begin{equation}
  \Pi_{\rm vac}^\dagger = \Pi_{\rm vac}\,, \quad \Pi_{\rm
    vac}\,a^\dagger = 0\,, \quad \Pi_{\rm vac}^2 =
  \Pi_{\rm vac}\,,
\end{equation}
one can determine it as
\begin{equation}
  \Pi_{\rm vac}(a,a^\dagger) = \sum_{n=0}^\infty \frac{(-1)^n}{n!}\,
  (a^\dagger)^n\,a^n\,.
\end{equation}
Therefore, the $T^{m,n;p,q}$ basis is related to the $X^{m,n;p,q}$
basis as an infinite linear combination,
\begin{equation}\label{lincombHSgen}
  T^{m,n;p,q} = \sum_{s,t=0}^\infty
  \frac{(-1)^{s+t}}{\sqrt{C^{m+s}_s\, C^{p+s}_s\, C^{n+t}_t\,
      C^{q+t}_t}}\, X^{m+s,n+t;p+s,q+t}\,,
\end{equation}
where $C^m_n$ is the binomial coefficient. If we restrict the
higher-spin algebra to all finite linear combinations of $X^{m,n;p,q}$
--- hence polynomials in the oscillators --- then the basis
$T^{m,n;p,q}$ does not belong to the higher-spin algebra. In other
words, the finite linear combinations of $X^{m,n;p,q}$ and
$T^{m,n;p,q}$ give two distinct endomorphism algebras.
 This subtlety
arises due to the fact that we are dealing with infinite-dimensional
spaces.

Having this subtlety in mind, let us proceed further.  From the
viewpoint of the endomorphisms in the $T^{m,n;p,q}$ basis, one would
expect the adjoint module to be the tensor product of a singleton and
its anti-singleton (as first pointed out in \cite{Iazeolla:2008ix}):
\begin{equation} \label{=?}
  {\rm Adj}\overset{?}{=}{\rm Sng}\otimes \overline{{\rm Sng}}\,,
\end{equation}
where we put the question mark at the equality because of an
inconsistency we shall face soon below.  The anti-singleton,
denoted by ``$\overline{\rm Sng}$'', is a highest-weight module with
maximal energy $-E_0$, whereas the singleton ``Sng'' is a
lowest-weight module with minimal energy $E_0$ (see the section
\ref{sec: gen d} for additional comments on the definition of
anti-singletons).  From the clear relation between ${\rm Sng}$ and
$\overline{\rm Sng}$\,, we can relate the character of the
anti-singletons to that of the singletons as
\begin{equation}\label{identity2+3}
  \chi^{so(2,3)}_{\overline{\rm Sng}}(\b,\a) = \chi^{so(2,3)}_{\rm
    Sng}(-\b,-\a) = \chi^{so(2,3)}_{\rm Sng}(\b,\a)\,.
\end{equation}
The last equality holds because the singleton characters are even
functions of $\b$ and $\a$ (see \eqref{DiRac char}).  If all the above
discussions were free from subtleties, we should be able to reproduce
the character of the adjoint module \eqref{Adj char} as the product of
the singleton and anti-singleton characters.  However, the identity
\eqref{identity2+3} already shows that it cannot be so, because the
adjoint and twisted adjoint modules have different characters (since
they are not isomorphic). More explicitly, we find the following
discrepancies
\begin{eqnarray}
  && \chi^{so(2,3)}_{\rm Adj}(\b,\a) = \frac{\cosh^2\frac\b2 +
    \cos^2\frac\a2}{(\cosh\b - \cos\a)^2}\,\nn && \neq
  \chi^{so(2,3)}_{\rm Rac}(\b,\a)\, \chi^{so(2,3)}_{\overline{\rm
      Rac}}(\b,\a) = \frac{\cosh^2\frac\b2}{(\cosh\b - \cos\a)^2} \nn
  &&\neq \chi^{so(2,3)}_{\rm Di}(\b,\a)\,
  \chi^{so(2,3)}_{\overline{\rm Di}}(\b,\a) =
  \frac{\cos^2\frac\a2}{(\cosh\b-\cos\a)^2}\,.
  \label{discrep}
\end{eqnarray}
What went wrong? There are several potential sources of discrepancies.
First, it might be due to the problem of change of basis between
$X^{m,n;p,q}$ and $T^{m,n;p,q}$\,. Second, it might be a problem of
characters: the lowest-weight modules and the highest-weight modules
have different radius of convergence for $q$ or, equivalently, for
$\b$\,. The former one converges for $\b>0$ while the latter one does
so for $\b<0$\,. Once the infinite series are evaluated in the
convergent region of $\b$, this region can be analytically continued
to the outer region.  However, there might be subtleties in handling
the characters of lowest-weight modules and highest-weight modules
simultaneously. To give away the bottom line already, various
considerations (that are presented below) indicate that the relation
\eqref{=?} itself, namely the naive twisted Flato-Fronsdal theorem,
should be modified.

In order to understand better this discrepancy, let us redo the
character computations using the oscillator realization.  The
higher-spin algebra is spanned by the elements $X^{m,n;p,q}$ with
$m+n+p+q \in 2\,\N$, as defined in \eqref{basisHS}, which also form a
basis. In the present case, the $so(2,3)$ subalgebra acts on the
elements of the higher-spin algebra through the adjoint action. The
generators $X^{m,n;p,q}$ are also eigenvectors of $E$ and $J_3$:
\begin{equation}
  [E, X^{m,n;p,q}] = \frac{m+n-p-q}2\, X^{m,n;p,q} \, , \qquad [J_3,
    X^{m,n;p,q}] = \frac{m-n-p+q}2\,X^{m,n;p,q} \, .
\end{equation}
This implies that the character associated to the adjoint module is:
\begin{equation}
  \chi^{so(2,3)}_{\rm Adj}(\beta,\alpha) = \tr_{\rm Adj}
  \left(e^{-\beta\, E +i\,\alpha J_3}\right) = \sum_{m+n+p+q\,\in2\,\N}
  e^{-\beta\,\frac{m+n-p-q}2 \,+\, i\,\alpha\,\frac{m-n-p+q}2}\,.
  \label{E J eq}
\end{equation}
Note that the above series is not well-defined because of the infinite
degeneracy for a given eigenvalue $E$ and $J_3$. However, we can still
make some formal manipulations on it. Let us start by separating
\eqref{E J eq} into two parts:
\begin{equation}
  \chi^{so(2,3)}_{\rm Adj}(\beta,\alpha) = \sum_{\substack{m+n\in2\N
      \\ p+q\in 2\N}} e^{-\beta\,\frac{m+n-p-q}2\, +\,
    i\,\alpha\,\frac{m-n-p+q}2}+ \sum_{\substack{m+n\in2\N+1 \\ p+q\in
      2\N+1}} e^{-\beta\,\frac{m+n-p-q}2 \,+\,
    i\,\alpha\,\frac{m-n-p+q}2}.
\end{equation}
The first series factors as
\begin{eqnarray}
  && \sum_{\substack{m+n\in2\N}} e^{-\frac\beta2 \,(m+n+1) +
    i\,\frac\alpha2\, (m-n)}\sum_{\substack{p+q\in 2\N}}
  e^{+\frac\beta2 \,(p+q+1) - i\,\frac\alpha2\, (p-q)} \nonumber \\ &&
  = \chi^{so(2,3)}_{\rac}(\beta,\alpha) \times
  \chi^{so(2,3)}_{\rac}(-\beta,-\alpha) =
  \chi^{so(2,3)}_{\rac}(\beta,\alpha) \times
  \chi^{so(2,3)}_{\arac}(\beta,\alpha)\, ,
  \label{firstSum}
\end{eqnarray}
whereas the second series factors as
\begin{eqnarray}
  && \sum_{\substack{m+n\in2\N+1 }} e^{-\frac\beta2\,(m+n+1) +
    i\,\frac\alpha2\, (m-n)} \sum_{\substack{ p+q\in 2\N+1}}
  e^{+\frac\beta2\,(p+q+1) - i\,\frac\alpha2\, (p-q)} \nonumber \\ &&
  = \chi^{so(2,3)}_{\di}(\beta,\alpha) \times
  \chi^{so(2,3)}_{\di}(-\beta,-\alpha) =
  \chi^{so(2,3)}_{\di}(\beta,\alpha) \times
  \chi^{so(2,3)}_{\adi}(\beta,\alpha)\,.
\end{eqnarray}
Therefore, we find
\begin{equation}
  \chi^{so(2,3)}_{\rm Adj} = \chi^{so(2,3)}_{\rac} \times
  \chi^{so(2,3)}_{\arac} + \chi^{so(2,3)}_{\di} \times
  \chi^{so(2,3)}_{\adi},
  \label{Adj tff}
\end{equation}
which is in accordance with \eqref{Adj char}. Note again that the
above manipulation is formal and can be understood only as a
regularization procedure. The generators $T^{m,n;p,q}$ are also
eigenvectors, of identical eigenvalues, than the generators
$X^{m,n;p,q}$. However, the basis elements $T^{m,n;p,q}$ of the space
${\rm Rac}\otimes \overline{{\rm Rac}}$ are such that
$m+n\in2\,\mathbb N$ and $p+q\in2\,\mathbb N$, therefore the
corresponding character is equal to the first sum \eqref{firstSum} and
one finds
\begin{equation}
  \chi^{so(2,3)}_{{\rm Rac} \otimes \overline{{\rm Rac}}} =
  \chi^{so(2,3)}_{\rac} \times \chi^{so(2,3)}_{\arac}\,.
\end{equation}
Similarly, the basis elements $T^{m,n;p,q}$ of the space ${\rm
  Di}\otimes \overline{{\rm Di}}$ are such that $m+n\in2\,\mathbb N+1$
and $p+q\in2\,\mathbb N+1$, leading to
\begin{equation}
  \chi^{so(2,3)}_{{\rm Di}\otimes \overline{{\rm Di}}} =
  \chi^{so(2,3)}_{\di} \times \chi^{so(2,3)}_{\adi}\,.
\end{equation}
These computations suggest a neat conclusion in four dimensions: the
heuristic equality \eqref{=?} should be replaced with
\begin{equation} \label{=!}
  {\rm Adj}=({\rm Rac} \otimes \overline{{\rm Rac}}) \oplus ({\rm Di}
  \otimes \overline{{\rm Di}})\,,
\end{equation}
as suggested from the change of basis \eqref{lincombHSgen} if one
properly takes into account the range of the indices.

Let us summarize what we have observed.  First, we have seen that
the adjoint module of the higher-spin algebra is actually larger than that
of the Rac and anti-Rac tensor-product module.  This was manifest in
the oscillator analysis and the complementary vector space was
identified with the $\di\otimes \adi$ module.  In fact, as we shall
see in below, the tensor-product module $\rac\otimes \arac$ fails to
cover the entire adjoint module also in higher dimensions.  However,
the complementary space cannot be interpreted as $\di\otimes \adi$
except in four dimensions.  This should be related to the fact that
only in four dimensions Rac and Di have the same endomorphism algebra. In other
words, the type-A and type-B higher-spin algebras coincide with each
other only in four dimensions.
 
We can also regard the complementary space as a ``permuted'' Rac
module, in the sense that
\begin{equation}\label{di-rac-continuation}
  \chi^{so(2,3)}_{\di}(i\,\alpha_1,\alpha_2) =
  -\,\chi^{so(2,3)}_{\rac}(i\,\alpha_2,\alpha_1)\,.
\end{equation}
Then, the result \eqref{Adj tff} can be viewed as the symmetrization,
\begin{eqnarray}
  \chi^{so(2,3)}_{\rm Adj}(i\,\a_1,\a_2) \eq \chi^{so(2,3)}_{\rm
    Rac}(i\,\a_1,\a_2)\, \chi^{so(2,3)}_{\overline{\rm
      Rac}}(i\,\a_1,\a_2) + (1\leftrightarrow 2)\nn \eq
  \chi^{so(2,3)}_{\rm Di}(i\,\a_1,\a_2)\,
  \chi^{so(2,3)}_{\overline{\rm Di}}(i\,\a_1,\a_2) + (1\leftrightarrow
  2)\,.
  \label{4d tff}
\end{eqnarray}
In fact, the character of the adjoint module is clearly symmetric
under the exchange of $i\,\b$ and $\a$ as it is the sum of the characters of
$\cD(1-s,s-1)$ given in \eqref{(r,r)} having this property.  On the
other hand, the product of the singleton and anti-singleton characters
is generically asymmetric as we can see in \eqref{discrep}.  Hence,
the simplest way to relate this asymmetric function to the symmetric
one would be by the symmetrization of \eqref{4d tff}.  
The relation \eqref{4d tff} at the level of the characters 
can be translated back to the modules as
\begin{eqnarray}\label{reftwFF}
  {\rm Adj}\eq \left({\rm Rac} \otimes \overline{{\rm
      Rac}}\right)\oplus \left(\t({\rm Rac}) \otimes \t(\overline{{\rm
      Rac}})\right) \nn \eq \left( \di\otimes \overline{\di}
  \right)\oplus \left( \t(\di) \otimes \t(\overline{\di}) \right).
\end{eqnarray}
where $\t$ is the weight-space map exchanging the two Cartan generators,
and hence can be viewed as an element of the Weyl group of $so(2,3)$ (quotiented
by the normalizer subgroup of $\rac\otimes\overline{{\rm Rac}}$ or $\di \otimes \overline{{\rm Di}}$). In order to obtain the second equality in \eqref{reftwFF}, we used the relations
\begin{eqnarray}
  \t({\rm Rac}) \otimes \t(\overline{{\rm Rac}}) = \di\otimes
  \overline{\di}\,, \nn \t(\di) \otimes \t(\overline{\di}) = {\rm Rac}
  \otimes \overline{{\rm Rac}}\,,
\end{eqnarray}
which can also be used in order to relate \eqref{=!} and
\eqref{reftwFF}. Since the ``symmetrized'' tensor product in
\eqref{reftwFF} can be generalized to higher dimensions, we propose it
as a refined version for the twisted Flato-Fronsdal
theorem.\footnote{Let us stress that the twisted Flato-Fronsdal
  theorem \eqref{reftwFF} essentially relies on the change of basis in
  the higher-spin algebra (more precisely, a suitable completion
  thereof). In other words, our proof does not actually relies on
  characters.}  Interestingly, the idea of symmetrization works in
higher dimensions as well as for the higher order singletons, as we
shall show in the following sections.

Before moving to general dimensions, let us comment on the unitarity
of modules.  The mere tensor product of two unitary modules (${\rm
  Sng}$ and $\overline{\rm Sng}$) should not result in a non-unitary
module (the adjoint module) in general. Hence, at first glance, this
indicates that a refinement is needed in the naive twisted
Flato-Fronsdal theorem \eqref{=?}. But, the issue is in fact more
subtle: the new additional term in the refined twisted Flato-Fronsdal
theorem can be written either as $\di\otimes\overline{\di}$ in
\eqref{=!}  or as $\t({\rm Rac}) \otimes \t(\overline{{\rm Rac}})$ in
\eqref{reftwFF}. Since $\di$ is unitary while $\t({\rm Rac})$ is not,
the (non-)unitarity of the refinement term is not clear. This subtlety
can be related to the possibility that the relation \eqref{=!} or
\eqref{reftwFF} may require a suitable \textit{completion} of the
corresponding vector spaces. Indeed, the change of basis
\eqref{lincombHSgen} relating the two modules expresses the generator
$T^{m,n;p,q}$ as an infinite linear combination of the generators
$X^{m,n;p,q}$, and the norm of the former may diverge even though each
summand has a finite norm. This subtle point will be left somewhat
implicit in expressions like \eqref{=!} and \eqref{reftwFF}. This
issue may be related to the regularization of the adjoint module
character provided by the twisted Flato-Fronsdal theorem.

%%%%%%%%%%%%%%%%%%%%%%%%%%%%
\section{General Dimensions}
\label{sec: gen d}
%%%%%%%%%%%%%%%%%%%%%%%%%%%%

In this section, we shall provide more evidences of the
``symmetrization'' prescription for the relation between singleton
  and adjoint module character, by examining the type-A and type-B
  models in any dimension. 

For a smooth demonstration, let us provide here some details about the
$so(d)$ and $so(2,d)$ characters.  A unitary irreducible
representation of $so(d)$ is entirely determined by a highest weight
$\bm \ell=(\ell_1,\ldots,\ell_r)$ with $r=[d/2]$ the integer part of
$d/2$ which is also the rank of $so(d)$, and
$\ell_1\geqslant\cdots\geqslant\ell_{r-1}\geqslant|\ell_r|$ (the last
number $\ell_r$ can be negative only for $so(2r)$\,) are either all integers or all half-integers. Its character is
given by
\begin{equation}
  \chi^{so(d)}_{\bm\ell}(\bm x) =
  \tr_{\bm\ell}\!\left[x_{1}^{M^{12}}\cdots x_{r}^{M^{2r-1\,
        2r}}\right] =\left\{
  \begin{array}{cc}
    \frac{\det\left[x_{i}{}^{k_{j}} - x_{i}{}^{-k_{j}}\right]}
         {\Delta^{\sst (r)}(\bm x)\,
           \prod_{i=1}^{r}\left(x_{i}{}^{\frac12}-x_{i}{}^{-\frac12}\right)}
         \qquad & [d=2r+1] \\ \\ \frac{ \det\left[x_{i}{}^{k_{j}} +
             x_{i}{}^{-k_{j}}\right] + \det\left[x_{i}{}^{k_{j}} -
             x_{i}{}^{-k_{j}}\right]} {2\, \Delta^{\sst (r)}(\bm x)}
         \qquad & [d=2r]
  \end{array}
  \right.\,,
\end{equation}
with $\bm x=(x_1,\ldots,x_r)$ and $k_{i}=\ell_{i}+\frac d2-i$. Here
$\Delta^{\sst (r)}(\bm x)$ is the Vandermonde determinant,
\begin{equation}
  \Delta^{\sst (r)}(\bm x)=\prod_{1\leqslant i<j \leqslant
    r}\left(x_{i}+x_{i}^{-1}-x_{j}-x_{j}^{-1}\right).
\end{equation}
Notice that the character formulae displayed in the previous section
can be recovered after setting $x_k = e^{i\,\alpha_k}$ with $k=1,
\dots, r$ (and setting $r=1$ since $d=3$ there).
        
Turning now to the non-compact Lie algebra $so(2,d)$, any of its
irreducible lowest-weight modules can be described in terms of
(quotients of) lowest-weight generalized Verma modules
$\cV(\Delta,\bm\ell)$ (see Appendix \ref{app: detail} for conventions and
  technical details). The character of the latter module is given
by
\begin{equation}
  \chi^{so(2,d)}_{\cV(\Delta,\bm\ell)}(q,\bm x) =
  \tr_{\cV(\Delta,\bm\ell)}\!\left[q^{M^{0'0}}\,x_{1}^{M^{12}}\cdots
    x_{r}^{M^{2r-1\,2r}}\right] = q^{\Delta}\,\Pd{d}(q,\bm
  x)\,\chi^{so(d)}_{\bm\ell}(\bm x)\,,
  \label{ch V}
\end{equation}
where the function $\Pd{d}(q,\bm x)$ defined as
\begin{equation}
  \Pd{d}(q,\bm x) = \frac1{(1-q)^{d-2r}} \prod_{i=1}^{r}
  \frac1{\left(1-q\,x_{i}\right)\left(1-q\,x_{i}^{-1}\right)} =
  \chi^{so(2,d)}_{\cV(0,\bm 0)}(q,\bm x)\,,
  \label{Pd}
\end{equation}
is the character of the module associated with the trivial weight.

The contragredient representation carried by 
the dual module of a module $M$
has the opposite quantum numbers\,\footnote{Recall that, given a
  representation $(V,\rho)$ of a semisimple Lie algebra $\mathfrak g$,
  the contragredient representation $(V^*, \rho^*)$ is defined as
  $(\rho^*(x) \cdot \phi)(v) = \phi (\rho \circ \tau(x) \cdot v)$ for
  $x\in\mathfrak g$, $v\in V$, $\phi\in V^*$ and $\tau$ the Chevalley
  involution. This automorphism of $\mathfrak g$ acts on the Cartan
  subalgebra generators $H_i$ and the ladders operators $E_\alpha$ as
  $\tau(H_i)=-H_i$ and $\tau(E_\alpha)=-E_{-\alpha}$.} 
  with respect to $M$.
For a given lowest-weight module $M$, there exists
a highest-weight module $\overline{M}$  with exactly the opposite quantum numbers.
We shall refer to this highest-weight module as ``anti-$M$''
and equate it with the dual module, disregarding potential subtleties of infinite dimensional vector space.
Then,
the characters of the anti-module $\overline M$ is
simply related to that of the module $M$ as
\begin{equation}
  \chi^{so(2,d)}_{\overline M}(q, \bm x) =
  \chi^{so(2,d)}_{M}(q^{-1},\bm x^{-1})\, ,
\end{equation}
where $\bm x^{-1}=(x_1^{-1},x_2^{-1},\ldots,x_r^{-1})$.

For future use, let us enlist a few properties of the function $\Pd d$
and of the $so(d)$ characters. First, $\Pd d(q, \bm x)$ satisfies
\begin{equation}
  \Pd d(q^{-1},{\bm x}^{-1}) = \Pd d(q^{-1},{\bm x}) = (-q)^d\,\Pd
  d(q,\bm x)\,,
  \label{prop 1}
\end{equation}
and can be expressed as a series of $so(d)$ character as \cite{Dolan:2005wy}
\begin{equation}
      \Pd d (q, \bm x) = \sum_{s,n= 0}^\infty q^{s+2n}\,
      \chi^{so(d)}_s(\bm x)\,,
      \label{prop 2}
\end{equation}
where $\chi^{so(d)}_s$ denotes the spin-$s$ character of $so(d)$,
corresponding to the highest weight $(s,0,\ldots,0)$.  Finally, the
$so(2+d)$ character of the irrep $(\ell_0,\Bell)$ can be written in terms of the $so(d)$ character for the irrep $\Bell$ as
\begin{equation}
  \chi^{so(2+d)}_{(\ell_0,\bm\ell)}(x_0, \bm x) = \sum_{k=0}^r \,\Pd d
  (x_k, {\bm x}_k)\times \left\{
  \begin{array}{cc}
    \left(x_k^{-\ell_0}- x_k^{\ell_0+d}\right)
    \chi^{so(d)}_{\bm\ell}(\bm x_k)\, \qquad
    &[d=2r+1]\\ \\ \left[x_k^{-\ell_0}\, \chi^{so(d)}_{\bm\ell_-}({\bm
        x}_k) + x_k^{\ell_0+d}\,\chi^{so(d)}_{\bm\ell_+}({\bm x}_k)
      \right] \qquad & [d=2r]
  \end{array}\right.,
  \label{so 2+d}
\end{equation}
where ${\bm x}_k = (x_0, \dots, x_{k-1}, x_{k+1}, \dots, x_r)$ and
$\bm\ell_\pm=(\ell_1,\ldots,\ell_{r-1}, \pm \ell_r)$\,.  This identity
will play a key role in uncovering simple relations between the
singleton and adjoint module characters, and can be derived from
the Weyl character formula (see Appendix \ref{app:Weyl}).  It is
worth noting that the identity \eqref{so 2+d} can be also viewed as a
relation between the $so(2+d)$ character and the $so(2,d)$ character
of $\cV(\Delta,\bm\ell)$\,:
\begin{equation}
  \chi^{so(2+d)}_{(\ell_0,\bm\ell)}(x_0, \bm x) = \sum_{k=0}^r \left\{
    \begin{array}{cc}
    \chi^{so(2,d)}_{\cV(-\ell_0,\bm\ell)}(x_k,\bm x_k)
    -\chi^{so(2,d)}_{\cV(\ell_0+d,\bm\ell)}(x_k,\bm x_k) \qquad
    &[d=2r+1]\\ \\ \chi^{so(2,d)}_{\cV(-\ell_0,\bm\ell_-)}(x_k,\bm
    x_k) +\chi^{so(2,d)}_{\cV(\ell_0+d,\bm\ell_+)}(x_k,\bm x_k) \qquad
    & [d=2r]
    \end{array}\right..
\end{equation}
Another property of the $so(d)$ characters that will prove useful in
the subsequent sections is the following:
\begin{equation}
  \chi^{so(d)}_{(\ell_1, \dots, \ell_{j-1}, \ell_j-1,\ell_{j+1}+1,
    \ell_{j+2}, \dots, \ell_r)}(\bm x) = -\chi^{so(d)}_{(\ell_1,
    \dots, \ell_{j-1}, \ell_{j+1}, \ell_j, \ell_{j+2}, \dots,
    \ell_r)}(\bm x)\, ,
  \label{sym_prop_sod}
\end{equation}
as it implies in particular that the character with the label 
\begin{equation}
    (\ell_1, \dots,\ell_{j-1}, \ell-1, \ell, \ell_{j+2}, \dots, \ell_r)
\end{equation}
identically vanishes.
Equipped with the above identities, let us consider the type-A
and type-B higher-spin theories in any dimension, which are based on
the scalar and spinor singletons respectively.

%*********************************************%
\subsection{Type A}
\label{typA}
%*********************************************%

\begin{figure}[!h]
  \center
  \begin{tikzpicture}
    \draw[thick,->] (-1.5,0) -- (5.8,0) node[anchor=north west] {$s$};
    \draw[thick,->] (0,-3.8) -- (0,3.8) node[anchor=south east] {$E$};

    \draw (-0.2, 0) node[below] {\small $0$};
    \draw (-0.4, 1.5pt) -- (-0.4, -1.5pt);  
    \draw (-0.8, 1.5pt) -- (-0.8, -1.5pt); 
    \draw (-1.2, 1.5pt) -- (-1.2, -1.5pt); 
    \draw (0.4, 1.5pt) -- (0.4, -1.5pt) node[below=0.2] {\small $1$}; 
    \draw (0.8, 1.5pt) -- (0.8, -1.5pt) node[below=0.2] {\small $2$}; 
    \draw (1.2, 1.5pt) -- (1.2, -1.5pt) node[below=0.2] {\small $3$}; 
    \draw (1.6, 1.5pt) -- (1.6, -1.5pt) node[below=0.2] {\small $4$}; 
    \draw (2, 1.5pt) -- (2, -1.5pt); 
    \draw (2.4, 1.5pt) -- (2.4, -1.5pt) ; 
    \draw (2.8, 1.5pt) -- (2.8, -1.5pt) node[below=4] {$\dots$}; 
    \draw (3.2, 1.5pt) -- (3.2, -1.5pt); 
    \draw (3.6, 1.5pt) -- (3.6, -1.5pt); 
    \draw (4, 1.5pt) -- (4, -1.5pt); 
    \draw (4.4, 1.5pt) -- (4.4, -1.5pt); 
    \draw (4.8, 1.5pt) -- (4.8, -1.5pt); 

    \draw (-1.5pt, 0.4) -- (1.5pt, 0.4); 
    \draw (-1.5pt, 0.8) -- (1.5pt, 0.8) node[left=5] {\small $\ez \ \ $}; 
    \draw (-1.5pt, 1.2) -- (1.5pt, 1.2) node[left=5] {\small $\ez+1$}; 
    \draw (-1.5pt, 1.6) -- (1.5pt, 1.6)node[left=5] {\small $\ez+2$}; 
    \draw (-1.5pt, 2) -- (1.5pt, 2) node[left=5] {\small $\ez+3$}; 
    \draw (-1.5pt, 2.4) -- (1.5pt, 2.4); 
    \draw (-1.5pt, 2.8) -- (1.5pt, 2.8) node[left=8] {$\vdots$}; 
    \draw (-1.5pt, 3.2) -- (1.5pt, 3.2); 
    \draw (-1.5pt, 3.6) -- (1.5pt, 3.6); 
    \draw (-1.5pt, -0.4) -- (1.5pt, -0.4); 
    \draw (-1.5pt, -0.8) -- (1.5pt, -0.8) node[left=5] {\small $-\ez \ \ \ $}; 
    \draw (-1.5pt, -1.2) -- (1.5pt, -1.2) node[left=5] {\small $-(\ez+1)$}; 
    \draw (-1.5pt, -1.6) -- (1.5pt, -1.6) node[left=5] {\small $-(\ez+2)$}; 
    \draw (-1.5pt, -2) -- (1.5pt, -2) node[left=5] {\small $-(\ez+3)$}; 
    \draw (-1.5pt, -2.4) -- (1.5pt, -2.4); 
    \draw (-1.5pt, -2.8) -- (1.5pt, -2.8) node[left=8] {$\vdots$}; 
    \draw (-1.5pt, -3.2) -- (1.5pt, -3.2); 
    \draw (-1.5pt, -3.6) -- (1.5pt, -3.6); 

    \draw (0, 0.8) node[color=blue] {$\boldsymbol{\times}$};
    \draw (0.4, 1.2) node[color=blue] {$\boldsymbol{\times}$};
    \draw (0.8, 1.6) node[color=blue] {$\boldsymbol{\times}$};
    \draw (1.2, 2) node[color=blue] {$\boldsymbol{\times}$};
    \draw (1.6, 2.4) node[color=blue] {$\boldsymbol{\times}$};
    \draw (2, 2.8) node[color=blue] {$\boldsymbol{\times}$};
    \draw (2.4, 3.2) node[color=blue] {$\boldsymbol{\times}$};
    \draw (2.6, 3.4) node[color=black] {$\boldsymbol{\cdot}$};
    \draw (2.8, 3.6) node[color=black] {$\boldsymbol{\cdot}$};
    \draw (3, 3.8) node[color=black] {$\boldsymbol{\cdot}$};

    \draw (0, -0.8) node[color=red] {$\boldsymbol{\times}$};
    \draw (0.4, -1.2) node[color=red] {$\boldsymbol{\times}$};
    \draw (0.8, -1.6) node[color=red] {$\boldsymbol{\times}$};
    \draw (1.2, -2) node[color=red] {$\boldsymbol{\times}$};
    \draw (1.6, -2.4) node[color=red] {$\boldsymbol{\times}$};
    \draw (2, -2.8) node[color=red] {$\boldsymbol{\times}$};
    \draw (2.4, -3.2) node[color=red] {$\boldsymbol{\times}$};
    \draw (2.6, -3.4) node[color=black] {$\boldsymbol{\cdot}$};
    \draw (2.8, -3.6) node[color=black] {$\boldsymbol{\cdot}$};
    \draw (3, -3.8) node[color=black] {$\boldsymbol{\cdot}$};
  \end{tikzpicture}
  \caption{Weight diagram of the scalar singleton, or \it Rac \rm
    (blue crosses) and of the scalar anti-singleton, or \it anti-Rac
    \rm (red crosses).}
    \label{fig1}
\end{figure}
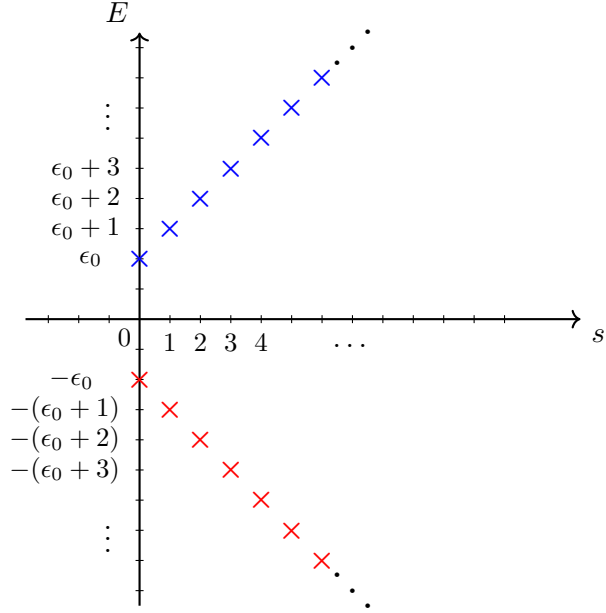

Let us begin with the type-A massless higher-spin gravity in $d+1$
dimensions. It is expected to be dual to the $U(N)$ free scalar
CFT in $d$ dimensions. The free conformal scalar field 
carries nothing but
the spin-$0$ singleton representation of $so(2,d)$\,:
\begin{equation}\label{Racterminology}
  \cD\!\left(\tfrac{d-2}2, \bm 0 \right) \equiv \frac{\V(\frac{d-2}2,
    \bm 0)}{\D(\frac{d+2}2, \bm 0)}\, ,
\end{equation}
that, from now on, we shall refer to  as ``Rac'', thereby extending the four-dimensional
terminology to any $d$\,. Note that $\bm 0$ stands for the trivial
weight $(0,\dots,0)$. Using \eqref{ch V}, one computes the character
of the quotient in \eqref{Racterminology}:
\begin{equation}
  \chi^{so(2,d)}_{\rm Rac}(q, \bm x) = q^{\frac{d-2}2} \left(1-q^2\right) \Pd d
  (q, \bm x).
\end{equation}
The energy eigenvalues of this representation is bounded from below,
hence it is a lowest-weight representation. One can define an
analogous representation whose energy is now bounded from above. We
refer to the corresponding module as anti-singleton $\arac$ and its
character is simply related to that of the singleton as
\begin{equation}
  \chi_{\arac}^{so(2,d)}(q, \bm x) = \chi_{\rac}^{so(2,d)}(q^{-1}, \bm
  x^{-1}) = \chi_{\rac}^{so(2,d)}(q^{-1}, \bm x)\,.
 \end{equation}
Notice that the character of the $\rac$ and $\arac$ modules are simply
related to each other, using \eqref{prop 1}, by a sign depending on
the parity of $d$, namely,
\begin{equation}
  \chi_{\arac}^{so(2,d)}(q, \bm x) = (-1)^{d+1}
  \chi^{so(2,d)}_{\rac}(q, \bm x)\,.
  \label{relation_sign_rac}
\end{equation}
Using the property \eqref{prop 2}, the character of $\rac$ and $\arac$
can be also expressed as
\begin{equation}
  \chi_{\rac}^{so(2,d)}(q, \bm x) = \sum_{s=0}^\infty q^{\e_0+s}\,
  \chi_s^{so(d)}(\bm x)\,, \qquad \chi_{\arac}^{so(2,d)}(q, \bm x)
  = \sum_{s=0}^\infty q^{-\e_0-s}\, \chi_s^{so(d)}(\bm x)\,,
  \label{weight_rac}
\end{equation}
where $\e_0 := (d-2)/2$\,. From the above formulae, the weight space
of the $\rac$ and $\arac$ representations respectively can be read
off. Their weights are depicted in Fig \ref{fig1}.

Now, let us see how the character of the adjoint module can be related
to the Rac module. The higher-spin algebra of type-A theory is the
collection of the $so(2+d)$ irrep $(s-1,s-1):=(s-1, s-1, 0, \dots,
0)$\, for $s=1, 2, \dots, \infty$. Applying the identity \eqref{so
  2+d} to these irreps, we obtain
\begin{equation}
  \chi^{so(2+d)}_{(s-1,s-1)}(x_0,\bm x) = \sum_{k=0}^r \, \Pd d
  (x_k,\bm x_k) \left(x_k^{1-s} +(-1)^d\, x_k^{s+d-1}\right)
  \chi^{so(d)}_{s-1}(\bm x_k)\,.
\end{equation}
In the right-hand-side of the equation, the last two factors can be
summed over $s$ by using the properties \eqref{prop 1} and \eqref{prop
  2} as
\begin{eqnarray}
  \sum_{s=1}^\infty \left(x_k^{1-s} +(-1)^d\,x_k^{s+d-1}\right)
  \chi_{s-1}^{so(d)}(\bm x_k) = \left
  (x_k^{-1}-x_k\right)\left(x_k-x_k^{-1}\right) \Pd d (x_k^{-1}, \bm
  x_k)\,.
\end{eqnarray}
Notice that when summing the above expression, we used the identity
\eqref{prop 2} as if it was valid simultaneously in both domains of
convergence $\rvert x_k \rvert < 1$ and $\rvert x_k \rvert > 1$ for all $k=0,1,\cdots,r$.
Multiplying the above equation by $\Pd d(x_k,\bm x_k)$ and
symmetrizing over $k$, we finally obtain
\begin{empheq}[box=\eqbox]{equation}
  \chi^{so(2+d)}_{\rm Adj}(x_0,\bm x)= \sum_{s=1}^\infty
  \chi^{so(2+d)}_{(s-1, s-1)}(x_0,\bm x) = \sum_{k=0}^r
  \chi_{\rac}^{so(2,d)}(x_k, \bm x_k) \, \chi_{\arac}^{so(2,d)}(x_k,
  \bm x_k)\,, 
  \label{twisted_flato_fronsdal}
\end{empheq}
where one should remember that ${\bm x}_k = (x_0, \dots, x_{k-1},
x_{k+1}, \dots, x_r)$.  Hence, the sum over the characters
corresponding to all the $so(2+d)$ two-row rectangular Young diagrams
(i.e. the diagrams \eqref{KT} for $s=1, 2, \dots, \infty$) is equal to
the product of the characters of a Rac and an anti-Rac symmetrized
over all variables.

%*****************%
\subsection{Type B}
%*****************%
The type-B massless higher-spin gravity in $d+1$ dimensions is conjectured to be dual to the free
CFT with Dirac spinor in $d$ dimensions.

\subsubsection*{Even $d+1$ Dimensions}

For $d+1$ even (that is $d$ odd), this free Dirac spinor carries the
spin-$\tfrac12$ singleton representation,
\be
\di:=\cD\Big(\frac{d-1}2,\bm{\frac12}\,\Big)=\frac{\cV(\frac{d-1}2,\bm{\frac12})}{\cV(\frac{d+1}2,\bm{\frac12})}
\ee
with $\bm{\frac12}=(\frac12,\dots,
\frac12)$\,.  The character of Di has the form \cite{Dolan:2005wy},
\begin{equation}
  \chi^{so(2,d)}_{\di}(q, \bm x) = q^{\frac{d-1}2}\,(1 - q)\,\Pd d (q,
  \bm x)\, \chi^{so(d)}_{\bm{\frac12}}(\bm x)\,,
\end{equation}
and the character of anti-Di is simply related to the above as
\begin{equation}
  \chi^{so(2,d)}_{\adi}(q, \bm x) = \chi^{so(2,d)}_{\di}(q^{-1}, {\bm
    x}^{-1}) = \chi^{so(2,d)}_{\di}(q^{-1}, \bm x)\,,
\end{equation}
since $\chi^{so(d)}_{\bm{\frac12}}({\bm
  x}^{-1})=\chi^{so(d)}_{\bm{\frac12}}(\bm x)$.  From the identity
\eqref{prop 2}, we derive another useful identity,
\begin{equation}
  (1 - q)\,\Pd d (q, \bm x)\, \chi^{so(d)}_{\bm{\frac12}}(\bm x) =
  \sum_{s = 0}^\infty q^{s}\,
  \chi^{so(d)}_{(s+\frac12,{\frac12}^{r-1})}(\bm x)\,,
  \label{Di id}
\end{equation}
where the notation $c^m$ in a weight stands for a sequence of $m$
identical entries $c$\,: for instance,
$(s+\tfrac12,{\tfrac12}^{r-1})=(s+\tfrac12,\tfrac12,\dots,\tfrac12)$.
The identity \eqref{Di id} implies that, similarly to the Rac, the
character of Di can be written as
\begin{equation}
  \chi_{\di}^{so(2,d)}(q, \bm x) = \sum_{s = 0}^\infty q^{\frac{d-1}2
    + s}\, \chi^{so(d)}_{(s+\frac12,{\frac12}^{r-1})}(\bm x)\,.
\end{equation}
Notice that the character of the $\di$ singleton is 
actually identical to that of its anti-singleton $\adi$ for $d=2r+1$: 
\begin{equation}
    \chi_{\adi}^{so(2,2r+1)}(q, \bm x) = \chi_{\di}^{so(2,2r+1)}(q, \bm x)\,,
\end{equation}
similarly to the Rac case.

We want to relate this character to that of the adjoint module of
type-B higher-spin algebra. In the section \ref{typA}, we started from
the adjoint module and showed that its character can be written in
terms of the Rac and anti-Rac characters.  In the type-B case, its
higher-spin algebra was identified in \cite{Vasiliev:2004cm}, so we
can proceed, in principle, in the same way. However, the utility of
the twisted Flato-Fronsdal theorems is actually to identify the
higher-spin algebra directly from the underlying singleton modules.
Hence, for type-B theory let us proceed in the opposite way to the
type-A case: we begin with the Di and anti-Di character and find the
character of the adjoint module.

The starting point is the product of the Di and anti-Di characters,
\begin{equation}
 \chi_\di^{so(2,d)}(q, \bm x)\, \chi_{\adi}^{so(2,d)}(q, \bm x)
 =\left(1-q+1-q^{-1}\right) \Pd d (q, \bm x)\, \Pd d (q^{-1}, \bm x)\,
 \chi^{so(d)}_{\bm{\frac12}}(\bm x) \,\chi^{so(d)}_{\bm{\frac12}}(\bm
 x)\,.
\end{equation}
Using \eqref{prop 1} and \eqref{Di id}, the above can be expressed as
\begin{eqnarray}
  &&\chi_\di^{so(2,d)}(q, \bm x)\, \chi_{\adi}^{so(2,d)}(q, \bm x) \nn
  &&= \Pd d (q, \bm x)\sum_{s=0}^\infty \left(q^{-s} -q^{s+d}\right)
  \chi^{so(d)}_{(s+\frac12,\frac12^{r-1})}(\bm x)\,
  \chi^{so(d)}_{\bm{\frac12}}(\bm x) \nn && =\Pd d (q, \bm
  x)\sum_{s=0}^\infty \sum_{m=0}^{r-1} \left(q^{-s} -q^{s+d}\right)
  \left[\chi^{so(d)}_{(s,1^{m})}(\bm x)
    +\chi^{so(d)}_{(s+1,1^{m})}(\bm x)\right].
\end{eqnarray}
In the last line, the product of two $so(d)$ characters is decomposed
in terms of other $so(d)$ characters using the decomposition rule (see e.g. \cite{King1971} for the general decomposition rule of the tensor product of two $so(d)$-modules, recalled in \cite{Bekaert:2006py}):
\begin{equation}
  \big(s+\tfrac12,\tfrac12^{r-1}\big) \otimes
  \big(\tfrac12^r\big) = \bigoplus_{m=0}^{r-1}\,
  \Big[\,(s,1^m)\oplus (s+1,1^{m})\,\Big]\,.
  \label{s111}
\end{equation}
Note that the modules with improper weight labels do not contribute,
namely the first term in the summand on the right-hand-side of the
above decomposition is absent, when $m\neq0$, for $s=0$. This can be also viewed from
the fact that the characters associated with the improper labels that appear
in \eqref{s111} identically vanish, due to the identity
\eqref{sym_prop_sod}. Finally, symmetrizing over the variables, we get
\begin{equation}
  \sum_{k=0}^r \chi_{\di}^{so(2,d)}(x_k, \bm x_k)\,
  \chi_{\adi}^{so(2,d)}(x_k, \bm x_k) =
  \sum_{s=0}^\infty \sum_{m=0}^{r-1}
  \left[\chi^{so(2+d)}_{(s,s,1^{m})}(x_0,\bm x)
  +\chi^{so(2+d)}_{(s,s+1,1^{m})}(x_0,\bm x)\right].
\end{equation}
Once again, the characters with improper labels identically vanish
(i.e. the second term on the right-hand-side), hence we derive in the
end,
\begin{empheq}[box=\eqbox]{equation}
  \sum_{k=0}^r \chi_{\di}^{so(2,d)}(x_k, \bm x_k) \times
  \chi_{\adi}^{so(2,d)}(x_k, \bm x_k) = \chi^{so(2+d)}_{0}(x_0, \bm
  x) + \sum_{s=2}^\infty \sum_{m=0}^r \chi^{so(2+d)}_{(s-1,s-1,
    1^m)}(x_0, \bm x)\,,
  \label{Bevend+1}
\end{empheq}
where $\chi^{so(2+d)}_{0}=1$
corresponds to the identity 
of the higher-spin algebra.
The content of the right-hand-side precisely coincides with the
adjoint module of the type-B higher-spin algebra identified in
\cite{Vasiliev:2004cm}.

\subsubsection*{Odd $d+1$ Dimensions}

For even values $d=2r$ of the boundary dimension, one can consider a
chiral (or anti-chiral) spinor singleton, i.e. whose spin is either $\half_+$ or
$\half_-$. The corresponding $so(2,d)$ module is
\cite{Shaynkman:2004vu}
\begin{equation}\label{Dipm}
  \di_\pm := \D\big( \tfrac{d-1}2, \half_\pm\big) = \frac{\V\big(
    \tfrac{d-1}2, \half_\pm\big)}{\D\big( \tfrac{d+1}2,
    \half_\mp\big)}\, ,
\end{equation}
and its character reads
\begin{equation}
  \chi^{so(2,2r)}_{\di_\pm}(q, \bm x) = q^{\frac{2r-1}2} \big(
  \chi^{so(2r)}_{\bm{\frac12}_\pm}(\bm x) - q\,
  \chi^{so(2r)}_{\bm{\frac12}_\mp}(\bm x) \big) \Pd {2r} (q, \bm x)\,.
\end{equation}
Using the expression \eqref{prop 2} of $\Pd d$, we can rewrite the
character as
\begin{equation}
  \chi^{so(2,2r)}_{\di_\pm}(q, \bm x) = \sum_{s=0}^\infty
  q^{\frac{2r-1}2+s}\, \chi^{so(2r)}_{\big(s+\frac12,
    \frac12^{r-1}_\pm\big)}(\bm x)\,.
  \label{decompo_di_pm}
\end{equation}
The characters of the anti-Di modules are given, by definition, as
\begin{equation}
  \chi^{so(2,2r)}_{\adi_\pm}(q, \bm x) =
  \chi^{so(2,2r)}_{\di_\pm}(q^{-1}, {\bm x}^{-1})\,.
\end{equation}
But now differently from the previous cases, the symmetry property of
the $so(2r)$ character $\chi^{so(2r)}_{\big(s+\frac12,
  \frac12^{r-1}_\pm\big)}(\bm x)$ depends on the parity of $r$\,:
\begin{equation}
  \chi^{so(2r)}_{\big(s+\frac12, \frac12^{r-1}_\pm\big)}({\bm x}^{-1})
  = \left\{
  \begin{array}{cc}
    \chi^{so(2r)}_{\big(s+\frac12, \frac12^{r-1}_\pm\big)}({\bm
      x})&\quad [{\rm even}\ r]
    \medskip\\ \chi^{so(2r)}_{\big(s+\frac12,
      \frac12^{r-1}_\mp\big)}({\bm x})& \quad [{\rm odd}\ r]
  \end{array} 
  \right.\,.
\end{equation}
As a consequence, the relation between the characters of the anti-Di
and the Di modules also depend on the parity of $r$\,:
\begin{equation}
  \chi^{so(2,2r)}_{\adi_\pm}(q, \bm x) = - \left\{
  \begin{array}{cc}
    \chi^{so(2,2r)}_{{\di}_\mp}(q, \bm x)&\quad [{\rm even}\ r]
    \medskip\\
    \chi^{so(2,2r)}_{{\di}_\pm}(q, \bm x)& \quad [{\rm odd}\ r]
  \end{array}
  \right.\,.
\end{equation}
Due to the possible chirality of the $\di$ singleton, only a subset of
the mixed-symmetry fields, present in the even-dimensional twisted
Flato-Fronsdal theorem \eqref{Bevend+1}, will appear. Let us start by
recalling the generalized Flato-Fronsdal theorem (first derived in
\cite{Vasiliev:2004cm}), before deriving the corresponding twisted
version. To do so, we will need the following decomposition rules of
$so(2r)$,
\begin{equation}
  \half_\pm \otimes \half_\pm = \bigoplus_{m=0}^{\left[\frac r2
      \right]} 1_\pm^{r-2m}\, , \quad \text{and} \quad \half_+ \otimes
  \half_- = \bigoplus_{m=0}^{\left[\frac{r-1}2 \right]} 1^{r-1-2m}\, ,
\end{equation}
together with
\begin{equation}
  (s+\tfrac12, \tfrac12^{r-1}_\pm) \otimes \half_\pm =
  \bigoplus_{m=0}^{\left[ \frac{r-1}2 \right]} (s+1, 1^{r-1-2m}_\pm)
  \oplus \bigoplus_{m=0}^{\left[ \frac{r-2}2 \right]} (s,
  1^{r-2-2m})\, ,
\end{equation}
and
\begin{equation}
  (s+\tfrac12, \tfrac12^{r-1}_\pm) \otimes \half_\mp =
  \bigoplus_{m=0}^{\left[ \frac{r-2}2 \right]} (s+1, 1^{r-2-2m})
  \oplus \bigoplus_{m=0}^{\left[ \frac{r-1}2 \right]} (s,
  1^{r-1-2m}_\pm)\,.
\end{equation}
In the following, we will treat separately the case of odd and even
rank $r$:
\begin{itemize}
\item \underline{Even rank $r=2k$}: In this case, the tensor product
  of two singletons of the same chirality decomposes into a direct sum
  of hook-shaped massless fields whose first columns are of all even
  heights from $0$ to $r$, together with a collection of massive
  $p$-forms with $p$ taking all even values from $0$ to
  $r$. Explicitly,
  \begin{equation}
    \left( \chi^{so(2,d)}_{\di_\pm}\right)^2 = \sum_{m=0}^{k}
    \chi^{so(2,d)}_{\cD(d-1, 1_\pm^{2m})} + \sum_{s=2}^\infty
    \sum_{m=0}^{k-1} \chi^{so(2,d)}_{\cD(s+d-2, s,1_\pm^{2m+1})}\,.
          \label{ddi1}
  \end{equation}
  In particular, this decomposition contains the massive scalar
  $\D\big( d-1\,,\, \mathbf{0} \big)$ as well as massless fields whose
  first columns are of maximal height $r$ and of the same chirality as
  the $\di$ singletons. The totally symmetric fields are however
  absent from this spectrum, they are instead contained in the tensor
  product of two singletons of opposite chiralities, together with
  hook-shaped massless fields and massive $p$-forms whose first column
  is of odd height:
  \begin{equation}
    \chi_{\di_+}^{so(2,d)} \times \chi^{so(2,d)}_{\di_-}=
    \sum_{s=1}^\infty \chi^{so(2,d)}_{\cD(s+d-2,s)} +
    \sum_{s=2}^\infty \sum_{m=1}^{k-1} \chi^{so(2,d)}_{\cD(s+d-2, s,
      1^{2m})} + \sum_{m=1}^{k-1} \chi^{so(2,d)}_{\cD(d-1,
      1^{2m+1})}\,.
    \label{ddi2}
  \end{equation}
  Using the $so(d)$ tensor product rules recalled previously, as well
  as the decomposition \eqref{decompo_di_pm}, one can show that the
  tensor product of a spinor singleton of fixed chirality with its
  anti-singleton decomposes as
  \begin{empheq}[box=\eqbox]{equation}
    \sum_{j=0}^r \chi_{\di_\pm}^{so(2,d)}(x_j, \bm x_j) \times
    \chi^{so(2,d)}_{\overline{\di_\pm}}(x_j, \bm x_j) =
    \chi^{so(2+d)}_{0}(x_0, \bm x)+ \sum_{s=2}^\infty
    \sum_{m=0}^{k-1} \chi^{so(2+d)}_{(s-1,s-1,1^{2m})}(x_0, \bm x)\, ,
    \label{di1}
  \end{empheq}
  whereas the tensor product of the $\di_+$ singleton with the
  $\overline{\di_-}$ anti-singleton yields
  \begin{empheq}[box=\eqbox]{equation}
    \sum_{j=0}^r \chi_{\di_\pm}^{so(2,d)}(x_j, \bm x_j) \times
    \chi^{so(2,d)}_{\overline{\di_\mp}}(x_j, \bm x_j) =
    \sum_{s=2}^\infty\sum_{m=0}^{k-1}
    \chi^{so(2+d)}_{(s-1,s-1,1^{2m+1}_\pm)}(x_0, \bm x)\,.
    \label{di2}
  \end{empheq}
  The modules appearing in the same/opposite-chirality twisted
  Flato-Fronsdal theorem \eqref{di1}/\eqref{di2} correspond to the
  Killing tensors associated to the massless fields appearing in the
  opposite/same-chirality Flato-Fronsdal theorem
  \eqref{ddi2}/\eqref{ddi1}.  This crossed correspondence may look
  problematic if we consider the (anti-)chiral projection, but it is
  in fact consistent since, in the non-minimal type-B theory, we have to take
  the tensor product of $\di_\pm$ and its complex conjugate for the
  bulk spectrum.  In $d=4k$ dimensions, the complex conjugate flips
  the chirality, hence the bulk spectrum is \eqref{ddi2}, which is
  compatible with \eqref{di1} \cite{Vasiliev:2004cm,Giombi:2016pvg, Gunaydin:2016amv}.
 
\item \underline{Odd rank $r=2k+1$}: In this case, the tensor product
  of two singletons of the same chirality decomposes into a direct sum
  of hook-shaped massless fields whose first columns are of any odd
  height, together with a collection of massive $p$-forms with $p$
  taking all odd values from $1$ to $r$. Explicitly, \be \left(
  \chi^{so(2,d)}_{\di_\pm} \right)^2 = \sum_{s=1}^\infty
  \chi^{so(2,d)}_{\cD(s+d-2,s)}+ \sum_{s=2}^\infty \sum_{m=1}^{k}
  \chi^{so(2,d)}_{\cD(s+d-2, s,1^{2m}_\pm)}+ \sum_{m=1}^{k}
  \chi^{so(2,d)}_{\cD(d-1,1_\pm^{2m+1})}\, .  \ee Notice that
  contrarily to the case of odd rank, this tensor product contains the
  tower of totally symmetric fields of arbitrary spin but does not
  contain the massive scalar $\D\big( d-1\,,\, \mathbf 0 \big)$. The
  latter is instead part of the tensor product decomposition of two
  $\di$ singletons of opposite chiralities, together with hook-shaped
  massless fields whose first columns are of any even height as well
  as massive $p$-forms with $p=2,4,\dots,r-1$:
  \begin{equation}
    \chi_{\di_+}^{so(2,d)}\times \chi^{so(2,d)}_{\di_-} =
    \sum_{s=2}^\infty \sum_{m=1}^{k} \chi^{so(2,d)}_{\cD(s+d-2, (s,
      1^{2m-1}))} + \sum_{m=0}^{k} \chi^{so(2,d)}_{\cD(d-1, 1^{2m})}\,
    .
    \label{didiodd}
  \end{equation}
  A computation similar to the previous case shows that the tensor
  product of a spinor singleton of fixed chirality with its
  anti-singleton can be decomposed as follows:
  \begin{empheq}[box=\eqbox]{equation}
    \sum_{j=0}^r \chi_{\di_\pm}^{so(2,d)}(x_j, \bm x_j) \times
    \chi^{so(2,d)}_{\overline{\di_\pm}}(x_j, \bm x_j) =
    \chi^{so(2,d)}_{0}(x_0, \bm x) + \sum_{s=2}^\infty
    \sum_{m=0}^{k} \chi^{so(2+d)}_{(s-1,s-1,1^{2m}_\pm)}(x_0, \bm
    x)\,,
    \label{didiodd2}
  \end{empheq}
  i.e. as the direct sum of the Young diagram describing the Killing
  tensors associated to each massless field appearing in
  \eqref{didiodd}. Finally, the tensor product $\di_+ \otimes \overline{\di_-}$,
  as well as the tensor product $\di_- \otimes \overline{\di_+}$, both contain
  the same $so(2+d)$ diagrams, i.e. those associated with the Killing
  tensor of the massless fields appearing in \eqref{didiodd2}, namely,
  \begin{empheq}[box=\eqbox]{align}
    \sum_{j=0}^r \chi_{\di_\pm}^{so(2,d)}(x_j, \bm x_j) \times
    \chi^{so(2,d)}_{\overline{\di_\mp}}(x_j, \bm x_j) = \sum_{s=2}^\infty
    \sum_{m=0}^{k-1} \chi^{so(2+d)}_{(s-1,s-1,1^{2m+1})}(x_0, \bm x)\,
    .
  \end{empheq}
\end{itemize}
If one instead consider a spinor singleton which is a Dirac fermion,
i.e. contains both chiralities, then the corresponding $\di$ module is
given by the direct sum of the two chiral modules:
\begin{equation}\label{chiralities}
 \di:= \D(\tfrac{d-1}2,\half) = \D(\tfrac{d-1}2,\half_+) \oplus
  \D(\tfrac{d-1}2,\half_-)\, ,
\end{equation}
whose character reads
\begin{equation}
  \chi^{so(2,d)}_{\di}(q, \bm x) = q^{\tfrac{d-1}2}\, (1-q)\, \Big(
  \chi^{so(d)}_{\bm{\frac12}_+}(\bm x) +
  \chi^{so(d)}_{\bm{\frac12}_-}(\bm x) \Big) \Pd d (q, \bm x)\, .
\end{equation}
Notice that in this case, the characters of the parity-invariant $\di$
and $\adi$ modules are also simply related by a dimension dependent
sign, namely,
\begin{equation}
  \chi^{so(2,d)}_{\adi}(q, \bm x) = (-1)^{d+1} \chi^{so(2,d)}_{\di}(q,
  \bm x)\, .
\end{equation}
The endomorphism algebra of this parity-invariant singleton admits a
similar decomposition to the previously covered odd-$d$ case, except
for the fact that most diagrams have a multiplicity $2$:
\begin{empheq}[box=\eqbox]{align}
  \sum_{k=0}^r \chi_{\di}^{so(2,d)}(x_k, \bm x_k) &\times
  \chi_{\adi}^{so(2,d)}(x_k, \bm x_k) = 2\,
  \chi^{so(2+d)}_{0}(x_0, \bm x) \,+ \nn
  & +2  \sum_{s=2}^\infty  \sum_{m=0}^{r-1}  \chi^{so(2+d)}_{(s-1,s-1,
    \1^m)}(x_0, \bm x) \,+\nn
  &+\sum_{s=2}^\infty\Big( \chi^{so(2+d)}_{(s-1,s-1, \1^{r-1}_+)}(x_0, \bm x) +
  \chi^{so(2+d)}_{(s-1,s-1, \1^{r-1}_-)}(x_0, \bm x)\,\Big)\,. 
\end{empheq}
The appearance of those extra degeneracies with respect to the odd-$d$ case
\eqref{Bevend+1} is caused by the fact that we include both
chiralities in \eqref{chiralities}, hence the representations for
which the last $so(d)$ weight vanishes (i.e. $\ell_r=0$) come twice.

%******************%
\subsection{Type AB}
%******************%
Although one of the appealing features of higher-spin holography is
the fact that these dualities do not require supersymetry, the four
dimensional higher-spin gravity admits a supersymmetric extension with
arbitrary $\cal N$: see \cite{Sezgin:2012ag} for a review (as well as
the recent paper \cite{Pang:2016ofv} where several one-loop tests of
these extensions were performed, together with \cite{Gunaydin:2016amv}
for the $6$-dimensional case). Supersymmetric higher-spin
algebras\,\footnote{In dimensions $3,4$ and $6$, the higher symmetries
  of super-Laplacians were studied in \cite{Howe:2016iqw}, thereby
  extending Eastwood's approach to the supersymmetric case.}  were
studied in four dimensions in \cite{Gunaydin:1989um,Konstein1989,
  Konstein:1989ij}, an analysis later extended to any dimension in
\cite{Vasiliev:2004cm}, where it was also shown that the spectrum of
these supersymmetric higher-spin theories is given by the tensor
product of the direct sum of the $\rac$ and $\di$ singletons (possibly
decorated with Chan-Paton factor, that we will not consider
here).\footnote{See also \cite{Govil:2013uta, Govil:2014uwa,
    Fernando:2014pya, Fernando:2015tiu, Gunaydin:2016amv} for the
  quasiconformal approach to higher-spin (super)algebras.}
		
In four dimensions, the ${\cal N}=1$ supersymmetric extension of the
algebra $so(2,3)\cong sp(4,{\mathbb R})$ is the superalgebra
$osp(1|4)$, of which the sum ${\rm Di}\oplus{\rm Rac}$ is a
supermultiplet. The tensor product of a $\rac$ with a $\di$ decomposes
into an infinite tower of totally symmetric massless fields of all
half-integer spin $s=\tfrac12, \tfrac32,\dots$, and therefore the
tensor product of the irreducible $osp(1|4)$-module $\di \oplus \rac$
with itself contains all totally symmetric fields of integer and
half-integer spins (as well as the mixed-symmetry fields appearing in
the tensor product of two $\di$ singletons in higher dimensions). The
${\cal N}=1$ higher-spin superalgebra extending $osp(1|4)$
can be realized in terms of the oscillators $a$ and $b$ introduced in
the section \ref{sec: 4d} by relaxing the constraint of parity. By
extending the computations of $so(2,3)$ characters, one can check at
the level of characters the isomorphism \cite{Iazeolla:2008ix}:
\begin{equation}
  {\rm Adj} =\bigoplus\limits_{s=1,\frac32, 2, \frac52,\dots} {\cal
    D}(1-s,s-1)\, =\,({\rm Di}\oplus{\rm Rac})\otimes (\overline{{\rm
      Di}}\oplus\overline{{\rm Rac}})\,,
  \label{AB=!}
\end{equation}
which is the supersymmetric extension of \eqref{=!}. Here, ``Adj'' stands for the adjoint module of the four-dimensional ${\cal N}=1$ higher-spin superalgebra.

In higher dimensions, the Di and Rac do not form a supermultiplet on
their own, due to the fact that the $AdS_{d+1}$ isometry algebra
  $so(2,d)$ admits a supersymmetric extension (i.e. a Lie superalgebra
  which contains the latter in its bosonic subsector), only in
  dimensions $d+1=4,5, 6$ and $7$, superalgebras which are
  respectively $osp({\cal N}|4)$, $sl({\cal N}|4)$, $F(4)$ and
  $osp({\cal N}|8)$.  Nevertheless, let us investigate the twisted
Flato-Fronsdal theorem for this pair of modules. Using
\eqref{decompo_di_pm}, the product of the character of a $\rac$
singleton with that of the $\adi$ anti-singleton can be written as
\begin{equation}
  \chi_{\rac}^{so(2,d)}(q, \bm x) \times \chi^{so(2,d)}_{\adi}(q, \bm
  x) = q^{-\frac12}\,(1-q^2)\, \Pd d (q, \bm x) \sum_{s=0}^\infty
  q^{-s}\, \chi^{so(d)}_{(s+\frac12,\frac12 ^{r-1})}(\bm x)\, ,
\end{equation}
whereas for $\arac$ with $\di$:
\begin{equation}
  \chi_{\arac}^{so(2,d)}(q, \bm x)\times \chi^{so(2,d)}_{\di}(q, \bm
  x) = q^{\frac12}\,(1-q^{-2}) \,\Pd d (q^{-1}, \bm x)
  \sum_{s=0}^\infty q^{s} \,\chi^{so(d)}_{(s+\frac12,\frac12
    ^{r-1})}(\bm x)\, ,
\end{equation}
Symmetrizing the $r+1$ variables of the above expression and using
\eqref{so 2+d}, we end up with the following sum of $so(2+d)$
characters:
\begin{eqnarray}
  && \sum_{k=0}^r \left(\chi_{\rac}^{so(2,d)}(x_k, \bm x_k)\times
  \chi^{so(2,d)}_{\adi}(x_k, \bm x_k) + \chi_{\arac}^{so(2,d)}(x_k,
  \bm x_k) \times \chi^{so(2,d)}_{\di}(x_k, \bm x_k) \right)\\ && =
  \sum_{s=0}^\infty \chi^{so(d+2)}_{(s+\frac12,s+\frac12,\frac12
    ^{r-1})}(x_0, \bm x) - \sum_{s=0}^\infty
  \chi^{so(d+2)}_{(s-\frac32,s+\frac12,\frac12 ^{r-1})}(x_0, \bm x)\,
  . \nonumber
\end{eqnarray}
Using the symmetry property \eqref{sym_prop_sod}, the characters
appearing in the second sum can be expressed as the characters of bona fide
$so(2+d)$ Young diagrams, given that
\begin{equation}
  \chi^{so(d+2)}_{(s-\frac32,s+\frac12,\frac12 ^{r-1})}= -
  \chi^{so(d+2)}_{(s-\frac12,s-\frac12,\frac12 ^{r-1})}\, , \qquad
  \chi^{so(d+2)}_{(-\frac12,-\frac12,\frac12 ^{r-1})} = 0\, ,
\end{equation}
hence we end up with
\begin{empheq}[box=\eqbox]{align}
  \sum_{k=0}^r \chi_{\di\oplus\rac}^{so(2,d)}(x_k, \bm x_k) & \times
  \chi^{so(2,d)}_{\adi\oplus\arac}(x_k, \bm x_k) = \sum_{s=1}^\infty
  \sum_{m=1}^{r-1} \chi^{so(2+d)}_{(s-1,s-1,1^m)}(x_0, \bm x) \nn & +
  \sum_{s=1}^\infty \left(\chi^{so(d+2)}_{(s-1,s-1)}(x_0, \bm x) +
  2\,\chi^{so(d+2)}_{(s-\frac12,s-\frac12,\frac12 ^{r-1})}(x_0, \bm
  x)\right)\,.
  \label{Di-Rac}
\end{empheq}
Notice that we are considering a parity-invariant spin-$\tfrac12$
singleton here. Strictly speaking, the formula \eqref{Di-Rac} holds
for odd $d$. For $d=2r$, we have $\di = \di_+ \oplus \di_-$, which
leads to the appearance of a multiplicity $2$ (that we leave implicit)
for all diagrams appearing in the above identity, except
those of maximal height (namely for $m=r-1$ in the first sum on the
left hand side) which appear once with each chirality.  Again, this
result agrees with the gauge fields present in the type-AB
higher-spin gravity.

%%%%%%%%%%%%%%%%%%%%%%%%%%
\section{Lower Dimensions}
\label{low-dim}
%%%%%%%%%%%%%%%%%%%%%%%%%%

%*************************%
\subsection{Two Dimensions}\label{Twodim}
%*************************%
Although the $d=1$ case may appear\footnote{Actually, higher-spin extensions of two-dimensional Jackiw-Teitelboim gravity have been considered \cite{Alkalaev:2014qpa,Grumiller:2013swa,Alkalaev:2013fsa} in the context of AdS$_2$/CFT$_1$ holography.} somewhat degenerate from the point of view
of the higher-spin interpretation of its representations, the
characters of the conformal algebra $so(2,1)$ (see
e.g. \cite{Barut1965, Klimyk1995, Kitaev:2017hnr} for details on the
representations of this algebra) provide a useful toy model for seeing
explicitly the subtleties related to their domain of convergences.

%*******************************%
\subsubsection*{Relevant modules}
%*******************************%
The character of an $so(2,1)$ Verma module ${\cal V}_\Delta$ of
lowest-weight $\Delta$ is
\begin{equation}
  \chi^{so(2,1)}_{{}_{{\cal V}_\Delta}}(q) = \frac{q^\Delta}{1-q}
  \,\stackrel{\scriptsize |q|<1}{=}\, \sum\limits_{n=0}^\infty
  q^{\Delta+n}\,.
  \label{so21Verma}
\end{equation}
The domain of convergence of the power series in the variable $q$
around the origin $q=0$, is the disk $|q|<1$. This meromorphic
function admits an analytic continuation in the domain $|q|>1$ where
it has a convergent power series expansion in the variable $q^{-1}$
around the point at infinity $q=\infty$:
\begin{equation}
  \chi^{so(2,1)}_{{}_{{\cal V}_\Delta}}(q) =
  -\,\chi^{so(2,1)}_{{}_{{\cal V}_{1-\Delta}}}(q^{-1}) =
  -\,\frac{q^{\Delta-1}}{1-q^{-1}} \,\stackrel{\scriptsize |q|>1}{=}\,
  -\sum\limits_{n=0}^\infty q^{\Delta-1-n}\,.
\end{equation}
However, the coefficients of this power series in $q^{-1}$ are
negative integers, which prohibit the interpretation of this function
as the character of a highest-weight module.

The lowest-weight case \eqref{so21Verma} should be compared with the
character of the $so(2,1)$ Verma module $\overline{{\cal V}_\Delta}$
of highest-weight $-\Delta$:
\begin{equation}
  \chi^{so(2,1)}_{{}_{\overline{{\cal V}_\Delta}}}(q) =
  \chi^{so(2,1)}_{{}_{{\cal V}_\Delta}}(q^{-1}) =
  \frac{q^{-\Delta}}{1-q^{-1}} \, \stackrel{\scriptsize |q|>1}{=}\,
  \sum\limits_{n=0}^\infty q^{-\Delta-n}\,.
\end{equation}
The domain of convergence of this power series around the point at
infinity $q=\infty$ is the domain $|q|>1$. One should stress that it
is this power series in $q^{-1}$ with positive integer coefficients
that justifies the interpretation of this meromorphic function as the
character of a highest-weight module. However, notice the identity,
\begin{equation}\label{identityVbarV}
  \chi^{so(2,1)}_{{}_{\overline{{\cal V}_\Delta}}} =
  -\chi^{so(2,1)}_{{}_{{\cal V}_{1-\Delta}}}\,,
\end{equation}
which is valid everywhere (except at $q=0,\infty$).

The Verma module $\cV_{\Delta}$ is unitary for $\Delta>0$.  For
  $\Delta\leqslant 0$, the module is non-unitary and becomes reducible for
  non-positive half-integer $\Delta=-j$: the non-unitary module $\cV_{-j}$
  contains an invariant submodule $\cV_{j+1}$ which is unitary.  The
  irreducible module ${\cal D}_j={\cal V}_{-j}/{\cal V}_{j+1}$ is
  nothing but the $(2j+1)$-dimensional spin-$j$ module, which is non-unitary
  for $so(2,1)$ but becomes unitary for $so(3)$.  The character of the
  irreducible module ${\cal D}_j$ is
\begin{eqnarray}\label{so21spinj}
  \chi^{so(2,1)}_{{}_{{\cal D}_j}}(q) & = & \chi^{so(2,1)}_{{}_{{\cal
        V}_{-j}}}(q) - \chi^{so(2,1)}_{{}_{{\cal V}_{j+1}}}(q) =
  \sum\limits_{n=-j}^j q^n\\ & = & \frac{q^{j+\frac12} -
    q^{-j-\frac12}}{q^{\frac12} - q^{-\frac12}} = \frac{\sinh
    (j+\frac12)\beta}{\sinh\frac{\beta}2} \equiv
  \chi_j^{so(3)}(i\beta) \,.
\end{eqnarray}
Since this $so(2,1)$-module is finite-dimensional, it is both
lowest-weight and highest-weight, which translates into the property
$\chi^{so(2,1)}_{{}_{{\cal D}_j}}(q^{-1})=\chi^{so(2,1)}_{{}_{{\cal
      D}_j}}(q)$. The Laurent series in the right of \eqref{so21spinj}
contains negative powers, seen either as a power series in $q$ or as a
power series in $q^{-1}$, but it converges for any $q\neq 0,\infty$.

%*************************************%
\subsubsection*{Flato-Fronsdal theorem}
%*************************************%
The Clebsch-Gordan decomposition of these finite-dimensional $so(2,1)$-modules is the celebrated rule
\begin{equation}
  \chi^{so(2,1)}_{{}_{{\cal D}_{j_1}}}(q) \times
  \chi^{so(2,1)}_{{}_{{\cal D}_{j_2}}}(q) =
  \sum_{j=|j_1-j_2|}^{j_1+j_2} \chi^{so(2,1)}_{{}_{{\cal D}_j}}(q)
  \label{CG_decompo}
\end{equation}
as can be checked by an explicit computation of the product of
characters of the form \eqref{so21spinj}. One may also consider the
tensor product of two lowest-weight Verma modules,
\begin{equation}\label{d=1FF}
  \chi^{so(2,1)}_{{}_{{\cal V}_{\Delta_1}}}(q) \times
  \chi^{so(2,1)}_{{}_{{\cal V}_{\Delta_2}}}(q) =
  \frac{q^{\Delta_1+\Delta_2}}{(1-q)^2} \,\stackrel{\scriptsize
    |q|<1}{=}\, \sum_{n=0}^{\infty} \chi^{so(2,1)}_{{}_{{\cal
        V}_{\Delta_1+\Delta_2+n}}}(q)
\end{equation}
where we used the formula \eqref{so21Verma} and expanded $(1-q)^{-1}$
in power series around the origin (valid for $|q|<1$) to obtain the
result,
\begin{equation}\label{d=1FF!}
  {\cal V}_{\Delta_1}\otimes{\cal V}_{\Delta_2} =
  \bigoplus_{n=0}^{\infty} {\cal V}_{\Delta_1+\Delta_2+n}\,,
\end{equation}
which we will use later on.

The formula \eqref{d=1FF!} is a sort of $d=1$ analogue of the
Flato-Fronsdal theorem, in the sense that it is a decomposition of the
tensor product of two lowest-weight infinite-dimensional modules.
However, it becomes much more delicate to manipulate the tensor
product of infinite-dimensional modules where one module is
lowest-weight and the other one is highest-weight (as in the twisted
Flato-Fronsdal theorem) because the resulting module is neither lowest
nor highest weight. Therefore, its character involves Laurent series
which should be treated with care.

%*********************************************%
\subsubsection*{Twisted Flato-Fronsdal theorem}
%*********************************************%
Treating the characters as meromorphic functions and performing the
power series expansion, one can write the equality as
\begin{eqnarray}\label{d=1twFF}
  \chi^{so(2,1)}_{{}_{{\cal V}_{\Delta_1}}}(q) \times
  \chi^{so(2,1)}_{{}_{\overline{{\cal V}_{\Delta_2}}}}(q) & = &
  -\,\chi^{so(2,1)}_{{}_{{\cal V}_{\Delta_1}}}(q) \times
  \chi^{so(2,1)}_{{}_{{\cal V}_{1-\Delta_2}}}(q) =
  \frac{-\,q^{1+\Delta_1-\Delta_2}}{(1-q)^2}
  \nonumber\\ &\stackrel{\scriptsize |q|<1}{=}& - \sum_{n=0}^{\infty}
  \chi^{so(2,1)}_{{}_{{\cal V}_{1+\Delta_1-\Delta_2+n}}}(q)\,.
\end{eqnarray}
Unfortunately, the last line has negative coefficients as a power
series (in $q$ or in $q^{-1}$), which confirms that it lacks any sound
group-theoretical intepretation as lowest or highest weight module.

In fact, the group-theoretical interpretation of the factors in the
left-hand-side of \eqref{d=1twFF}, as characters of lowest vs
highest weight modules, holds in two distinct domains ($|q|<1$ versus
$|q|>1$). The same remark holds for the infinite sum
$\bigoplus_{j=0}^\infty {\cal D}_j$ of all irreducible
finite-dimensional $so(2,1)$ modules, which could be considered as a
$d=1$ analogue of the adjoint module: it is tempting (and will be justified below) to compute
the formal sum $\sum_{j=0}^\infty\chi^{so(2,1)}_{{}_{{\cal D}_j}}(q)$
via the formula \eqref{so21spinj} as the difference of the two series
in the right-hand-sides of
\begin{equation}
  \sum_{j=0}^\infty\chi^{so(2,1)}_{{}_{{\cal V}_{-j}}}(q)
  \,\stackrel{\scriptsize |q|<1}{=}\,\frac{-q}{(1-q)^2}\quad
%\end{equation}
  \mbox{and} 
%\begin{equation}
  \quad\sum_{j=0}^\infty\chi^{so(2,1)}_{{}_{{\cal V}_{j+1}}}(q)
  \,\stackrel{\scriptsize |q|>1}{=}\,\frac{q}{(1-q)^2}\,,
\end{equation}
to obtain the meromorphic function
\begin{equation}\label{magicidentity}
  \sum_{j=0}^\infty\chi^{so(2,1)}_{{}_{{\cal D}_{j}}}(q) =
  -\frac1{2\sinh^2\frac{\beta}2} = -\frac{2\,q}{(1-q)^2}\,,
\end{equation}
which has negative coefficients as a power series, whether in $q$ or
$q^{-1}$. Again, this fact precludes any clear group-theoretical
interpretation.  Nevertheless, the equality \eqref{magicidentity} is
valid. Indeed, the following trigonometric series is divergent but can
be evaluated via Cesaro's resummation:\footnote{See e.g. \cite{Knopp}
  (Chap. XIII, Sect. 60, Ex. 7).}
\begin{equation}
  \sum_{n=0}^\infty\sin[(n+\tfrac12)\,x] =
  \frac{1}{2\,\sin\frac{x}2}\,,
\end{equation}
from which one deduces the character formula,
\begin{equation}
  \sum_{j=0}^\infty\chi_j^{so(3)}(\alpha) =
  \frac1{2\sin^2\frac{\alpha}2}\,,
\end{equation}
which holds as a distribution and whose Wick rotation is \eqref{magicidentity}. Therefore, one can write
the relation,
\begin{empheq}[box=\eqbox]{align}
  \label{d=1twistFF}
  \sum\limits_{j=0}^\infty\chi^{so(2,1)}_{{}_{{\cal D}_{j}}}(q) = 2\,
  \chi^{so(2,1)}_{{}_{{\cal V}_{\Delta}}}(q) \times
  \chi^{so(2,1)}_{{}_{\overline{{\cal V}_{\Delta}}}}(q)\,,
\end{empheq}
as a $d=1$ analogue of the twisted Flato-Fronsdal
theorem.\footnote{The tensor product of two discrete series
  representations of respectively highest and lowest weight kind has
  been studied in \cite{Repka1978} (see \cite{Kitaev:2017hnr} for a
  recent review): its usual decomposition contains in fact principal
  as well as complementary series representations on top of discrete
  series ones.  Therefore, naively the tensor product decomposition of
  \cite{Repka1978} is not reproduced by our $so(2,1)$ character
  analysis. However, these two approaches are difficult to compare because
  continuous (principal and complementary) series representations have
  unbounded spectrum of $E$. Nevertheless, the decomposition
  \eqref{d=1twistFF} will be justified below.}

In order to provide a concrete realization of the isomorphism,
\begin{equation}\label{isom}
  {\cal V}_{\Delta} \otimes\overline{{\cal V}_{\Delta}} =
  \bigoplus_{j=0}^\infty {\cal D}_j\,,
\end{equation}
let us consider the elements $L_+^m\,|\Delta\rangle$ of the Verma
module $\cV_\Delta$ of $so(2,1)$ generated by the lowest-weight vector
$|\Delta\rangle$. Mimicking the discussion of the oscillator
realization in Section \ref{sec: 4d}, one can introduce two concrete
realizations of the generators of ${\rm End}(\cV_\Delta)$, and compare the
$so(2,1)$ decomposition in the two cases:
\begin{itemize}
\item Firstly, as elements of $\cV_\Delta \otimes
  \overline{\cV_\Delta}$\,,
  \begin{equation}
    T^{m;n}\,:=\,L_+^m|\Delta\ra\la\Delta|L_-^n\,\,.
  \end{equation}
  with $m,n\in\mathbb N$\,. The identities
  \begin{eqnarray}{}
    [L_-,T^{m;n}] &=& m(2\,\Delta+m-1)\, T^{m-1;n}-T^{m;n+1} \,,
    \nn {}
    [L_+,T^{m;n}] &=& T^{m+1;n}-n(2\,\Delta+n-1)\,T^{m;n-1}\,.
  \end{eqnarray}
  allows to identify the lowest (or highest) weight states of the
  spin-$j$ submodule $\cD_j\subset {\rm End}(\cV_\Delta)$: they are the
  elements
  \begin{equation}
    Y^j_{-j} := \sum_{m=0}^\infty \frac{\Gamma(2\Delta-1)}{m!\,
      \Gamma(2\Delta+m)}\,T^{m;m+j}\,, \quad Y^j_j :=
    \sum_{m=0}^\infty \frac{\Gamma(2\Delta-1)}{m!\,
      \Gamma(2\Delta+m)}\,T^{m+j;m}
  \end{equation}
  of ${{\cal V}_{\Delta}}\otimes\overline{{\cal V}_{\Delta}}$\,.
\item Secondly, as elements in the universal enveloping
  algebra\,\footnote{This is motivated by the fact that there exists a
    natural inclusion $\U\big(\mathfrak{g}\big) / {\rm Ann}(M)
    \hookrightarrow End(M)$ for $\mathfrak g$ a Lie algebra and
    $M$ a $\mathfrak g$-modules with annihilator ${\rm Ann}(M)$
    \cite{Joseph1980}.} ${\cal U}\big(so(2,1)\big)$\,. Let us recall
  that the Lie algebra $so(2,1)$ is spanned by the generators
  $\{E,L_+,L_-\}$ obeying the commutation relations
  \begin{equation}
    [E,L_\pm] = \pm L_{\pm}\, ,  \qquad [L_-,L_+] =
    2E\, .
  \end{equation}
  Following closely the presentation of
  the universal enveloping
  algebra of $so(2,1) \cong sl(2,\R)$ \cite{Fradkin:1990ki,
    Fradkin:1990ir} (see also \cite{Hewitt2012, Dixmier1977} for more
  details), we can consider the basis given by
  \begin{equation}
    X^{j,k}_{m} := c_{j,m}\, {\Ca}^k\, {\rm ad}_{L_-}^{m} L_+^j\, ,
    \quad \text{with} \quad 0 \leqslant m \leqslant 2j\, , \qquad k
    \in\N\, ,
    \label{basisUEA}
  \end{equation}
  and where $\Ca:=E^2-\tfrac12(L_+ L_- + L_- L_+)=E(E-1)+L_+L_-$ is
  the Casimir operator of $so(2,1)$ generating the center of the
  universal enveloping algebra, while the coefficients $c_{j,m}$ are
  normalization factors. The decomposition of $\U\big(so(2,1)\big)$ in
  terms of finite-dimensional modules under the adjoint action of
  $so(2,1)$ can be read more easily, as
  \begin{equation}
    {\rm ad}_{L_+} L_+^j = 0\, , \quad \text{and} \quad {\rm ad}_{E}
    L_+^j = j\, L^j_+\, , \qquad \forall j \in \N\, ,
  \end{equation}
  i.e. $L_+^j$ defines a highest-weight vector of weight $j$, and thus
  the various power of ${\rm ad}_{L_-}^m$ for $m=0,\dots,2j$ define
  the elements of this spin-$j$ module. However, each of those modules
  have an infinite multiplicity in $\U\big( so(2,1) \big)$ since they
  appear dressed with arbitrary power of the quadratic Casimir
  operator of $so(2,1)$ according to \eqref{basisUEA}. Considering the
  quotient,
  \begin{equation}\label{gllambda}
    gl[\lambda] := \frac{\U\big(so(2,1)\big)}{\langle \Ca -
      \mu_\lambda \id \rangle}\, , \quad \text{with} \quad \mu_\lambda
    := \frac{\lambda^2-1}4\, ,
  \end{equation}
  i.e. by modding out the ideal ${\cal I}_\lambda = \big( \Ca -
  \mu_\lambda \id \big)\, \U\big(so(2,1)\big)$ of the universal enveloping
  algebra, the vector
  space of the resulting algebra decomposes as the direct sum \cite{Feigin1988}
  \begin{equation}
    gl[\lambda] = \bigoplus_{j=0}^\infty {\cal D}_{j}\, .
    \label{decompo_gl_lambda}
  \end{equation}
  In other word, by fixing the value of the Casimir operator, one
  lifts the (infinite) degeneracy of the finite-dimensional modules. A
  basis of $gl[\lambda]$ is therefore given by:
  \begin{equation}
    V_n^j := (-1)^{j+n}\, \frac{(j+n)!}{(2j)!}\, {\rm
      ad}_{L_-}^{j-n} L^{j}_+\, , \qquad \text{with} \quad |n|
    \leqslant j\, ,
  \end{equation}
  where the generators $\{V^j_n\}_{-j \leqslant n \leqslant j}$ span the spin $j$ module $\cD_j$ in the decomposition \eqref{decompo_gl_lambda}. One can also describe this decomposition in more covariant terms by
  making use of the generators $\tilde{L}^A :=
  \frac12\epsilon^{ABC}L_{BC}$ of $so(2,1)$. All elements of ${\cal
    U}\big(so(2,1)\big)$ can be written as linear combinations of
  elements of the form $P({\cal C}_2)\,\tilde{L}_{\{A_1}\cdots
  \tilde{L}_{A_j\}}$ where $P({\cal C}_2)$ is a polynomial in the quadratic
  Casimir ${\cal C}_2 = -\tilde{L}^A\tilde{L}_A$ and the brackets over the
  indices indicates total symmetrization over all indices and
  traceless projection. Therefore, ${\cal U}\big(so(2,1)\big)$
  branches in spin-$j$ submodules. Moreover, the quadratic Casimir
  operator takes the value ${\cal C}_2 = \Delta(\Delta-1)$ on the Verma
  module $\cV_\Delta$ with $\Delta = \frac{1\pm \lambda}2$. Therefore,
  the elements $\tilde{L}_{\{A_1}\cdots \tilde{L}_{A_j\}}$ provide a
  covariant basis of ${\rm End}(\cV_\Delta)$\,.
\end{itemize}

%***************************%
\subsection{Three Dimensions}
%***************************%

The $d=2$ conformal algebra is a direct sum of two $d=1$ conformal
algebras: $so(2,2)=so(2,1)\oplus so(2,1)$. Accordingly, the $so(2,2)$
Verma module $\cV(\Delta, s)$ is related to the that of $so(2,1)$ as
\be \cV(\Delta,s)=\cV_{\frac{\Delta+s}2} \otimes
\cV_{\frac{\Delta-s}2}.  \ee Note that the spin $s$ can take negative
values here since they are eigenvalues of $so(2)$.  Introducing the
variables,
\begin{equation}\label{zzbar}
  z=q\,x=e^{-\b+i\,\a}\,, \qquad \bar z=q\,x^{-1}=e^{-\b-i\,\a}\,,
\end{equation}
for the $so(2,2)$ weights,
the character of $\cV(\Delta,s)$ 
is given by
\be
 \chi^{so(2,2)}_{{\cal V}(\Delta,s)}(z,\bar z) =
  \chi^{so(2,1)}_{\cV_{\frac{\Delta+s}2}}(z)\,
  \chi^{so(2,1)}_{\cV_{\frac{\Delta-s}2}}(\bar z)=\frac{z^{\frac{\Delta+s}2}\,{\bar z}^{\frac{\Delta-s}2}}{(1-z)(1-\bar
    z)}\,.
\ee

%*******************************%
\subsubsection*{Relevant modules}
%*******************************%
The character of a scalar field of lowest energy $E_0=\Delta$ is
\begin{equation}
  \chi^{so(2,2)}_{\cD(\Delta,0)}(z,\bar z) =
  \chi^{so(2,1)}_{\cV_{\frac\Delta2}}(z)\,
  \chi^{so(2,1)}_{\cV_{\frac\Delta2}}(\bar z) = \frac{(z\,\bar
    z)^{\Delta/2}}{(1-z)(1-\bar z)}\,.
\end{equation}
In the limit when the conformal weight of the scalar field goes to the
unitarity bound, $\Delta\to \tfrac{d-2}2=0$, one finds
\begin{equation}\label{limitzero}
  \chi^{so(2,2)}_{\cD(\Delta\to 0,0)}(z,\bar z)=\frac{1}{(1-z)(1-\bar
    z)} =1+\frac{z}{1-z}+\frac{\bar z}{1-\bar
    z}+\chi^{so(2,2)}_{\cD(2,0)}(z,\bar z)\,,
\end{equation}
which can be understood from the property of the $so(2,1)$-module,
\begin{equation}
  \cV_{0}=1\oplus \cV_1\,,
\end{equation}
where $1$ denotes the trivial representation of $so(2,1)$.  The last
term in \eqref{limitzero} reflects the appearance of a submodule,
$\cD(2,0)\subset \cV(0,0)$. The appearance of such a submodule holds
in any dimension: $\cD(\tfrac{d+2}2,\bm 0)\subset \cV(\tfrac{d-2}2,\bm
0)$, and the Rac has been defined uniformly for all dimensions as the
quotient $\cV(\tfrac{d-2}2,\bm 0)/\cD(\tfrac{d+2}2,\bm 0)$, cf
\eqref{Racterminology}. However, in $d=2$ the
Rac\,=\,$\cV(0,0)/\cD(2,0)$ is a \textit{reducible} module: 
its
character reads
\begin{equation}\label{charRac}
  \chi^{so(2,2)}_{\rac}(z,\bar z) = 1 + \frac{z}{1-z} + \frac{\bar
    z}{1-\bar z}\,,
\end{equation}
where each of the three terms correspond to the characters of
different irreducible modules.  The first term is the character of the
trivial module, ${\cal D}(0,0)$, which corresponds to a zero-mode in
field-theoretical terms. In fact, the Rac always describes a conformal
scalar in dimension $d$ but the zero-mode has canonical conformal
weight (i.e. $\tfrac{d-2}2=0$) only for $d=2$.  The last two terms are
the characters of the modules $\cD(1,+1)$ and $\cD(1,-1)$\,, which
correspond to the left and right moving scalar fields living on the
$d=2$ conformal boundary, respectively.

Let us introduce the notations, 
\begin{equation}
  \cV(\Delta,s)_0:=\cV(\Delta,+s) \oplus
  \cV(\Delta,-s)\,,\quad\mbox{and}\quad
  \cD(\Delta,s)_0:=\cD(\Delta,+s) \oplus \cD(\Delta,-s)\,,
\end{equation}
for parity-invariant combinations.  The Di module, describing the
$d=2$ conformal spinor, is $\di = \cD(\frac12, \frac12)_0 = \di_+
\oplus \di_-$ with $ \di_\pm=\cD(\frac12,\pm\frac12)$, cf \eqref{Dipm}
for $d=2$.  For spin $s\geqslant 1$, the irreducible module $\cD(s,\pm
s)$ is given by
\begin{eqnarray}
  \cD(s,+s)\eq {\cal V}(s,s)/{\cal V}(s+1,s-1)=(\cV_{s}\otimes
  \cV_0)/(\cV_{s}\otimes \cV_1)=\cV_{s}\otimes 1,\nn \cD(s,-s)\eq
     {\cal V}(s,-s)/{\cal V}(s+1,-s+1)=(\cV_0\otimes
     \cV_s)/(\cV_1\otimes \cV_s) =1\otimes \cV_s\,.
\end{eqnarray}
The corresponding character reads
\begin{equation}
  \chi^{so(2,2)}_{\cD(s,+s)}(z,\bar z)=\frac{z^s}{1-z}\,,
  \qquad
  \chi^{so(2,2)}_{\cD(s,-s)}(z,\bar z)=\frac{\bar z^s}{1-\bar z}\,.
  \label{char_ms}
\end{equation}
For $s\geqslant 1$, the parity-invariant  module $\cD(s,s)_0=\cV(s,s)_0/\cV(s+1,s-1)_0$ describes a conserved spin-$s$
conformal current in $d=2$ dimensions or, equivalently, a massless
spin-$s$ $AdS_3$ field. Their holomorphic decomposition as a direct sum,
$\cD(s,+s)\oplus \cD(s,-s)$, reflects the standard lore that massless
fields do not have propagating degrees of freedom in three
dimensions. In this sense one may consider all these fields as
spin-$s$ singletons (as pointed out in e.g. \cite{Vasiliev:2004cm,
  Boulanger:2014vya}).  For $s=1$, there is a subtlety in the
interpretation: the parity-invariant combination of the two
irreducible modules, $\cD(1,1)_0=\cD(1,+1)\oplus \cD(1,-1)$, can be obtained as
the quotient $\cV(1,1)_0/\big(2\,\cD(2,0)\big)$,
whereas the quotient 
\be
\mbox{Max}:=\cV(1,1)_0\,/\,\cD(2,0)\,,
\ee which
would be the analogue of the higher-dimensional case of spin-$1$ massless field, is
isomorphic to the reducible module $\cD(1,1)_0\inplus
\cD(2,0)$.\footnote{The notation $U=V\inplus W$ indicates thats $U$ is the semidirect sum of the modules $V$ and $W$ thereby indicating that the $W$ is a submodule of $U$ and that $V=U/W$.} The former %(irreducible)
module corresponds to the
$U(1)\times U(1)$ Chern-Simons theory whereas the latter %(reducible)
module corresponds to the Maxwell theory.  The latter is Hodge dual to
a massless scalar in three dimensions without zero-mode.  In
group-theoretical terms, this equivalence translates into the
isomorphisms 
\be
\cV(0,0) =\cD(0,0)\inplus \mbox{Max}=\rac\inplus\cD(2,0)\,,
\ee
since $\mbox{Max}=\cD(1,1)_0\inplus
\cD(2,0)$ and $\rac=\cD(0,0)\inplus\cD(1,1)_0$.

The list of the relevant parity-invariant $so(2,2)$ modules and their field-theoretical
interpretations are summarized in Table \ref{list}.  Note that we
also included the finite-dimensional irreducible modules
$\cD(1-s,s-1)=(\cD_{s-1}\otimes 1)\oplus (1\otimes \cD_{s-1})$
describing conformal Killing tensors of rank $s-1$.

\begin{table}[h]
  \resizebox{0.99\textwidth}{!}{
    \begin{tabular}{|c|c|c|c|}
      \hline Modules & AdS$_3$ & CFT$_2$ & Equivalent descriptions
      \\ \hline\hline $\cD(0,0)$ & Vacuum & Constant zero-mode &
      $1\otimes 1$ \\\hline $\cD(0,0)\inplus\cD(1,1)_0$ & Rac & Conformal
      scalar & ${\cal V}(0,0)/\cD(2,0)$\\\hline $\cD(\frac12,\frac12)_0$
      & Di & Conformal spinor & $(\cV_{\frac12}\otimes 1)\oplus
      (1\otimes \cV_{\frac12})$ \\\hline $\cD(1,1)_0$ & $U(1)^{\otimes
        2}$ Chern-Simons & Chiral bosons & $(\cV_1\otimes 1)\oplus
      (1\otimes \cV_1)$ \\\hline $\cD(1,1)_0\inplus\cD(2,0)$ & Maxwell
      field & Conserved current & ${\cal V}(1,1)_0/\cD(2,0)$ \\\hline
      $\cD(s,s)_0$ & Massless spin-$s$ field & Conserved spin-$s$
      current & $(\cV_{s}\otimes 1)\oplus (1\otimes \cV_s)$ \\\hline
      $\cD(1-s,s-1)$ & Killing tensor & Conformal Killing tensor &
      $(\cD_{s-1}\otimes 1)\oplus (1\otimes \cD_{s-1})$ \\\hline
    \end{tabular}
  }
  \label{list}
  \caption{List of relevant $so(2,2)$ modules and their
    field-theoretical interpretations}
\end{table}

%*************************************%
\subsubsection*{Flato-Fronsdal theorem}
%*************************************%
Given the identities,
\begin{equation}
  \sum_{s=2}^\infty\chi^{so(2,2)}_{\cD(s,s)_0}(z,\bar z) =
  \left(\frac{z}{1-z}\right)^2 + \left(\frac{\bar z}{1-\bar
    z}\right)^2\,,
\end{equation}
and
\begin{equation}
  \sum_{s=1}^\infty\chi^{so(2,2)}_{\cD(s,s)_0}(z,\bar z) =
  \left(\frac{z^\frac12}{1-z}\right)^2 + \left(\frac{\bar
    z^\frac12}{1-\bar z}\right)^2\,,
\end{equation}
it is tempting to write the Flato-Fronsdal theorem for $so(2,2)$ as 
\begin{equation}
  \left(\chi^{so(2,2)}_{\cD(1,1)_0}(z,\bar z)\right)^2 = 2\,
  \chi^{so(2,2)}_{\cD(2,0)}(z,\bar z) + \sum_{s=2}^\infty
  \chi^{so(2,2)}_{\cD(s,s)_0}(z,\bar z)\,,
  \label{RacRac}
\end{equation}
and
\begin{equation}
  \left(\chi^{so(2,2)}_{\di}(z,\bar z)\right)^2 = 2\,
  \chi^{so(2,2)}_{\cD(1,0)}(z,\bar z) +
  \sum_{s=1}^\infty\chi^{so(2,2)}_{\cD(s,s)_0}(z,\bar z)\,.
  \label{DiDi}
\end{equation}
The equation \eqref{DiDi} is consistent with the generalized
Flato-Fronsdal theorem in general dimensions \cite{Vasiliev:2004cm},
while the case \eqref{RacRac} is consistent with the tensor product of
spin-$1$ singleton, namely the type-C case, in general dimensions
\cite{Dolan:2005wy}. In fact, the Flato-Fronsdal theorem in the scalar
case is rather
\begin{eqnarray}
  \left(\chi^{so(2,2)}_{\rac}(z,\bar z)\right)^2\eq \left[1 +
    \chi^{so(2,2)}_{\cD(1,1)_0}(z,\bar z) +
    \chi^{so(2,2)}_{\cD(2,0)}(z,\bar z)\right]\, +\,
  \left[\chi^{so(2,2)}_{\cD(1,1)_0}(z,\bar z) +
    \chi^{so(2,2)}_{\cD(2,0)}(z,\bar z)\right]\nn && +\,
  \sum_{s=2}^\infty\chi^{so(2,2)}_{\cD(s,s)_0}(z,\bar z),
\end{eqnarray}
where the scalar field (i.e. the first term
between squared brackets) is described by the reducible module
$\cV(0,0)$ and contains a non-normalizable zero-mode, and the spin-$1$
field (i.e. the second term between squared brackets) corresponds to
Maxwell theory.
%(i.e. the reducible module $\cD(1,1)_0+\cD(2,0)$).  
This
version of Flato-Fronsdal theorem has been considered in
\cite{Giombi:2014iua} where the IR divergence caused by the zero-mode
has been thrown away.

Let us consider now the tensor product of two singletons of spin $s$
and $s'$. For the same chiralities, the $d=1$ formula \eqref{d=1FF}
implies that the tensor product decomposes into the direct sum of all
massless fields of spin $\sigma \geqslant s+s'$ and of chirality $\pm$
as
\begin{equation}
  \chi^{so(2,2)}_{\D(s,\pm s)}\, \chi^{so(2,2)}_{\D(s',\pm s')} =
  \sum_{\sigma=s+s'}^\infty \chi^{so(2,2)}_{\D(\sigma,\pm \sigma)}\,.
\end{equation}
For the opposite chiralities, 
the tensor product reduces to a single
massive field of spin $s-s'$ and of minimal energy $s+s'$\,:
\begin{equation}
  \chi^{so(2,2)}_{\D(s,+s)}\, \chi^{so(2,2)}_{\D(s',-s')}
  = \chi^{so(2,2)}_{\D(s+s',s-s')}\, .
\end{equation}
Collecting the previous decompositions, we can write the tensor
product of two parity-invariant spin-$s$ and spin-$s'$ singletons, for
$s\neq s'$ as
\begin{equation}
  \chi^{so(2,2)}_{\D(s,s)_0}(z,\bar z)\,
  \chi^{so(2,2)}_{\D(s',s')_0}(z,\bar z) =
  \chi^{so(2,2)}_{\D(s+s',|s-s'|)_0}(z,\bar z) +
  \sum_{\sigma=s+s'}^\infty \chi^{so(2,2)}_{\D(\sigma,\sigma)_0}(z, \bar
  z)\, ,
\end{equation}
and for $s=s'\neq0$ as
\begin{equation}
 \left( \chi^{so(2,2)}_{\D(s,s)_0}(z,\bar z)\right)^2 =
 2\,\chi^{so(2,2)}_{\D(2s,0)}(z,\bar z) + \sum_{\sigma=2s}^\infty
 \chi^{so(2,2)}_{\D(\sigma,\sigma)_0}(z, \bar z)\, ,
 \label{spin s FF3}
\end{equation}
which, in particular, reproduces the results for the $\di$ in
\eqref{DiDi} and for the spin-$1$ singleton in \eqref{RacRac}.

%*********************************************%
\subsubsection*{Twisted Flato-Fronsdal theorem}
%*********************************************%

The finite-dimensional irreps of $so(4)$ are characterised by weights
$(s_1,s_2)$ where $s_1$ and $|s_2|$ are both non-negative integers (or
both half-integers) but $s_2$ can be negative.  It is isomorphic to
the tensor product of two $so(3)$ finite dimensional modules,
$\cD_{\frac{s_1+s_2}2}\otimes \cD_{\frac{s_1-s_2}2}$.  In terms of the
characters, this is
\begin{equation}
  \chi^{so(4)}_{(s_1, s_2)}(q, x) = \chi^{so(3)}_{\frac{s_1+s_2}2}(z)
  \, \chi^{so(3)}_{\frac{s_1-s_2}2}(\bar z)\,,
\end{equation}
and in particular,
\begin{equation}
  \chi_{(s-1,s-1)}^{so(4)}(q, x) = \chi^{so(3)}_{s-1}(z) \, , \qquad
  \chi_{(s-1,1-s)}^{so(4)}(q, x) = \chi^{so(3)}_{s-1}(\bar z)\, .
\end{equation}
The characters of the chiral spin-$s$ anti-singletons read
\begin{equation}
  \chi^{so(2,2)}_{\overline{\D(s,+s)}}(z) =
  \chi^{so(2,2)}_{\D(s,+s)}(z^{-1})\, , \quad \text{and} \quad
  \chi^{so(2,2)}_{\overline{\D(s,-s)}}(\bar z) =
  \chi^{so(2,2)}_{\D(s,-s)}(\bar z^{-1})\,.
\end{equation}
The formula \eqref{d=1twistFF} implies the identities:
\begin{empheq}[box=\eqbox]{align}\label{d=2twistFF}
  \qquad &2\, \chi^{so(2,2)}_{\D(s,+s)}(z) \,
  \chi^{so(2,2)}_{\overline{\D(s,+s)}}(z) = \sum_{\s=1}^\infty
  \chi^{so(4)}_{(\s-1,\s-1)}(z)\,, \qquad \nn \qquad &2\,
  \chi^{so(2,2)}_{\D(s,-s)}(\bar z) \,
  \chi^{so(2,2)}_{\overline{\D(s,-s)}}(\bar z) = \sum_{\s=1}^\infty
  \chi^{so(4)}_{(\s-1,1-\s)}(\bar z)\,, \qquad
\end{empheq}
which can be seen as the twisted Flato-Fronsdal in the chiral (or
antichiral) sector.  Notice however that the massless fields having
the Killing tensors $\cD(\sigma-1, \pm(\sigma-1))$ with $\sigma=1, \dots,
2s-1$ are not present in the field content given by the Flato-Fronsdal
theorem \eqref{spin s FF3}.  In fact, these additional modules can be
factorized as
\begin{equation}
  \sum_{\sigma=1}^{2s-1} \chi^{so(4)}_{(\sigma-1,\pm(\sigma-1))} =
  \left( \chi^{so(4)}_{(s-1,\pm(s-1))}\right)^2=\chi^{so(4)}_{(s-1,\pm(s-1))}\chi^{so(4)}_{\overline{(s-1,\pm(s-1))}}\,,
  \label{d=2 factorization}
\end{equation}
and can be interpreted as the endomorphism algebra of the finite-dimensional module $\cD(s-1,s-1)=\cD_{s-1}\otimes 1$ (or,
respectively, $\cD(s-1,1-s)=1\otimes\cD_{s-1}$).  This reflects that the massless
spin-$s$ module $\cD(s,s)=\cV_s\otimes 1$ appears as a submodule of
$\cV(1-s,1-s)=\cV_{1-s}\otimes 1$, whose irreducible part corresponds to $\cD_{s-1}=\cV_{1-s}/\cV_{s}$.   The
appearance of the submodule should be related to the fact that the
spin-$s$ singletons (for $s \geqslant 1$) possess gauge symmetries.
We will see in the next section that a similar phenomenon takes place
for higher-spin singletons in $d+1=5$ dimensions.
  
Let us now consider the twisted Flato-Fronsdal theorem for the type-A
and type-B models.  We note first that the Rac and Di characters satisfy
\begin{equation}
  \chi^{so(2,2)}_{\arac/\adi}(z,\bar z) =
  \chi^{so(2,2)}_{\rac/\di}(z^{-1},\bar z^{-1})
  =-\chi^{so(2,2)}_{\rac/\di}(z,\bar z).
  \label{2d sym prop}
\end{equation}
For the Di module, the above is due to the symmetry property of
$\chi^{so(2,2)}_{\D(\frac12,\pm\frac12)}$.  For the Rac module, this
is possible only when we include the zero-mode.  By taking the product
of singleton and anti-singleton characters, we obtain
\begin{eqnarray} 
  \chi^{so(2,2)}_{\rac}(z,\bar z)\, \chi^{so(2,2)}_{\arac}(z,\bar
  z)\eq -\left(1+\frac{z}{1-z}+\frac{\bar z}{1-\bar z}\right)^2,\nn
  \chi^{so(2,2)}_{\di}(z,\bar z)\, \chi^{so(2,2)}_{\adi}(z,\bar z)\eq
  -\left(\frac{z^\frac12}{1-z}+\frac{\bar z^\frac12}{1-\bar
    z}\right)^2.
\end{eqnarray}
Analogously to the higher dimensional cases, we take the
symmetrization prescription. The exchange $q\leftrightarrow x$
translates into $(z,\bar z)\leftrightarrow (z,1/{\bar z})$ according
to the definition \eqref{zzbar}.  Explicit computation leads to
\begin{eqnarray}
  \chi_{\rac}(z,\bar z)\, \chi_{\arac}(z,\bar z) +\chi_{\rac}(z,\bar
  z^{-1})\, \chi_{\arac}(z,\bar z^{-1}) \eq
  -1-\frac{2\,z}{(1-z)^2}-\frac{2\,\bar z}{(1-\bar z)^2}\,,\nn
  \chi_{\di}(z,\bar z)\, \chi_{\adi}(z,\bar z)+ \chi_{\di}(z,\bar
  z^{-1})\, \chi_{\adi}(z,\bar z^{-1}) \eq
  -\frac{2\,z}{(1-z)^2}-\frac{2\,\bar z}{(1-\bar z)^2}\,,
\end{eqnarray}
where we suppressed the superscript $so(2,2)$ for compactness of the
expressions.  Comparing these results with the $so(4)$ characters, we
find
\begin{empheq}[box=\eqbox]{align}\label{tFFd=2Rac}
  &\chi^{so(2,2)}_{\rac}(q, x)\, \chi^{so(2,2)}_{\arac}(q, x) +
  \chi^{so(2,2)}_{\rac }(x, q)\, \chi^{so(2,2)}_{\arac }(x, q) \nn
  &=\,\chi^{so(4)}_{(0,0)}(q,x)+ \sum_{s=2}^\infty
  \chi^{so(4)}_{(s-1,s-1)_0}(q,x)\,,
\end{empheq}
for the type-A model, and
\begin{empheq}[box=\eqbox]{align}\label{tFFd=2Di}
  &\chi^{so(2,2)}_{\di}(q, x)\, \chi^{so(2,2)}_{\adi}(q, x) +
  \chi^{so(2,2)}_{\di}(x, q)\, \chi^{so(2,2)}_{\adi}(x, q) \nn
  &=\,2\,\chi^{so(4)}_{(0,0)}(q,x)+ \sum_{s=2}^\infty
  \chi^{so(4)}_{(s-1,s-1)_0}(q,x)\,,
\end{empheq}
for the type-B model.  Here, the $so(4)$ module $(r,r)_0$ means the
direct sum of the $(r,r)$ and $(r,-r)$ modules.  Remark that the
type-A model contains the trivial module $(0,0)$ once whereas the
type-B model has it twice.  They correspond to the Killing tensors of
the Maxwell and $U(1)\times U(1)$ Chern-Simons theory, respectively.\\

In AdS$_3$/CFT$_2$, the higher-spin holography \cite{Gaberdiel:2010ar,
  Gaberdiel:2010pz, Gaberdiel:2012uj} involves more models than in
higher dimensions: in fact, there is a one-parameter family of models
which includes the type-A and type-B models as particular points in
the parameter space. This parameter (corresponding to the 't Hooft
coupling in the AdS/CFT context) is often denoted by $\lambda$
\cite{Feigin1988, Fradkin:1990ir} (see also \cite{Bergshoeff:1989ns,
  Bordemann:1989zi}) (or sometimes $\nu$ \cite{Vasiliev:1989re,
  Prokushkin:1998bq}) and the chiral part of the underlying higher
spin algebra is referred to as $hs[\lambda]$ and its asymptotic
extension as $\cW_\infty[\lambda]$ \cite{Campoleoni:2010zq,
  Campoleoni:2011hg, Henneaux:2010xg, Henneaux:2012ny}.  The former
higher-spin algebra is the simple\footnote{Except for $\lambda=N$ a
  positive integer, in which case it contains an infinite-dimensional
  ideal as described in \eqref{I[N]}.} subalgebra of the Lie algebra
\eqref{gllambda}: \be gl[\lambda]={\mathbb R}\oplus hs[\lambda]\,, \ee
which is the endomorphism algebra of the modules
$\cV_{\frac{1\pm\lambda}2}$ (the two modules of different signs have
the same symmetry $gl[\lambda]$, as the latter depends on $\lambda$
only through its square $\lambda^2$).  In the region $0\leqslant
\lambda <1$, both modules are unitary and irreducible.  However, when
$\lambda$ becomes a positive integer, say $N$, then
$\cV_{\frac{1+N}2}$ is a unitary irreducible submodule of the
non-unitary reducible module $\cV_{\frac{1-N}2}$.  Moreover, the
higher-spin algebra decomposes as the semidirect sum,\footnote{For a
  Lie algebra $\mathfrak{g}$ and the following semidirect sum of
  $\mathfrak{g}$-modules $U=V\inplus W$ (where $W\subset U$ is the
  submodule and $V=U/W$ is the quotient module), the algebra ${\cal
    A}:=\U(\mathfrak{g})/{\rm Ann}(U) \subset {\rm End}(U)$ preserves
  the submodule $W$ (i.e. ${\cal A}\,W\subset W$).  Moreover, $\cA$
  decomposes as a semidirect sum, $$\U(\mathfrak{g})/{\rm Ann}(U)
  =[\,\U(\mathfrak{g})/{\rm Ann}(V)\,]\inplus {\cal I}\,,$$ where
  ${\cal I}\subset{\cal A}$ is the ideal spanned by the elements with
  image in $W$ (i.e. ${\cal I}\,U\subset W$). Notice that the latter
  property also holds in the simpler case when $U, V$ and $W$ are
  vector spaces (not necessarily with a $\mathfrak g$-module
  structure) in the sense that the subalgebra ${\cal A}\subset {\rm
    End}(U)$ of endomorphisms of $U$ preserving the subspace $W$
  decomposes as a semidirect sum, ${\cal A}={\rm End}(V)\inplus {\cal
    I},$ where ${\cal I}:={\cal A}\cap {\rm Hom}(U,W)$.}
\begin{equation}
  hs[N]=sl(N)\inplus {\cal J}_N\label{I[N]}\,,
\end{equation}
where ${\cal J}_N$ is an infinite-dimensional ideal of $hs[N]$
decomposing in irreducible modules of $so(2,1)$ as
\begin{equation}
  {\cal J}_N= \bigoplus_{j=N}^\infty {\cal D}_{j}\, ,
\end{equation}
while $sl(N)$ is a finite-dimensional higher-spin algebra which
appears here as the symmetry of the irreducible module
$\cD_{\frac{N-1}2} = \cV_{\frac{1-N}2}/\cV_{\frac{1+N}2}$ and which
decomposes as
\begin{equation}
  sl(N)=\bigoplus_{j=1}^{N-1} {\cal D}_{j}\, .
\end{equation}
This allows to shed some light on the comments below the twisted
Flato-Fronsdal \eqref{d=2twistFF} for the spin-$s$ singleton in the
chiral sector: the character \eqref{d=2 factorization} corresponds to
the symmetry algebra $sl(2s-1)$ of the Killing tensor $\cD(s-1,s-1) =
\cD_{1-s} \otimes 1$ while the symmetry algebra of the $d=2$ spin-$s$
singleton $\cD(s,s) = \cV_s \otimes 1$ is isomorphic to the ideal
${\cal J}_{2s-1}\,$.

Following the discussion at the end of section \eqref{Twodim} on
$gl[\lambda]$, one may say that the result \eqref{d=1twistFF} can be
viewed as the twisted Flato-Fronsdal theorem relevant for the
description of $hs[\lambda]$.  In this sense, for a generic value of
$\lambda$\,, one (or a combination) of the modules $\cV_{\frac{1\pm
    \lambda}2}$ ought to play the role of singleton.  However, it does
not seem possible to realize this picture in terms of a
parity-invariant twisted Flato-Fronsdal theorem, except for the type-A
and type-B models (cf \eqref{tFFd=2Rac}-\eqref{tFFd=2Di}\,) which
correspond respectively to $\lambda=1$ and $\lambda=0$ cases.
Technically, it is because the character of the latter module does not
have property similar to \eqref{2d sym prop}.  In fact, it is known
that the underlying CFT has a free field description only for
$\lambda=0,1$.

%%%%%%%%%%%%%%%%%%%%%%%%%%%%%%%%%%%
\section{Extensions and Exceptions}
\label{sec: exex}
%%%%%%%%%%%%%%%%%%%%%%%%%%%%%%%%%%%

%***********************%
\subsection{Type A$_\ell$}
\begin{figure}[!ht]\label{fig2}
  \center
  \begin{tikzpicture}
    \draw[thick,->] (-1.8,0) -- (5.8,0) node[anchor=north west] {$s$};
    \draw[thick,->] (0,-4.8) -- (0,4.8) node[anchor=south east] {$E$};

    \draw (-0.2, 0) node[below] {\small $0$};
    \draw (-0.4, 1.5pt) -- (-0.4, -1.5pt);  
    \draw (-0.8, 1.5pt) -- (-0.8, -1.5pt); 
    \draw (-1.2, 1.5pt) -- (-1.2, -1.5pt); 
    \draw (0.4, 1.5pt) -- (0.4, -1.5pt) node[below=0.2] {\small $1$}; 
    \draw (0.8, 1.5pt) -- (0.8, -1.5pt) node[below=0.2] {\small $2$}; 
    \draw (1.2, 1.5pt) -- (1.2, -1.5pt) node[below=0.2] {\small $3$}; 
    \draw (1.6, 1.5pt) -- (1.6, -1.5pt) node[below=0.2] {\small $4$}; 
    \draw (2, 1.5pt) -- (2, -1.5pt); 
    \draw (2.4, 1.5pt) -- (2.4, -1.5pt) ; 
    \draw (2.8, 1.5pt) -- (2.8, -1.5pt) node[below=4] {$\dots$}; 
    \draw (3.2, 1.5pt) -- (3.2, -1.5pt); 
    \draw (3.6, 1.5pt) -- (3.6, -1.5pt); 
    \draw (4, 1.5pt) -- (4, -1.5pt); 
    \draw (4.4, 1.5pt) -- (4.4, -1.5pt); 
    \draw (4.8, 1.5pt) -- (4.8, -1.5pt); 

    \draw (-1.5pt, 0.4) -- (1.5pt, 0.4); 
    \draw (-1.5pt, 0.8) -- (1.5pt, 0.8) node[left=5] {\small $\ez -\ell+1\ \ $}; 
    \draw (-1.5pt, 1.2) -- (1.5pt, 1.2); 
    \draw (-1.5pt, 1.6) -- (1.5pt, 1.6)  node[left=5] {\small $\ez-\ell+3$}; 
    \draw (-1.5pt, 2) -- (1.5pt, 2); 
    \draw (-1.5pt, 2.4) -- (1.5pt, 2.4)  node[left=5] {$\vdots$}; 
    \draw (-1.5pt, 2.8) -- (1.5pt, 2.8); 
    \draw (-1.5pt, 3.2) -- (1.5pt, 3.2)  node[left=8] {\small $\ez+\ell-1$}; 
    \draw (-1.5pt, 3.6) -- (1.5pt, 3.6);
    \draw (-1.5pt, 4) -- (1.5pt, 4);
    \draw (-1.5pt, 4.4) -- (1.5pt, 4.4);
    \draw (-1.5pt, -0.4) -- (1.5pt, -0.4); 
    \draw (-1.5pt, -0.8) -- (1.5pt, -0.8) node[left=5] {\small $-\ez+\ell-1 \ \ \ $}; 
    \draw (-1.5pt, -1.2) -- (1.5pt, -1.2); 
    \draw (-1.5pt, -1.6) -- (1.5pt, -1.6) node[left=5] {\small $-\ez+\ell-3$}; 
    \draw (-1.5pt, -2) -- (1.5pt, -2); 
    \draw (-1.5pt, -2.4) -- (1.5pt, -2.4) node[left=5] {$\vdots$}; 
    \draw (-1.5pt, -2.8) -- (1.5pt, -2.8);
    \draw (-1.5pt, -3.2) -- (1.5pt, -3.2) node[left=8] {\small
      $-\ez-\ell+1$};
    \draw (-1.5pt, -3.6) -- (1.5pt, -3.6);
    \draw (-1.5pt, -4) -- (1.5pt, -4);
    \draw (-1.5pt, -4.4) -- (1.5pt, -4.4);

    \draw (0, 0.8) node[color=blue] {$\boldsymbol{\times}$};
    \draw (0, 1.6) node[color=blue] {$\boldsymbol{\times}$};
    \draw (0, 2.4) node[color=blue] {$\boldsymbol{\times}$};
    \draw (0, 3.2) node[color=blue] {$\boldsymbol{\times}$};    
    \draw (0.4, 1.2) node[color=blue] {$\boldsymbol{\times}$};
    \draw (0.4, 2) node[color=blue] {$\boldsymbol{\times}$};
    \draw (0.4, 2.8) node[color=blue] {$\boldsymbol{\times}$};
    \draw (0.8, 1.6) node[color=blue] {$\boldsymbol{\times}$};
    \draw (0.8, 2.4) node[color=blue] {$\boldsymbol{\times}$};
    \draw (0.8, 3.2) node[color=blue] {$\boldsymbol{\times}$};
    \draw (1.2, 2) node[color=blue] {$\boldsymbol{\times}$};
    \draw (1.2, 2.8) node[color=blue] {$\boldsymbol{\times}$};
    \draw (1.6, 2.4) node[color=blue] {$\boldsymbol{\times}$};
    \draw (1.6, 3.2) node[color=blue] {$\boldsymbol{\times}$};
    \draw (2, 2.8) node[color=blue] {$\boldsymbol{\times}$};
    \draw (2.4, 3.2) node[color=blue] {$\boldsymbol{\times}$};
    \draw (1.2, 3.6) node[color=blue] {$\boldsymbol{\times}$};
    \draw (0.4, 3.6) node[color=blue] {$\boldsymbol{\times}$};
    \draw (2, 3.6) node[color=blue] {$\boldsymbol{\times}$};
    \draw (2.8, 3.6) node[color=blue] {$\boldsymbol{\times}$};
    \draw (1.6, 4) node[color=blue] {$\boldsymbol{\times}$};
    \draw (0.8, 4) node[color=blue] {$\boldsymbol{\times}$};
    \draw (2.4, 4) node[color=blue] {$\boldsymbol{\times}$};
    \draw (3.2, 4) node[color=blue] {$\boldsymbol{\times}$};
    \draw (3.4, 4.2) node[color=black] {$\boldsymbol{\cdot}$};
    \draw (3.6, 4.4) node[color=black] {$\boldsymbol{\cdot}$};
    \draw (3.8, 4.6) node[color=black] {$\boldsymbol{\cdot}$};
    \draw (2.6, 4.2) node[color=black] {$\boldsymbol{\cdot}$};
    \draw (2.8, 4.4) node[color=black] {$\boldsymbol{\cdot}$};
    \draw (3, 4.6) node[color=black] {$\boldsymbol{\cdot}$};
    \draw (1.8, 4.2) node[color=black] {$\boldsymbol{\cdot}$};
    \draw (2, 4.4) node[color=black] {$\boldsymbol{\cdot}$};
    \draw (2.2, 4.6) node[color=black] {$\boldsymbol{\cdot}$};
    \draw (1, 4.2) node[color=black] {$\boldsymbol{\cdot}$};
    \draw (1.2, 4.4) node[color=black] {$\boldsymbol{\cdot}$};
    \draw (1.4, 4.6) node[color=black] {$\boldsymbol{\cdot}$};
    
    \draw (0, -0.8) node[color=red] {$\boldsymbol{\times}$};
    \draw (0, -1.6) node[color=red] {$\boldsymbol{\times}$};
    \draw (0, -2.4) node[color=red] {$\boldsymbol{\times}$};
    \draw (0, -3.2) node[color=red] {$\boldsymbol{\times}$};    
    \draw (0.4, -1.2) node[color=red] {$\boldsymbol{\times}$};
    \draw (0.4, -2) node[color=red] {$\boldsymbol{\times}$};
    \draw (0.4, -2.8) node[color=red] {$\boldsymbol{\times}$};
    \draw (0.8, -1.6) node[color=red] {$\boldsymbol{\times}$};
    \draw (0.8, -2.4) node[color=red] {$\boldsymbol{\times}$};
    \draw (0.8, -3.2) node[color=red] {$\boldsymbol{\times}$};
    \draw (1.2, -2) node[color=red] {$\boldsymbol{\times}$};
    \draw (1.2, -2.8) node[color=red] {$\boldsymbol{\times}$};
    \draw (1.6, -2.4) node[color=red] {$\boldsymbol{\times}$};
    \draw (1.6, -3.2) node[color=red] {$\boldsymbol{\times}$};
    \draw (2, -2.8) node[color=red] {$\boldsymbol{\times}$};
    \draw (1.2, -3.6) node[color=red] {$\boldsymbol{\times}$};
    \draw (0.4, -3.6) node[color=red] {$\boldsymbol{\times}$};
    \draw (2, -3.6) node[color=red] {$\boldsymbol{\times}$};
    \draw (2.8, -3.6) node[color=red] {$\boldsymbol{\times}$};
    \draw (1.6, -4) node[color=red] {$\boldsymbol{\times}$};
    \draw (0.8, -4) node[color=red] {$\boldsymbol{\times}$};
    \draw (2.4, -4) node[color=red] {$\boldsymbol{\times}$};
    \draw (3.2, -4) node[color=red] {$\boldsymbol{\times}$};
    \draw (2.4, -3.2) node[color=red] {$\boldsymbol{\times}$};
    \draw (3.4, -4.2) node[color=black] {$\boldsymbol{\cdot}$};
    \draw (3.6, -4.4) node[color=black] {$\boldsymbol{\cdot}$};
    \draw (3.8, -4.6) node[color=black] {$\boldsymbol{\cdot}$};
    \draw (2.6, -4.2) node[color=black] {$\boldsymbol{\cdot}$};
    \draw (2.8, -4.4) node[color=black] {$\boldsymbol{\cdot}$};
    \draw (3, -4.6) node[color=black] {$\boldsymbol{\cdot}$};
    \draw (1.8, -4.2) node[color=black] {$\boldsymbol{\cdot}$};
    \draw (2, -4.4) node[color=black] {$\boldsymbol{\cdot}$};
    \draw (2.2, -4.6) node[color=black] {$\boldsymbol{\cdot}$};
    \draw (1, -4.2) node[color=black] {$\boldsymbol{\cdot}$};
    \draw (1.2, -4.4) node[color=black] {$\boldsymbol{\cdot}$};
    \draw (1.4, -4.6) node[color=black] {$\boldsymbol{\cdot}$};
  \end{tikzpicture}
  \caption{Weight diagram of the scalar, \it order $\ell$, \rm
    singleton (blue crosses) and of the scalar, \it order $\ell$, \rm
    anti-singleton (red crosses).}
\end{figure}
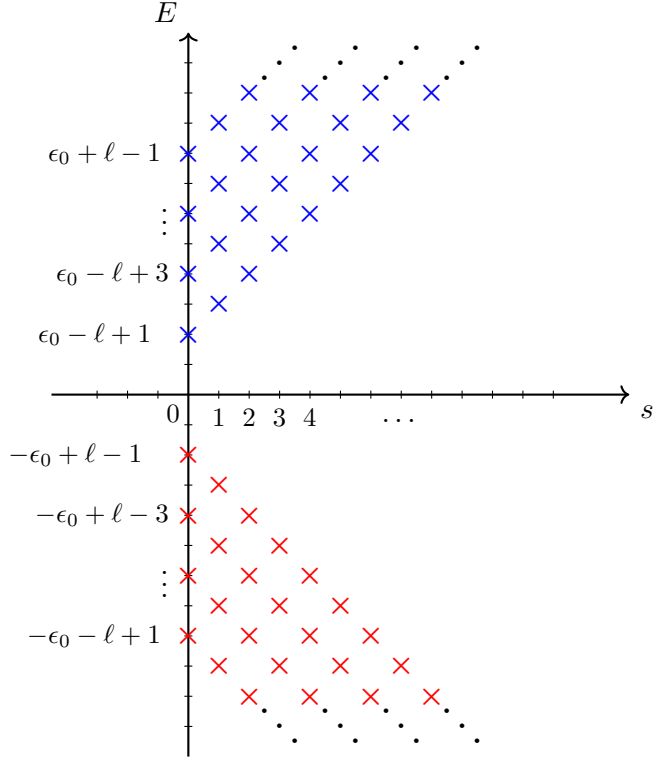

%***********************%
Let us generalize the previous analysis to the type-A$_\ell$
partially-massless higher-spin theory \cite{Skvortsov:2006at,
  Bekaert:2013zya, Alkalaev:2014nsa, Brust:2016zns,
  Brust:2016xif}. This family of theories, parametrized by a positive
integer $\ell$, involves not only infinitely many massless fields but
also partially-massless fields with odd depth
$t=1,3,\ldots,2\ell-1$. Its higher-spin algebra contains the
corresponding Killing tensors, given by $so(2+d)$ Young diagrams of
the form \cite{Bekaert:2013zya,Joung:2015jza},
\begin{equation}
  {\footnotesize \gyoung(_6{s-1},_4{s-t})}\, 
  \label{(r,t)}
\end{equation}
for odd $t=1,3,\ldots,2\ell-1$ and integer $s= t,\,t+1,\,\ldots$
The type-A$_\ell$ partially-massless higher-spin gravity in $d+1$
dimensions has been conjectured to be dual to the
higher-derivative scalar CFT in $d$ dimensions with the polywave
equation,
\begin{equation}
  \Box^\ell\,\phi = 0\,,
  \label{ho s}
\end{equation}
and the partially massless higher-spin algebra is the algebra of
symmetries of the above equation. In other words, it is the
endomorphism algebra of the solution space of \eqref{ho s}, as showed
in \cite{Eastwood2008} for $\ell=2$ and generalized to arbitrary 
values of $\ell$ in \cite{Gover2012} and \cite{Michel2011}. This space 
carries an irreducible (but non-unitary for $\ell \geqslant 2$) representation,
\begin{equation}
  {\rm Rac}_{\ell} := \cD(\tfrac{d-2\ell}2,\bm 0) =
  \cV(\tfrac{d-2\ell}2,\bm 0)\,/\,\cV(\tfrac{d+2\ell}2,\bm 0)\,,
\end{equation}
of the conformal algebra $so(2,d)$. Its character reads
\begin{equation}
  \chi^{so(2,d)}_{\rac_\ell}(q, \bm x) = q^{d/2} \left(q^{-\ell} -
  q^\ell\right) \Pd d (q, \bm x)\, .
\end{equation}
Using the property \eqref{prop 2}, this can be rewritten as
\begin{equation}
  \chi^{so(2,d)}_{\rac_\ell}(q, \bm x) = \sum_{s=0}^\infty
  \sum_{k=0}^{\ell-1} q^{\frac{d-2\ell}2+s+2k}\, \chi^{so(d)}_s(\bm x)
  = \sum_{s=0}^\infty\, \sum_{t=1,3,\dots}^{2\ell-1}
  q^{\frac{d-2\ell}2+s+t-1}\, \chi^{so(d)}_s(\bm x)\,.
  \label{weight_space_rac_l}
\end{equation}
The weight diagram of this representation can be immediately read off
from the above formula, and is composed of $\ell$ lines\footnote{For
  this reason, ${\rm Rac}_{\ell}$ is sometimes referred to as
  ``multipleton'' \cite{Angelopoulos:1999bz}, ``$\ell-$lineton''
  \cite{Iazeolla:2008ix} or ``multilineton'' \cite{Basile:2014wua}.}
similar to the one constituting the weight diagram of the original
Dirac singleton (recovered in the case $\ell=1$) as depicted in Fig
\ref{fig2}.

To derive, in the partially massless case, a twisted-Flato-Fronsdal
theorem analogous to \eqref{twisted_flato_fronsdal}, we begin with the
formula \eqref{so 2+d} applied to $(s-1,s-t)$\,:
\begin{eqnarray}
  \chi^{so(2+d)}_{(s-1,s-t)}(\bm x) = \sum_{k=0}^r
  \left(x_k^{1-s}+(-1)^d\,x_k^{s+d-1}\right) \chi^{so(d)}_{s-t}(\bm
  x_k)\, \Pd d (x_k, \bm x_k)\,.
\end{eqnarray}
The summand of the above series satisfy  
\begin{equation}
  \sum_{t=1,3,\dots}^{2\ell-1} \sum_{s=t}^\infty
  \left(x_k^{1-s}+(-1)^d\,x_k^{s+d-1}\right) \chi^{so(d)}_{s-t}(\bm
  x_k) =(x_k^{-\ell}-x_k^\ell)(x_k^\ell-x_k^{-\ell}) \,\Pd d
  (x_k^{-1}, \bm x_k)\,,
\end{equation}
where both of the properties \eqref{prop 1} and \eqref{prop 2} are
used for the derivation with the same subtleties related to
convergence as in the previous sections. Collecting these results, we
finally obtain the twisted-Flato-Fronsdal theorem for type-A$_\ell$
partially massless higher-spin theory:
\begin{empheq}[box=\eqbox]{equation}
  \qquad \sum_{k=0}^r \chi^{so(2,d)}_{\rac_\ell}(x_k, \bm x_k) \,
  \chi^{so(2,d)}_{\arac_\ell}(x_k, \bm x_k) =
  \sum_{t=1,3,\dots}^{2\ell-1} \sum_{s=t}^\infty
  \chi^{so(2+d)}_{(s-1,s-t)}(x_0,\bm x)\,. \qquad
\end{empheq}
This result agrees with the collection of Killing tensors
\eqref{(r,t)} and, thus, with the Flato-Fronsdal theorem for
type-A$_\ell$ theory.

%***********************%
\subsection{Type B$_\ell$}
%***********************%
Similarly to the type-A$_\ell$ case, one can consider the
partially-massless extension of the type-B higher-spin algebra as the
symmetry algebra of the higher-order spinor singleton that we will
denote $\di_\ell$, corresponding to the module:
\begin{equation}
  \D(\tfrac{d+1-2\ell}2, \half) = \frac{\V(\tfrac{d+1-2\ell}2,
    \half)}{\D(\tfrac{d-1+2\ell}2, \half)}\, ,
\end{equation}
with the character:
\begin{equation}
  \chi_{\di_\ell}^{so(2,d)}(q, \bm x) = q^{\tfrac{d+1-2\ell}2}\,
  (1-q^{2\ell-1})\, \chi_{\bm{\frac12}}^{so(d)}(\bm x)\, \Pd d (q, \bm
  x)\, .
\end{equation}
This corresponds to a spin-$\frac12$ conformal field $\psi$, with
conformal weight $\tfrac{d+1-2\ell}2$ (non-unitary for $\ell\geqslant
2$ and corresponding to the Di for $\ell=1$) subject to the
higher-order Dirac equation:
\begin{equation}
  \slashed{\partial}^{2\ell-1} \psi = 0\, .
\end{equation}
The spectrum of possible bilinears in this fundamental field making up
the currents of the type-B$_\ell$ boundary theory was studied in
\cite{Alkalaev:2012rg,Basile:2014wua} and contains totally symmetric
as well as ``hook-shaped'' partially-conserved currents of all spins
(see \cite{Boulanger:2008up, Boulanger:2008kw, Skvortsov:2009zu,
  Skvortsov:2009nv} for more details on generic mixed-symmetry
partially massless fields and \cite{Alkalaev:2012ic} for the ones
relevant here).

Notice that in odd $d+1$ bulk dimensions, these higher-order
singletons can also be chiral as their unitary counter-part, i.e. one
can consider the modules:
\begin{equation}
  \di_{\ell\pm} := \D\big( \tfrac{d+1-2\ell}2\,,\, \half_\pm\big) =
  \frac{\V\big( \tfrac{d+1-2\ell}2\,,\, \half_\pm\big)}{\D\big(
    \tfrac{d-1+2\ell}2\,,\, \half_\mp\big)}\, ,
\end{equation}
whose character read:
\begin{equation}
  \chi_{\di_{\ell\pm}}^{so(2,d)}(q, \bm x) := q^{\frac{d+1-2\ell}2}\,
  \big(\chi_{\bm{\frac12}_\pm}^{so(d)}(\bm x) - q^{2\ell-1}
  \chi_{\bm{\frac12}_\mp}^{so(d)}(\bm x) \big)\, \, \Pd d (q, \bm
  x)\,.
\end{equation}
We will however only consider the parity-invariant singleton,
i.e. $\di_\ell := \di_{\ell+} \oplus \di_{\ell-}$, so as to be able to
treat both the odd and even dimensional cases on an equal footing. The
only subtlety to keep in mind when reading the decomposition hereafter
is that for $d=2r$, all diagrams of maximal height (namely $r+1$ for
$so(2+d)$) come with both chiralities, whereas all other diagrams come
with a multiplicity $2$.\\

\noindent Using the fact that the character of this higher-order
singleton can be expanded as
\begin{equation}
  \chi_{\di_\ell}^{so(2,d)}(q, \bm x) = \sum_{k=0}^{2(\ell-1)}
  \sum_{s=0}^\infty q^{\frac{d+1-2\ell}2+s+k}\, \chi_{(s+\frac12,
    \frac12 ^{r-1})}^{so(d)}(\bm x)\, ,
\end{equation}
and after a calculation similar to that of the previous section, one
can show that the tensor product of the $\di_\ell$ singleton and its
anti-singleton can be decomposed as
\begin{empheq}[box=\eqbox]{align}
  & \sum_{k=0}^r \chi_{\di_\ell}^{so(2,d)}(x_k, \bm x_k)\times
  \chi_{\adi_\ell}^{so(2,d)}(x_k, \bm x_k) \nn &=
  \sum_{t=1}^{2\ell-1}\left[ \chi^{so(2+d)}_{t-1}(x_0, \bm x)
    +\sum_{s=t+1}^\infty \sum_{m=0}^{r-1}
    \chi^{so(2+d)}_{(s-1,s-t,1^m)}(x_0, \bm x) \right]\nn & \qquad +
  \sum_{t=1}^{2\ell-2} \sum_{s=t+1}^\infty\sum_{m=0}^{r-1}
  \chi^{so(2+d)}_{(s-1,s-t,1^m)}(x_0, \bm x)\, .
  \label{HS_Bl}
\end{empheq}
Notice that the last triple sum in the above decomposition is absent
for $\ell=1$, in which case the second line reproduces the spectrum of
the type-B higher-spin algebra discussed in the previous subsection
(whereas the last line identically vanishes).\\

Notice also that the spectrum of the higher-spin theories so far
considered --- which is obtained by decomposing the tensor product of
the relevant singletons --- is closely related to the $so(2+d)$
decomposition of the underlying higher-spin algebra: the later is
composed of the Killing tensors of the all the (partially) massless
fields appearing in the former. We were able to recover these
decompositions from the tensor product of the relevant singleton and
its anti-singletons for the type-A$_\ell$ (with $\ell \geqslant 1$)
and type-B higher-spin algebras, however it seems to fail in the case
of the type-B$_\ell$ algebra (with $\ell > 1$). Indeed, the spectrum
of partially-massless fields appearing in the tensor product of two
$\di_\ell$ singletons reads \cite{Basile:2014wua}
\begin{eqnarray}
  && \di_\ell^{\otimes2}\supset \bigoplus_{t=1}^{2\ell-1}
  \left[\D\big( d-1, t \big) \oplus \bigoplus_{s=t+1}^\infty
    \bigoplus_{m=0}^{r-1} \D\big( s+d-t-1, s, 1^m \big)\right]\nn
  &&\qquad\quad +\, \bigoplus_{t=1}^{2\ell-2} \left[\D\big( d-1,t
    \big) \oplus \bigoplus_{s=t+1}^\infty \bigoplus_{m=0}^{r-1}\D\big(
    s+d-t-1,s, 1^m \big)\right]
\end{eqnarray}
and in particular contains two infinite towers of totally symmetric
partially massless fields, of all depths $t$ ranging from $1$ to
(respectively) either $2\ell-1$ or $2\ell-2$, and of all integer spins
$s\geqslant t$. It therefore seems that the set of Killing tensors
corresponding to totally symmetric partially-massless fields of spin
$s=t$ are missing for $t=1, 2, \dots, 2\ell-2$, i.e. one would expect
that the sum
\begin{equation}
  \sum_{t=1}^{2\ell-2} \chi^{so(2+d)}_{t-1}(x_0, \bm x)\,,
  \label{discr 1}
\end{equation}
should be added to \eqref{HS_Bl} in order make up the spectrum of the
type-B$_\ell$ higher-spin algebras.  
This is the first discrepancy that we find in our proposal.

\subsection{Type AB$_\ell$}

The type-AB$_\ell$ theory includes the cross terms of the Rac and Di
$\ell$-linetons, on top of the contents of the type-A$_\ell$ and
type-B$_\ell$ theories.  Since we have examined the latter cases in
the previous section, here we focus on the cross terms.

Using \eqref{decompo_di_pm}, the product of the character of a $\rac$
singleton with that of the $\adi$ anti-singleton can be written as
\begin{equation}
  \chi_{\rac_\ell}^{so(2,d)}(q, \bm x) \times
  \chi^{so(2,d)}_{\adi_\ell}(q, \bm x) = q^{-\frac12}\,(1-q^{2\ell})\,
  \Pd d (q, \bm x) \sum_{s=0}^\infty \sum_{k=0}^{2(\ell-1)} q^{-s-k}\,
  \chi^{so(d)}_{(s+\frac12,\frac12 ^{r-1})}(\bm x)\, ,
\end{equation}
whereas for $\arac$ with $\di$ as
\begin{equation}
  \chi_{\arac_\ell}^{so(2,d)}(q, \bm x)\times
  \chi^{so(2,d)}_{\di_\ell}(q, \bm x) = q^{\frac12}\,(1-q^{-2\ell})
  \,\Pd d (q^{-1}, \bm x) \sum_{s=0}^\infty \sum_{k=0}^{2(\ell-1)}
  q^{s+k} \,\chi^{so(d)}_{(s+\frac12,\frac12 ^{r-1})}(\bm x)\,.
\end{equation}
Their sum can be simplified to
\begin{eqnarray}
  &&\chi^{so(2,d)}_{\rac_\ell}(q, \bm x) \times 
  \chi_{\adi_\ell}^{so(2,d)}(q, \bm x) +
  \chi^{so(2,d)}_{\arac_\ell}(q, \bm x) \times
  \chi_{\di_\ell}^{so(2,d)}(q, \bm x) \nn
%   \sum_{s=0}^\infty
%  \sum_{k=0}^{2(\ell-1)} (q^{-s-1/2-k} + (-1)^d q^{s+d+1/2+k})\,
%  \chi^{so(d)}_{(s+\frac12,\frac12 ^{r-1})}(\bm x) \Pd d (q, \bm x)
%  \\ && \qquad + \sum_{s=0}^\infty \sum_{k=0}^{2(\ell-1)}
%  (q^{-s-1/2-k+2\ell} + (-1)^d q^{s+d+1/2+k-2\ell})\,
%  \chi^{so(d)}_{(s+\frac12,\frac12 ^{r-1})}(\bm x) \Pd d (q, \bm x)
  && = \sum_{t=1}^{2\ell-1} \sum_{s=t}^\infty
  \Big[\left(q^{-s+\frac12} + (-1)^d\,q^{s+d-\frac12}\right)
  \chi^{so(d)}_{(s-t+\frac12,\frac12 ^{r-1})}(\bm x)\,\Pd d (q, \bm x)
  \\ && \qquad \qquad -  \left(q^{-s+\frac12+t}
  + (-1)^d \,q^{s+d-\frac12-t}\right) \chi^{so(d)}_{(s+\frac12,\frac12
    ^{r-1})}(\bm x)\, \Pd d (q, \bm x)\Big]\,. \nonumber
\end{eqnarray}
Symmetrizing the $r+1$ variables of the above expression and using
\eqref{so 2+d}, we end up with the following sum of $so(2+d)$
characters
\begin{eqnarray}
  && \sum_{k=0}^r \left(\chi_{\rac_\ell}^{so(2,d)}(x_k, \bm x_k)\times
  \chi^{so(2,d)}_{\adi_\ell}(x_k, \bm x_k) + \chi_{\arac_\ell}^{so(2,d)}(x_k,
  \bm x_k) \times \chi^{so(2,d)}_{\di_\ell}(x_k, \bm x_k) \right)\nn && =
  \sum_{t=1}^{2\ell-1} \sum_{s=t}^\infty
  \chi^{so(d+2)}_{(s-\frac12,s-t+\frac12,\frac12 ^{r-1})}(x_0, \bm x)
  - \sum_{t=1}^{2\ell-1} \sum_{s=0}^\infty
  \chi^{so(d+2)}_{(s-t-\frac12,s+\frac12,\frac12 ^{r-1})}(x_0, \bm
  x)\, . 
\end{eqnarray}
Using the symmetry property \eqref{sym_prop_sod}, the characters
appearing in the second sum can be expressed as characters of bona fide
$so(2+d)$ Young diagrams, given that
\begin{equation}
  \chi^{so(d+2)}_{(s-t-\frac12,s+\frac12,\frac12 ^{r-1})}= -
  \chi^{so(d+2)}_{(s-\frac12,s-t+\frac12,\frac12 ^{r-1})}\, .
\end{equation}
Moreover, due to the same property, the sum of the characters for which
$s < t$ identically vanish. Hence, we end up with
\begin{empheq}[box=\eqbox]{equation}
  \sum_{k=0}^r \chi_{(\rac_\ell\otimes\adi_\ell) \oplus
    (\di_\ell\otimes\arac_\ell)}^{so(2,d)}(x_k, \bm x_k) = 2\,
  \sum_{t=1}^{2\ell-1}\, \sum_{s=t}^\infty
  \chi^{so(d+2)}_{(s-\frac12,s-t+\frac12,\frac12 ^{r-1})}(x_0, \bm
  x)\,,
\end{empheq}
which is
consistent with the corresponding Flato-Fronsdal theorem, giving
the decomposition of the tensor product $\rac_\ell\otimes\di_\ell$ \cite{Basile:2014wua}.

%*****************%
\subsection{Type J}
%*****************%
In even boundary dimensions $d=2\,r$, we have infinitely many
singleton representations \cite{Siegel:1988gd, Angelopoulos:1997ij} corresponding to
\begin{equation}
  \cS_{j\pm}= \cD\big(j+r-1, \mathbf{j}_\pm\big)\qquad \text{with}
  \qquad \mathbf{j}_\pm:=(j,\ldots,j,\pm j) \, .
\end{equation}
Their characters can be written in terms of the ones of the lowest-weight module
$\cV(\Delta,\Bell)$ as
\begin{eqnarray}
  \chi_{\cS_{j\pm}}(q,\bm x) \eq \sum_{m=0}^{r}
  (-1)^m\,\chi_{\cV\big(j+r-1+m,j^{r-m},(j-1)^{m}_\pm)\big)} (q,\bm x) \nn \eq
  q^{j+r-1}\,\Pd{2r}(q,\bm x)\sum_{m=0}^{r}
  (-q)^m\,\chi^{so(2r)}_{\big(j^{r-m},(j-1)^{m}_\pm\big)}(\bm x)\,.
\end{eqnarray}
Similarly to the other singletons that we have seen above, these
higher-spin singletons for even $d$ satisfy the property,
\begin{equation}
  \chi_{\cS_{j\pm}}(q,\bm x) =\sum_{s=0}^\infty
  q^{j+r+s-1}\,\chi^{so(d)}_{(s+j,j^{r-1}_\pm)}(\bm x)\,,
  \label{exp_spinj}
\end{equation}
from which the term ``singleton'' originated \cite{Ehrman1957}
(cf the historical comment in  \cite{Bekaert:2011js}). 
The anti-singleton is the highest-weight
counterpart of the singleton, and its character is
\begin{equation}
  \chi_{\overline{\cS_{j\pm}}}(q,\bm x) =
  \chi_{\cS_{j\pm}}(q^{-1},{\bm x}^{-1})
  =
  \left\{
  \begin{array}{cc}
  \chi_{\cS_{j\pm}}(q^{-1}, \bm x)&\quad [{\rm even}\ r]
  \medskip\\
  \chi_{\cS_{j\mp}}(q^{-1}, \bm x)& \quad [{\rm odd}\ r]
  \end{array} 
  \right..
  \label{anti j}
\end{equation}
Remark here that the singleton character $\chi_{\cS_{j\pm}}(q,\bm x)$
does not have a simple property under $q\to q^{-1}$, even in the
parity-invariant case, contrarily to the previously treated
$\rac_\ell$ and $\di_\ell$ singletons. We now consider the product of
these two characters:
\begin{equation}
  \chi_{\cS_{j\s}}(q,\bm x)\,\chi_{\overline{\cS_{j\t}}}(q,\bm x)
  =\Pd{2\,r}(q,\bm x)\,\sum_{s=0}^\infty \sum_{m=0}^{r} q^{m-s}
  (-1)^m\,\chi^{so(2\,r)}_{(j^{r-m},(j-1)^{m}_\s)}(\bm x)\,
  \chi^{so(2\,r)}_{(s+j,j^{r-1}_{\t})}(\bm x)\,,
  \label{SS}
\end{equation} 
where $\s$ and $\t$ stands for the signs $\pm$. 
To proceed, we need to decompose the representation,
\begin{equation}
  \left(j^{r-m},(j-1)^{m}_\s\right)\otimes\left(s+j,j^{r-1}_{\t}\right),
  \label{tensor_product}
\end{equation}
into $so(2r)$ irreps. Unfortunately this task is quite cumbersome
for generic value of $j,s,m,r$\,, hence we focus on the
  particular case $d=4$ in the rest of this section.

%*****************************************%
\subsubsection*{Type-J in five dimensions}
\label{par:typeJ}
%*****************************************%

For $d=4$, the Flato-Fronsdal
theorem has been derived in \cite{Dolan:2005wy} as
\ba  
    && \chi_{\cS_{j\pm}}\times \chi_{\cS_{j\pm}}
    = \sum_{k=0}^{2j} \chi^{so(2,4)}_{\cD(2j+2,k,\pm k)}(q, \bm x) + \sum_{s=2j+1}^{\infty} \chi^{so(2,4)}_{\cD(s+2,s,\pm2j)}(q, \bm x)\,,
    \label{FF same j}\\
    &&
     \chi_{\cS_{j\pm}}\times \chi_{\cS_{j\mp}}
    = \sum_{s=2j}^{\infty} \chi^{so(2,4)}_{\cD(s+2,s)}(q, \bm x)\,.
    \label{FF opposite j}
\ea
Let us consider the corresponding twisted Flato-Fronsdal theorem.
The tensor product \eqref{tensor_product} can be computed
more easily thanks to the low dimensional isomorphism $so(4) \cong
so(3) \oplus so(3)$. Consequently, an $so(4)$ irrep with highest
weight $(\ell_1,\ell_2)$ is equivalent to the direct sum of two
$so(3)$ irreps with highest-weights $j_\pm :=
\tfrac{\ell_1\pm\ell_2}2\,$, and the tensor product of two arbitrary
representations of $so(4)$ reduces to the tensor products of their two
respective $so(3)$ factors, for which we can use the well-known
Clebsch-Gordan decomposition \eqref{CG_decompo}. Applying it to the
formula \eqref{SS} with $\s=+$ and $ \t=-$\,, we obtain
  \begin{eqnarray}
    \chi_{\cS_{j+}}(q,\bm x) \,\chi_{\overline{\cS_{j-}}}(q,\bm x) &&\! =
    \bigg[ \sum_{s=0}^{2j-2} q^{2-s}\, \chi^{so(4)}_{(2j-1,-s-1)}(\bm x)
    - \sum_{s=0}^{2j-1} q^{1-s}\, \chi^{so(4)}_{(2j-1,-s)}(\bm x) \nn
       &&+ \sum_{s=2j}^\infty q^{-s}\, \chi^{so(4)}_{(s,-2j)}(\bm x) -
    \sum_{s=2j}^\infty q^{2-s}\, \chi^{so(4)}_{(s,-2j)}(\bm x) \bigg]\, \Pd
    4 (q, \bm x)\,.
   \label{ss+}
  \end{eqnarray}
The same for $\s=-, \t=+$ with variable $q^{-1}$ gives 
\begin{eqnarray}
  \chi_{\cS_{j+}}(q,\bm x) \,\chi_{\overline{\cS_{j-}}}(q,\bm x) &&\!=
  \bigg[ \sum_{s=0}^{2j-2} q^{s+2} \,\chi^{so(4)}_{(2j-1,s+1)}(\bm x)
    - \sum_{s=0}^{2j-1} q^{s+3} \,\chi^{so(4)}_{(2j-1,s)}(\bm x) \nn
    && + \sum_{s=2j}^\infty q^{s+4} \,\chi^{so(4)}_{(s,2j)}(\bm x) -
    \sum_{s=2j}^\infty q^{s+2}\, \chi^{so(4)}_{(s,2j)}(\bm x) \bigg]\,
  \Pd 4 (q, \bm x)\,.
  \label{ss-}
\end{eqnarray}
Symmetrizing the half sum of \eqref{ss+} and \eqref{ss-}, we find
\begin{equation}
  \sum_{k=0}^2 \chi_{\cS_{j\pm}}(x_k,\bm x_k)\,
  \chi_{\overline{\cS_{j\mp}}}(x_k,\bm x_k) = \sum_{s=2j+1}^\infty
  \chi^{so(6)}_{(s-1,s-1,\pm 2j)}(x_0, \bm x) + \sum_{s=0}^{2j-2}
  \chi^{so(6)}_{(2j-2,s,\pm s)}(x_0, \bm x)\,,
  \label{chichio}
\end{equation}
where we have included also the 
product $\cS_{j-}$ and $\overline{\cS_{j+}}$
using the relation \eqref{anti j}.
Let us comment about the two series in the right-hand-side of the equality.
Since the $so(6)$ irrep $(s-1,s-1,2j)$
is carried by the Killing tensor
of the spin-$(s,2j)$ gauge field, 
the first infinite series matches well 
the content of gauge fields in 
the Flato-Fronsdal theorem \eqref{FF opposite j}.
However, this spectrum does not include the gauge fields 
corresponding to the second finite series. 
In fact,
the second series can be written as a perfect square:
\begin{equation}
  \sum_{s=0}^{2j-2} \chi^{so(6)}_{(2j-2,s,\pm s)} = \left(
  \chi^{so(6)}_{(j-1,j-1,\pm(j-1))}\right)^2
  = \chi^{so(6)}_{(j-1,j-1,\pm(j-1))}\,
  \chi^{so(6)}_{\overline{(j-1,j-1,\mp(j-1))}}\,,
\end{equation}
where the $so(6)$ Young diagrams $(j-1,j-1,\pm(j-1))$ 
are the Killing tensor of the
spin-$\pm j$ singleton.
This factorization is analogous
to the $d=2$ case \eqref{d=2 factorization}.
It is interesting to note the identity,
\begin{equation}
  \chi^{so(6)}_{(j-1,j-1,\pm(j-1))}(q, \bm x) = \chi_{\cS_{j\pm}}(q, \bm x) +
  \chi_{\overline{\cS_{j\mp}}}(q, \bm x)\,,
\end{equation}
which is again somewhat analogous to the two-dimensional one \eqref{so21spinj}.
  
Similarly, the product of the character of a spin-$j$ singleton of
positive/negative chirality with the character of its own
anti-singleton can be decomposed into two different forms: firstly,
\begin{eqnarray}
  && \chi_{\cS_{j\pm}}(q,\bm x)\,\chi_{\overline{\cS_{j\pm}}}(q,\bm x)
  =\nn &&=\left[ \sum_{s=0}^\infty q^{-s} \,\chi^{so(4)}_s(\bm x) -
    \sum_{s=0}^\infty q^{1-s} \,\chi^{so(4)}_{(s+1)}(\bm x) +
    \sum_{s=1}^{2j-1} q^2\, \chi^{so(4)}_{(s,\pm s)}(\bm x) \right]
  \Pd 4 (q, \bm x) \, ,
\end{eqnarray}
and secondly,
\begin{eqnarray}
    && \chi_{\cS_{j\pm}}(q,\bm x) \,\chi_{\overline{\cS_{j\pm}}}(q,\bm
  x) = \chi_{\cS_{j\pm}}(q^{-1},{\bm x}^{-1})
  \,\chi_{\overline{\cS_{j\pm}}}(q^{-1},{\bm x}^{-1}) \nn &&=\left[
    \sum_{s=0}^\infty q^{s+4} \,\chi^{so(4)}_s(\bm x) -
    \sum_{s=0}^\infty q^{s+3}\, \chi^{so(4)}_{(s+1)}(\bm x) +
    \sum_{s=1}^{2j-1} q^2 \,\chi^{so(4)}_{(s,\pm s)}(\bm x) \right]
  \Pd 4 (q, \bm x)\, .
\end{eqnarray}
After symmetrization, the half sum of these two decompositions yields
\begin{eqnarray}
  \sum_{k=0}^{2} \chi_{\cS_{j\pm}}(x_k,\bm x_k)\,
  \chi_{\overline{\cS_{j\pm}}}(x_k,\bm x_k) \eq \sum_{s=1}^{2j-1}
  \chi^{so(6)}_{(s-1,s-1)}(x_0, \bm x) +\sum_{s=2j}^\infty
  \chi^{so(6)}_{(s-1,s-1)}(x_0, \bm x)\nn &&+\, \sum_{s=1}^{2j-1}
  \sum_{k=0}^2 \chi^{so(4)}_{(s,\pm s)}(\bm x_k)\,x_k^2\,\Pd 4
  (x_k,\bm x_k)\, .
  \label{chichis}
\end{eqnarray}
In the first line, the two-row Young diagrams $(s-1,s-1)$ for $s
\geqslant 2j$ correspond to the Killing tensors of the totally
symmetric massless fields that appear in the tensor product of two
spin-$j$ singletons of opposite chirality. The additional two-row
diagrams for $1\leqslant s \leqslant 2j-1$ can be interpreted as the
result of the tensor product of the Killing tensors of the spin-$\pm j$
singleton and its dual:
\begin{equation}
  \chi^{so(6)}_{(j-1,j-1,\pm(j-1))}(x_0, \bm x)\,
  \chi^{so(6)}_{\overline{(j-1,j-1,\pm(j-1))}}(x_0, \bm x) = \sum_{k=0}^{2j-2}
  \chi^{so(6)}_{(k,k)}(x_0, \bm x)\, .
  \label{fin j}
\end{equation}
If the second line of \eqref{chichis} were absent, the above result
matches well the symmetry of the $d=4$ spin-$j$ chiral singleton: the
modules $(s-1,s-1)$ with $s\geqslant2j$ correspond to the ideal part
of the symmetry, while the rest with $1\leqslant s\leqslant 2j-1$
correspond to the quotient part \cite{Boulanger:2011se,
  Manvelyan:2013oua, Joung:2014qya}. Since the character is not
sensitive to the indecomposability, it is natural that we get both the
ideal and quotient algebras here. However, the second line of
\eqref{chichis} does include additional terms.  We do not have clear
interpretation of these terms.
  
Let us conclude this section with the twisted Flato-Fronsdal of the
parity-invariant spin-$j$ singleton, having character $\chi_{\cS_{j}}
= \chi_{\cS_{j_+}} + \chi_{\cS_{j_-}}$.  By collecting the previous
results \eqref{chichio} and \eqref{chichis}, we obtain
\begin{empheq}[box=\eqbox]{align}
  & \sum_{k=0}^2 \chi_{\cS_{j}}(x_k, \bm x_k)\,
  \chi_{\overline{\cS_{j}}}(x_k, \bm x_k) \nn & = 2\sum_{s=2j}^\infty
  \chi^{so(6)}_{(s-1,s-1)}(x_0, \bm x) +\sum_{s=2j+1}^\infty
  \chi^{so(6)}_{(s-1,s-1,2j)_0}(x_0, \bm x) \nn &\quad +\big(
  \chi^{so(6)}_{(j-1,j-1,j-1)_0}(x_0, \bm x) \big)^2 +
  \sum_{s=1}^{2j-1} \chi^{so(6)}_{(s-1,s-1)}(x_0, \bm x)\,.
  \label{type j tff}
\end{empheq}
Here the subscript $0$ of the $so(6)$ modules signals that they are
the direct sum of the two chiral representations.  Remark that the
terms in the second line and the first term in the third line
correspond to the symmetry algebra of the parity-invariant spin-$j$
singleton.  The last term of the third line is from the additional
terms in \eqref{chichis}. By adding up two contributions from
$\chi_{\cS_{j+}}\,\chi_{\overline{\cS_{j+}}}$ and
$\chi_{\cS_{j-}}\,\chi_{\overline{\cS_{j-}}}$, such terms form the
$so(6)$ character written above.  These modules are in fact a part of
the generators of the quotient higher-spin algebra.  However, the quotient
algebra was already taken into account by the first term in the third
line.  Therefore, these modules are additional and do not match with
the symmetry algebra of the spin-$j$ singleton.

%%%%%%%%%%%%%%%%%%%%
\section{Discussion}
%%%%%%%%%%%%%%%%%%%%

In this paper, we have explored the relation between the $so(2,d)$
characters of the singletons and the adjoint module of higher-spin
algebras. Starting from the idea that the higher-spin algebra is the endomorphism algebra of the singleton module, we attempted to derive the character for the adjoint module as a product of the singleton character and its dual. We first noticed that a simple product of the characters cannot reproduce the adjoint module one because the latter is symmetric under the exchange of its arguments while the former lacks this symmetry. This lead to our symmetrization prescription of the character product. 

In Section \ref{sec: 4d}, we used the oscillator realization of the singleton and higher-spin algebra in four dimensions to relate the 
  extra term (arising from the symmetrization prescription) in the character to
  an extra piece (with respect to the naive tensor product) in the twisted Flato-Fronsdal theorem \eqref{=!}.
In Section \ref{sec: gen d}, we showed that the symmetrization prescription correctly reproduces the
  adjoint module character for the type-A and type-B models in any
  dimension.  This is based on several interesting identities of the
  $so(2,d)$ and $so(2+d)$ characters, which have their root in the
  Weyl character formula. In Section \ref{sec: exex}, the symmetrization prescription was shown to work 
  for the higher-order singleton case of type-A$_\ell$. However, in the
  type-B$_\ell$ theory, we found that the symmetrized product misses a few
  Killing tensor modules \eqref{discr 1}.  Moreover, for the
  higher-spin singletons, aka the type-J model, the symmetrized
  product contains more Killing tensor modules than necessary, cf
  \eqref{type j tff}.  In both counterexamples, the
  mismatch is by a finite number of modules.

The symmetrization prescription of the character arguments can be
viewed as an action of certain Weyl group elements.  Remember that the
Weyl group of a semisimple Lie algebra maps a Cartan subalgebra to
itself.  Since the variables that we symmetrize for the twisted
Flato-Fronsdal theorem are associated to the Cartan generators, the
symmetrization prescription can be induced by the action of the Weyl
group quotiented by its normalizer subgroup of the singleton and
anti-singleton tensor product.  Referring to such quotient group as
$\mathcal W'$, we can restate our prescription in terms of the modules
themselves as
\begin{equation}
  {\rm Adj}= \bigoplus_{w\in \mathcal W'} w({\rm Sng})\otimes
  w(\overline{\rm Sng})\,.
\end{equation}
In the case $d=3$, this prescription reproduces the twisted
Flato-Fronsdal theorem \eqref{reftwFF}.  In the singleton module ---
and in all other lowest-weight modules --- the energy generator $E$
plays a distinguished role with respect to the other Cartan
generators, which belong to the rotation subalgebra $so(d)$.  The
action of the aforementioned Weyl group elements symmetrize $E$ with
those other Cartan generators.  Here, it is interesting to note that
such an action will map the singleton module to a non-unitary module.

In contrast, the Killing tensors are already symmetric under this
action, hence should not be ``over-symmerized''. Maybe the application
of the symmetrization prescription to the type-J singleton mistreats
this subtle point and is the reason for the appearance of the
anomalous finite-dimensional module (the last term in \eqref{type j
  tff}).

To recapitulate, the heuristic prescription of symmetrization of the
character arguments works surprisingly well for type-A and type-B
models, as well as type-A$_\ell$, but we also found some
finite-dimensional discrepancy when the underlying singleton module is
a non-standard one and has more complicated structure.  This clearly
suggests that our prescription should have a more refined meaning and
asks for further investigations. One direction worth exploring would
be to analyze the type-A (or beyond: type-B, type-C, etc) higher-spin
algebra as the quotient of the universal enveloping algebra of
$so(2,d)$ by the Joseph ideal (or, respectively, more complicated
primitive ideals). By comparing the basis of such quotient space with
the lowest-weight module structure, we should be able to identify the
origin of the symmetrization and understand the finite-dimensional
mismatch in the cases beyond type-B. However, such a work is beyond
the scope of the current investigation and will be explored elsewhere.

\acknowledgments

  T.B. is grateful to N. Boulanger and C. Iazeolla for useful
  discussions on group theoretical issues discussed in the present
  work, as well as to D. Ponomarev for exchanges on
  anti-singletons. We are also grateful to an anonymous referee for
  insightful comments and suggestions. The research of T.B. and E.J.
  was supported by the National Research Foundation (Korea) through
  the grant 2014R1A6A3A04056670. The research of X.B. was supported by
  the Russian Science Foundation grant 14-42-00047 in association with
  the Lebedev Physical Institute.

\appendix

\section{Generalized Verma modules}
\label{app: detail}

Recall that the usual commutation relations of $so(2,d)$ read
\begin{equation}\label{comrels}
  [M_{AB}, M_{CD}] = i\,\Big( \eta_{BC}\,M_{AD} - \eta_{AC}\,M_{BD} -
  \eta_{BD}\, M_{AC} + \eta_{AD}\, M_{BC} \Big)\, ,
\end{equation}
where $A,B, \dots, = 0, 0', 1, \dots, d$, the generators are
antisymmetric and Hermitian, $M_{AB} = M_{AB}^\dagger = -M_{BA}$, and
$\eta := \text{diag}(-1,-1, 1, \dots, 1)$. We define
\begin{equation}
  E := M_{0'0}\, , \quad L_a^+ := M_{0a} - i M_{0'a}\, , \quad L_a^- :=
  M_{0a} + i M_{0'a}\, ,
\end{equation}
where $a,b=1,\dots,d$\,. In terms of these generators,
the above commutation relations \eqref{comrels}  can be rewritten:
\begin{equation}
  [E, L_a^\pm] = \pm L_a^\pm\, , \quad [L^-_a, L_b^+] = 2 \big(
  i\,M_{ab} + \delta_{ab} E \big)\, , \quad [M_{ab}, L_c^\pm] =
  2\,i\,\delta_{c[b}L_{a]}^\pm\, ,
  \label{usual_basis}
\end{equation}
together with the $so(d)$ subalgebra commutation relations
\begin{equation}
  [M_{ab}, M_{cd}] = i\Big( \delta_{bc} M_{ad} - \delta_{ac} M_{bd} -
  \delta_{bd} M_{ac} + \delta_{ad} M_{bc} \Big)\, .
\end{equation}
The $so(2,d)$ generalized Verma modules $\V(\Delta, \Bell)$ considered
in this work are the modules induced from finite-dimensional modules
$\mathbb{V}_{[\Delta;\, \Bell]}$ of the parabolic subalgebra 
%$so(2)\oplus iso(d) \equiv \text{span}
spanned by $E,M_{ab}$ and $L_c^-$ as follows:
\begin{itemize}
\item The finite-dimensional module $\mathbb{V}_{[\Delta;\,\Bell]}$
  carries a representation of $so(d)$ with highest weight $\Bell =
  (\ell_1, \dots, \ell_r)$ where $r=[\tfrac d2]$ is the rank of
  $so(d)$ and a (one-dimensional) representation of the $so(2)$
  algebra spanned by $E$ characterized by the weight $\Delta$. In
  other words, every element of $\mathbb{V}_{[\Delta;\,\Bell]}$ is an
  eigenvector of $E$ with eigenvalue $\Delta$. Finally, generators
  $L^-_a$ are represented trivially on
  $\mathbb{V}_{[\Delta;\,\Bell]}$, i.e. the module is annihilated by
  the action of these lowering operators.
\item The generalized Verma modules $\V(\Delta, \Bell)$ is freely
  generated by the action of the raising operators $L_a^+$, i.e. it is
  composed of elements of the form:
  \begin{equation}
    L_{a_1}^+ \dots L_{a_n}^+ \, \mathbb{V}_{[\Delta;\,\Bell]} \quad
    \in \quad \V(\Delta, \Bell) \quad \text{for} \quad n \in \N\, .
  \end{equation}
\end{itemize}

%\paragraph
\section{Weyl character formula}
\label{app:Weyl}

We make use of the notations introduced in Section \ref{sec: gen
    d} and we will give a derivation of formula \eqref{so 2+d} from
  the Weyl character formula.  This formula expresses the character
$\chi_\lambda$ of a finite-dimensional, irreducible representation of
a  complex semi-simple
Lie algebra $\mathfrak g$ as
\begin{equation}
  \chi_{\lambda} = \frac{\sum_{w\in\W}
    \varepsilon(w)\,e^{w(\lambda+\rho)-\rho}}{\prod_{\alpha \in \Phi_+}
    (1-e^{-\alpha})}\, ,
  \label{weyl_character}
\end{equation}
where $\lambda$ is the highest-weight labeling the representation,
$\W$ is the Weyl group of $\mathfrak g$, $\varepsilon(w)$ is the
signature of a Weyl group element and $\rho := \tfrac12 \sum_{\alpha
  \in \Phi_+} \alpha$ is the Weyl vector of $\mathfrak g$ defined as
the half-sum of all the positive roots (represented by the set
$\Phi_+$) of $\mathfrak g$. We are interested in $\mathfrak g =
so(2+d)$, for which the Weyl group is $\W \cong \Sn_{r+1} \ltimes
(\Z_2)^{r+1}$ for $d=2r+1$ and $\W \cong \Sn_{r+1} \ltimes
(\Z_2)^{r}$  for $d=2r$. In other words, the Weyl group acts as
the semi-direct product of the permutation group of $r+1$ elements
with a group of ``sign flips'' on the $r+1$ components of an $so(2+d)$
weight. More concretely, an element $w\in\W$ of the Weyl group
first flips the sign of a number of components of the $so(2+d)$ weight
(an arbitrary number of components for $d=2r+1$ and only an even number for $d=2r$) and then
permutes these $r+1$ components.\\

Formally, Lie algebra characters are maps from the weight space of the
algebra (which is isomorphic to the dual of the Cartan subalgebra
$\mathfrak h \subset \mathfrak g$) to the field of complex numbers:
\begin{equation}
  \chi_\lambda : \mathfrak{h}^* \rightarrow \C\, .
\end{equation}
The evaluation of expression \eqref{weyl_character} on an arbitrary
weight $\mu$ is defined through 
\begin{equation}
  e^{\lambda}(\mu) := e^{(\lambda,\, \mu)}\, ,
\end{equation}
where $(\,,)$ denotes the Killing form of $\mathfrak g$, which is
simply the Euclidean inner product on the weight space, $\mathfrak h^*
\cong \R^{r+1}$ for $so(2+d)$. As a consequence, the formula
\eqref{weyl_character}, when evaluated on a weight $\mu$, reads
\begin{equation}
  \chi_\lambda(\mu) = \frac{\sum_{w\in\W} \varepsilon(w)
    e^{(w(\lambda+\rho)-\rho,\,\mu)}}{\prod_{\alpha\in\Phi_+}\big(1-e^{-(\alpha,\,\mu)}\big)}\,
  .
\end{equation}

The Weyl character formula tells us that in order to compute the
character of a finite-dimensional, highest-weight irreducible
representation of a complex semi-simple Lie algebra $\mathfrak g$, we
should $(i)$ compute the product over the positive roots
$\prod_{\alpha\in\Phi_+} \frac{1}{1-e^{-\alpha}}$, and then $(ii)$
apply the whole Weyl group to the highest-weight $\lambda$ shifted by
the Weyl vector $\rho$. In the orthonormal basis $\uv_k$ (with $k=0,
\dots, r$) of $\R^{r+1}$, the set of positive roots of $so(2+d)$ is
given by
\begin{itemize}
\item When $d=2r$,
  \begin{equation}
    \Phi_+ = \big\{ \uv_i \pm \uv_j \quad \text{with} \quad 0
    \leqslant i < j \leqslant r \big\}\, ;
    \label{positive_roots_even}
  \end{equation}
\item When $d=2r+1$,
  \begin{equation}
    \Phi_+ = \big\{ \uv_i \pm \uv_j \quad \text{with} \quad 0
    \leqslant i < j \leqslant r \big\} \cup \big\{ \uv_k \quad
    \text{with} \quad k=0,\dots,r \big\}\, .
    \label{positive_roots_odd}
  \end{equation}
\end{itemize}
In the orthonormal basis, the components $\rho_k$ of the Weyl vector
read:
\begin{equation}
  \rho_k = \tfrac d2 - k\, , \qquad k=0,1, \dots, r\, .
\end{equation}
Notice that we have shifted the components numbering on purpose, so
that all the object defined above which do not have a $0$th component
can be reinterpreted as the same objects for the $so(d)$
subalgebra. In other words, the components $\rho_a$ for $a=1,\dots,r$
are those of the Weyl vector of $so(d)$, and the positive roots
previously enumerated which do not involve the unit vector $\uv_0$
make up the positive root system of $so(d)$ that we will denote
$\Phi_+^{so(d)}$. \\

Using \eqref{positive_roots_even} and \eqref{positive_roots_odd}, we
can express the Weyl denominator of \eqref{weyl_character} for
$so(2+d)$ in terms of the Weyl denominator of $so(d)$ as
\begin{equation}\label{DWeyl}
  \mathsf D_{\rm Weyl}^{so(2+d)}(\mu) := \prod_{\alpha \in \Phi_+}
  \frac{1}{1-e^{-(\alpha,\mu)}}\,.
\end{equation}
 In even dimensions,
$d=2r$, it becomes
\begin{eqnarray}
  \mathsf D_{\rm Weyl}^{so(2+2r)}(\mu) & = & \prod_{k=1}^r
  \frac{1}{(1-e^{-(\uv_0,\mu)}e^{-(\uv_k,\mu)})(1-e^{-(\uv_0,\mu)}e^{(\uv_k,\mu)})}
  \prod_{\alpha \in \Phi_+^{so(2r)}} \frac{1}{1-e^{-(\alpha,\mu)}}
  \\ & = & \prod_{k=1}^r \frac{1}{(1-x_0^{-1}x_k^{-1})(1-x_0^{-1}x_k)}
  \prod_{\alpha \in \Phi_+^{so(2r)}} \frac{1}{1-e^{-(\alpha,\mu)}}
  \\ & = & \Pd{2r}(x_0^{-1}, \bm x)\, \mathsf D_{\rm
    Weyl}^{so(2r)}(\mu)\, ,
  \label{den_even}
\end{eqnarray}
where we defined the formal variables $x_k := e^{\mu_k}$ for
$k=0,1,\dots, r$ and $\Pd{2r}(x_0, \bm x)$ is the function defined in
\eqref{Pd}. In odd dimensions, $d=2r+1$, a similar computation ---
taking into account the additional root $\uv_0 \in \Phi_+ \backslash
\Phi_+^{so(d)}$ with respect to the previous case --- yields the same
final result:
\begin{eqnarray}
  \mathsf D_{\rm Weyl}^{so(3+2r)}(\mu) & = &
  \frac{1}{1-e^{-(\uv_0,\mu)}}\prod_{k=1}^r
  \frac{1}{(1-e^{-(\uv_0,\mu)}e^{-(\uv_k,\mu)})(1-e^{-(\uv_0,\mu)}e^{(\uv_k,\mu)})}
  \\ && \qquad \qquad \times \prod_{\alpha \in \Phi_+^{so(2r+1)}}
  \frac{1}{1-e^{-(\alpha,\mu)}} \nonumber \\ & = & \Pd{2r+1}(x_0^{-1},
  \bm x)\, \mathsf D_{\rm Weyl}^{so(2r+1)}(\mu)\, .
  \label{den_odd}
\end{eqnarray}

\noindent Let us define
\begin{equation}\label{Clambda}
  \mathcal{C}_{\lambda} := \frac{e^{\lambda}}{\prod_{\alpha \in
      \Phi_+}(1-e^{-\alpha})}\, ,
\end{equation}
as well as the \textit{affine} action of a Weyl element $w$ on a
weight $\lambda$:
\begin{equation}
  w \cdot \lambda := w(\lambda+\rho) - \rho\, ,
\end{equation}
where $w(\lambda)$ still denote the \textit{linear} action of the Weyl
element $w$ on the weight $\lambda$.  Then we can rewrite the Weyl
character formula as
\begin{equation}
  \chi_\lambda = \sum_{w\in\W} \varepsilon(w)\,
  \mathcal{C}_{w\cdot\lambda}\, .
\end{equation}
It is furthermore possible to show that the following identity holds.
\begin{equation}
  \varepsilon(w)\, \mathcal{C}_{w\cdot\lambda} = w\big(
  \mathcal{C}_{\lambda} \big)\, ,
\end{equation}
and therefore \eqref{weyl_character} can be recasted as
\begin{equation}
  \chi_\lambda = \sum_{w\in\W} w\big( \mathcal{C}_\lambda \big)\, ,
\end{equation}
where the notation $w\big( \mathcal{C}_\lambda \big)$ represents the
action of the reflection $w$ on the variables which the final
character depends on, i.e. $\mu$. More concretely, in the case of
$\mathfrak g=so(2+d)$ of interest for us, the action of a generic
element $w\in\W$ on a weight $\lambda$ is to first flip the sign of a
number of components of $\lambda$ and then to permute those
components. To each component of the weight $\mu$, we associated a
formal variable, denoted above $x_k$ with $k=0, \dots,r$ for the
components of an $so(2+d)$ weight, which carries this component as an
exponent. As consequence, in the character formula the action of $w$
on a weight can be transfered as an operation on the variables $x_k$:
a sign flip of the $i$th component of a weight can be equivalently
represented as sending the corresponding variable $x_i$ to its inverse
$x_i^{-1}$, and the permutation of several components, say the $i$th
and the $j$th, of a weight are represented by the same permutation of
the corresponding variables $x_i$ and $x_j$. With that in mind, we can
simplify \eqref{weyl_character} by first summing on all elements of
the Weyl group of $so(d)$ (which we will denote $\W_{so(d)}$),
i.e. those reflections acting only on the last $r$ variables $x_i$
with $i=1, \dots, r$.  Using definitions \eqref{DWeyl} and
\eqref{Clambda} as well as formulae \eqref{den_even} and
\eqref{den_odd}, we can write
\begin{equation}
  \mathcal{C}^{so(2+d)}_\lambda(x_0, \bm x) = x_0^{\ell_0}\,
  \mathcal{C}_{\Bell}^{so(d)}(\bm x) \,\Pd d (x_0^{-1}, \bm x)
\end{equation}
where $\ell_0$ and $\Bell \equiv (\ell_1, \dots, \ell_r)$ are
respectively the $0$th and last $r$ components of the $so(2+d)$
highest weight $\lambda=(\ell_0,\ell_1, \dots, \ell_r)$. Considering
that the function $\Pd d (x_0^{-1}, \bm x)$ is invariant under any
$so(d)$ Weyl group element (it is unchanged under any permutation or
inversion of the variables $x_i$ with $i=1, \dots, r$), acting with
all elements of $\W_{so(d)}$ on $\mathcal{C}^{so(2+d)}_\lambda(x_0,
\bm x)$ will produce the character of the irreducible $so(d)$
representation with highest weight $\Bell$ out of the factor
$\mathcal{C}_{\Bell}^{so(d)}(\bm x)$:
\begin{equation}
  \sum_{w\in\W_{so(d)}} w\Big( \mathcal{C}^{so(2+d)}_\lambda(x_0, \bm
  x) \Big) = x_0^{\ell_0}\, \chi^{so(d)}_{\Bell}(\bm x) \Pd d
  (x_0^{-1}, \bm x)\, .
\end{equation}
After having accounted for elements of the subgroup $\W_{so(d)}$ of
$\W$, the character formula reads:
\begin{equation}
  \chi^{so(2+d)}_\lambda(x_0, \bm x) =
  \sum_{w\in\W\backslash\W_{so(d)}} w\Big( x_0^{\ell_0}
  \chi^{so(d)}_{\Bell}(\bm x) \Pd d (x_0^{-1}, \bm x) \Big)\, .
\end{equation}
Hence we need to take into account the elements of the Weyl group of
$so(2+d)$ that are not part of the subgroup $\W_{so(d)}$, i.e.
inversions of $x_0$ and permutations between $x_0$ and one of the
other variables $x_k$ for $k=1, \dots, r$. Using \eqref{prop 1}, the
character can finally be put into the same form as \eqref{so 2+d}:
\begin{equation}
  \chi^{so(2+d)}_\lambda(x_0, \bm x) = \sum_{k=0}^r \Big(
  x_k^{-\ell_0} \chi^{so(d)}_{\Bell_-}(\bm x_k) + (-)^d x_k^{\ell_0+d}
  \chi_{\Bell_+}^{so(d)}(\bm x_k) \Big) \Pd d (x_k, \bm x_k)\, ,
  \label{final_weyl}
\end{equation}
with
\begin{equation}
  \Bell_\pm \equiv (\ell_1, \dots, \ell_{r-1}, \pm \ell_r)\, ,
\end{equation}
for $d=2r$ and $\Bell_\pm = \Bell$ for $d=2r+1$. Indeed, remember that
the Weyl group for orthogonal algebras is a semi-direct product of the
group of sign flips with the group of permutations, which is why
\eqref{final_weyl} is composed a sum of two terms in which a variables
$x_k$ is singled out: those two terms correspond to the two
possibilities for $w\in\W\backslash\W_{so(d)}$, either to invert $x_0$
or not. The relative factor of $(-x_k)^d$ between those two terms
comes from the fact that the function $\Pd d (x_k, \bm x_k)$ obey
\eqref{prop 1}
\begin{equation}
  \Pd d (x_k^{-1}, \bm x_k) = (-x_k)^d \,\Pd d (x_k, \bm x_k)\, .
\end{equation}
Finally, the change of chirality from $\Bell_+$ to $\Bell_-$ in even
dimensions is due to the fact that in this case, any elements of the
Weyl group has to be composed of an \it even \rm number of sign flip
of the components of the weights. This means that if the $0$th
component is sent to minus itself (equivalently, $x_0$ is inverted),
then another of the $r$ remaining components has to also be
affected. As noticed above, the $\Pd d (q, \bm x)$ function is
invariant under any inversion of the variables $\bm x$, however one
can show that
\begin{equation}
  \chi^{so(2r)}_{\Bell_+}(x_1, \dots, x_k^{-1}, \dots, x_r) =
  \chi^{so(2r)}_{\Bell_-}(x_1, \dots, x_k, \dots, x_r)\,,
\end{equation}
i.e. inverting only one of the variables of the character of an
$so(2r)$ irreducible representation produces the character of the
$so(2r)$ irreducible representation with opposite
chirality,\footnote{Notice that this does not contradict the fact that
  the character of any irreducible representation of a compact Lie
  algebra $\mathfrak g$ is invariant under its Weyl group, as in the
  case of $so(2r)$ an element flipping the sign of an odd number of
  components of the highest weight is not part of the Weyl group
  $\W_{so(2r)}$} which explains formula \eqref{final_weyl}.

\bibliographystyle{JHEP}
\bibliography{biblio}

%\providecommand{\href}[2]{#2}\begingroup\raggedright\begin{thebibliography}{10}

%\end{thebibliography}\endgroup

\end{document}